\documentclass[3p, sort&compress, pdflatex]{elsarticle}

\usepackage{lineno,hyperref}
\modulolinenumbers[1]

\journal{Computer Methods in Applied Mechanics and Engineering}

\usepackage{dcolumn}  
\usepackage{amssymb, amsmath, amsfonts, amsthm, bm}  
\usepackage{mathtools}
\usepackage{lipsum}  
\usepackage{siunitx}  
\usepackage{accents}
\usepackage[inline]{enumitem}  
\usepackage{setspace, color, url}
\usepackage{comment}
\usepackage{here}
\usepackage{subfigure}
\usepackage{balance}
\usepackage{numprint}
\npthousandsep{\,}
\usepackage{algorithm}
\usepackage{algpseudocode}
\usepackage{todonotes}
\usepackage[normalem]{ulem}  
\usepackage{url}

\graphicspath{{./figures}}
\DeclareGraphicsExtensions{{.pdf}}

\newcommand{\bs}[1]{\boldsymbol{#1}}

\newcommand{\mdif}[1]{\dfrac{D #1}{Dt}}
\newcommand{\pdif}[2]{\dfrac{\partial #1}{\partial #2}}
\newcommand{\vect}[1]{\boldsymbol{#1}}
\newcommand{\mat}[1]{\mathbf{\uppercase{#1}}}
\newcommand{\lpnorm}[2]{\|{#1}\|_{#2}}

\newcommand{\defeq}{\vcentcolon=}

\newcommand{\rbrac}[1]{\left({#1}\right)}  
\newcommand{\sbrac}[1]{\left[{#1}\right]}  

\definecolor{deepred}{RGB}{230,0,0}

\newcommand\soutr{\bgroup\markoverwith{\textcolor{red}{\rule[.5ex]{2pt}{0.4pt}}}\ULon}
\newcommand\uliner{\bgroup\markoverwith {\textcolor{red}{\rule[-0.5ex]{2pt}{0.4pt}}}\ULon}

\begin{document}

\begin{frontmatter}

\title{4D topology optimization: Integrated optimization of the structure and self-actuation of soft bodies for dynamic motions}


\author{Changyoung Yuhn}
\author{Yuki Sato}
\author{Hiroki Kobayashi}
\author{Atsushi Kawamoto}
\author{Tsuyoshi Nomura \corref{cor}}
\ead{nomu2@mosk.tytlabs.co.jp}
\cortext[cor]{Corresponding author}
\address{Toyota Central R\&D Labs., Inc., Nagakute, Aichi 480-1192, Japan}

\begin{abstract}
Topology optimization is a powerful tool utilized in various fields for structural design.
However, its application has primarily been restricted to static or passively moving objects, mainly focusing on hard materials with limited deformations and contact capabilities.
Designing soft and actively moving objects, such as soft robots equipped with actuators, poses challenges due to simulating dynamics problems involving large deformations and intricate contact interactions.
Moreover, the optimal structure depends on the object's motion, necessitating a simultaneous design approach.
To address these challenges, we propose ``4D topology optimization,'' an extension of density-based topology optimization that incorporates the time dimension.
This enables the simultaneous optimization of both the structure and self-actuation of soft bodies for specific dynamic tasks.
Our method utilizes multi-indexed and hierarchized density variables distributed over the spatiotemporal design domain, representing the material layout, actuator layout, and time-varying actuation.
These variables are efficiently optimized using gradient-based methods.
Forward and backward simulations of soft bodies are done using the material point method, a Lagrangian--Eulerian hybrid approach, implemented on a recent automatic differentiation framework.
We present several numerical examples of self-actuating soft body designs aimed at achieving locomotion, posture control, and rotation tasks.
The results demonstrate the effectiveness of our method in successfully designing soft bodies with complex structures and biomimetic movements, benefiting from its high degree of design freedom.
\end{abstract}


\end{frontmatter}

\section{Introduction}\label{sec:intro}
The diverse physical forms of creatures can be attributed to a long evolutionary history.
Creatures have evolved to move effectively and efficiently in their surroundings, including the ground, water, and air, and their structures are closely tied to their movements~\cite{wilkinson2016restless}.
For example, humans have developed sophisticated feet that enable them to walk efficiently~\cite{cunningham2010influence}.
Their feet are composed of $26$ bones, plantar fascia, and ligaments and have a unique arch structure that provides both softness and stiffness depending on posture~\cite{venkadesan2020stiffness}.
Another example is the primate hand, which has evolved to facilitate movement.
Its opposable thumb helps primates in grasping branches, which is thought to be an adaptation to life in the trees~\cite{lemelin2007origins}.
Similarly, different shapes and positions of fish fins are suited for efficient swimming by generating vortex wakes~\cite{lauder2002forces}.
These creatures demonstrate that the structure influences the movement by changing the direction of deformation, degree of freedom, and contact angle with the environment.
Furthermore, the movement also affects the desirable structure by determining the load and direction of external forces.
Therefore, it is crucial to consider both structure and movement simultaneously when designing a dynamic object such as a soft robot.
However, this is challenging to accomplish using current optimization technologies and often requires numerous trials and errors, intuition, the expertise of experienced individuals, or biomimicry.

The co-design of structure and movement (control) of robots has been a long-standing challenge in the field of robotics, particularly for soft robots, whose flexibility offers significant design freedom in both structure and control~\cite{chen2020design}.
Nevertheless, extensive efforts have been made over the years.
Deimel~et~al.~\cite{deimel2017automated} used the particle swarm optimization method to co-design the structure and actuation of soft hands for grasping, showing that co-design consistently outperformed unilateral design, even when the design targets were swapped between the two during optimization.
Regarding soft robots for locomotion, Cheney~et~al.~\cite{cheney2013unshackling} conducted a pioneering study on co-designing the structure and actuation of soft robots for horizontal locomotion, using an evolutionary algorithm to optimize a voxel-wise combination of soft, hard, and actuating materials.
Cheney~et~al.~\cite{cheney2014evolved} further extended the method to incorporate the action potential model, which enabled the design of soft robots with more complex actuation.
More recently, Bhatia~et~al.~\cite{bhatia2021evolution} proposed an algorithm to design soft robots for various tasks, organizing the design of the structure and feedback control through neural networks as outer and inner loops, respectively.
Van~Diepen and Shea also designed the structure using voxels~\cite{vandiepen2019spatial} or geometric blocks~\cite{vandiepen2022codesign} and represented the actuation with the predefined patterns.
However, gradient-free optimization methods used in these studies have limited degrees of design freedom capacity because they explore the design space through stochastic trials.
Therefore, the studies typically limit the number of design variables with assumptions to efficiently solve the optimization problem, e.g., by coarsening the resolution of the structure or predefining a function for time-varying actuation.
Thus, there is a continued need for an integrated strategy to design the structure and control of soft robots simultaneously, with a redundant representation of both.

Topology optimization~\cite{bendsoe1988generating} has evolved into a powerful tool for determining the optimal material layout with high design freedom in various fields of structural design.
The basic idea behind topology optimization is to replace the structural optimization problem by optimizing the material distribution across the design domain.
The material distribution is usually represented by a continuous function, referred to as the normalized material density, which can be optimized to minimize or maximize a particular objective function using gradient-based methods.
The typical topology optimization process involves forward simulation to evaluate the objective function, gradient computation (or sensitivity analysis), and updating the design variables (i.e., the material distribution).
The main limitations in topology optimization result from the challenges of forward simulation and gradient computation.
Forward simulations cannot easily address large deformations and contacts with boundaries or between objects in a computationally efficient manner.
For example, in the finite element method (FEM), which is commonly used for topology optimization, large deformations often result in numerical instabilities caused by excessive mesh distortion with low stiffness~\cite{wang2014interpolation}.
Similarly, gradient computation can be difficult for highly nonlinear and sometimes discontinuous physical phenomena.
As a result, most applications of topology optimization have been limited to static objects or passively moving objects composed of hard materials, such as oscillating structures~\cite{yoon2010maximizing, silva2019critical, zhao2019efficient} and linkages~\cite{han2017topology, han2021topology}, despite ongoing efforts to extend the method~\cite{deaton2014survey, zargham2016topology}.
Applications of topology optimization to dynamic and contact problems are limited and still challenging.
In fluid applications, transient problems are dealt with laminar flows in \cite{yoon2022transient} and with unsteady incompressible Navier--Stokes flows in \cite{deng2011topology}.
As regards contact problems, advanced contact models have been introduced in \cite{fernandez2020topology} for handling large deformations, \cite{bluhm2021internal} for self-contact, and \cite{kristiansen2021topology} for contact involving frictions.

In recent years, mesh-free methods have been developed to simulate physical phenomena challenging to address by conventional mesh-based methods such as FEM.
One of these notable simulation techniques is the material point method (MPM)~\cite{sulsky1994particle}, which is a hybrid approach using Lagrangian particles and Eulerian grids.
The MPM is efficient in handling large deformations by using particles to represent objects and can automatically treat the contacts or collisions within its computational procedures~\cite{jiang2016material, vaucorbeil2020chapter}.
This makes it highly versatile, and it has been applied to a wide range of engineering fields~\cite{tao2018development, lei2021generalized, xiao2021dp, ma2022finite}, including topology optimization of structures with large static deformations~\cite{li2021lagrangian}.
The MPM has also been widely used in computer graphics, first introduced for the simulation of snow dynamics~\cite{stomakhin2013material}.
It has been used to efficiently simulate the behavior of various materials undergoing large deformations, contacts, splits, and merges~\cite{stomakhin2014augmented, jiang2016material, klar2016drucker, daviet2016semi}.
In density-based topology optimization, the MPM's Lagrangian--Eulerian hybrid nature offers advantages in addressing large deformations of void domains and design-dependent contacts.
While representing large deformations with particles, the governing equations can be solved stably using the physical quantities projected from the particles onto fixed grids.
Additionally, the particle--grid projections simplify contact handling, eliminating the need for contact boundary detection.
Thus, any particle within the design domain can serve as a contact surface, depending on the designed shape.

Furthermore, the field of numerical simulation has been revolutionized by introducing differentiable physics (DP).
DP incorporates the new differentiable programming technique into numerical simulations, allowing for more efficient gradient computation.
To apply topology optimization to dynamic problems, a backward simulation is required to compute the gradients of the objective function with respect to the design variables.
Despite the adjoint variable method's use for straightforward physical phenomena~\cite{haftka1989recent}, strong nonlinearity or discontinuity, such as large deformations and contacts, make it difficult.
In such cases, DP can be utilized to algorithmically differentiate the functions written in the simulation code, producing the gradients needed for the optimization process.
This automatic differentiation (AD) method is known as algorithmic differentiation, and it involves computing the derivative of a function by repeatedly applying the chain rule to its operations. 
DP has been demonstrated to be effective in controller optimization in the context of machine learning, and recent studies have shown that it can significantly accelerate optimization compared to model-free reinforcement learning algorithms~\cite{de2018end, hu2019chainqueen, hu2020difftaichi}.
Our recent study~\cite{sato2022topology} proposed a topology optimization method that incorporates the MPM and DP for locomoting soft bodies.
Despite the study's success in designing soft bodies that realize the target locomotion, it only focused on structural optimization and assumed a simple periodic actuation parameter and a pre-designed actuator layout.

\begin{figure}[tp]
    \centering
    \includegraphics[width=0.9\textwidth]{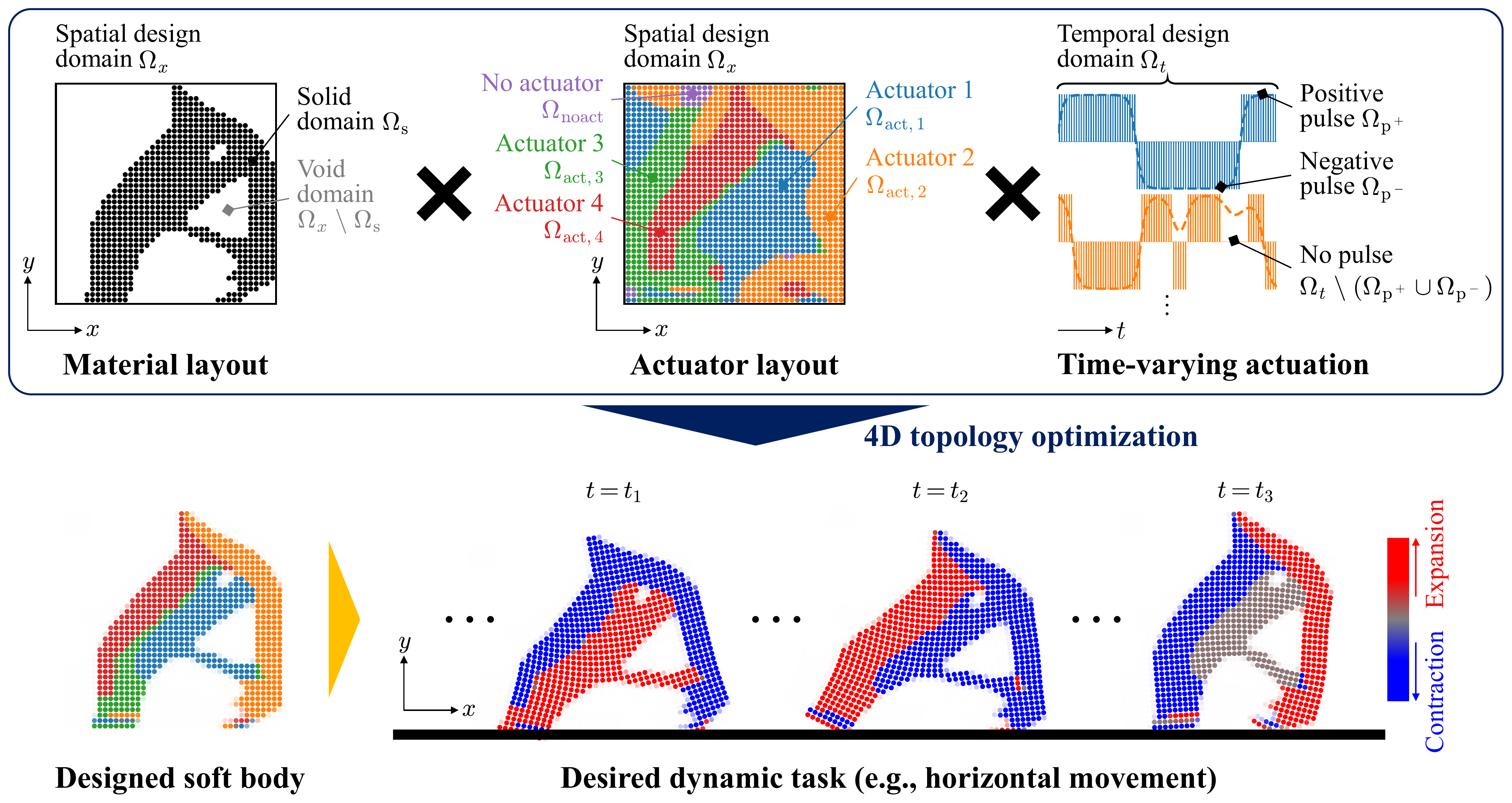}
    \caption{Schematic illustration of the four-dimensional (4D) topology optimization proposed in this study.}
    \label{fig:schematic}
\end{figure}

In this paper, we propose ``4D topology optimization,'' which extends conventional density-based topology optimization to include the time dimension (Fig.~\ref{fig:schematic}).
This method facilitates the simultaneous optimization of the structure and self-actuation of soft bodies for dynamic tasks.
In the proposed method, we consider the design of a soft body as an optimization problem in the spatiotemporal domain (4D), wherein we optimize the structure in the spatial domain (3D) and the actuation in the temporal domain (1D).
Spatial optimization involves optimizing the structure of the soft body and the placement of multiple actuators, represented by a multi-indexed density distributed over the design domain.
In the temporal domain, we optimize the time-varying actuation of each actuator, represented by a density of positive and negative pulses in the design domain.
By optimizing the spatial and temporal density distributions, we obtain the optimal structure and self-actuation of the soft body to perform a specific dynamic task with many degrees of design freedom.
The forward simulation and the associated gradient computation are performed using the MPM implemented in Taichi~\cite{hu2019taichi, hu2020difftaichi}, which is a programming language that supports high-performance computing and AD.
The optimization problem is formulated as minimizing an objective function for a target dynamic task, subject to constraints on the binarization of densities.
We solve this problem using the Adam optimizer~\cite{kingma2014adam} with the augmented Lagrangian method.
We present five numerical examples of designing soft bodies for specific dynamic tasks, including locomotion, posture control, and rotation.
The results indicate that our method is successful in designing self-actuating soft bodies with complex structures and biomimetic movements, taking advantage of the many degrees of design freedom.

The main findings of this study can be summarized as follows.
First, we developed a new method that integrates the design of the structure, actuator layout, and time-varying actuation of self-actuating soft bodies while still allowing a high degree of design freedom.
The method uses a gradient-based algorithm to optimize efficiently, even with much more design variables than the previous studies~\cite{cheney2013unshackling, cheney2014evolved, bhatia2021evolution, vandiepen2019spatial, vandiepen2022codesign, hu2020difftaichi, sato2022topology}.
As a result, we successfully designed soft bodies that dexterously utilize complex structures to perform animal-like movements.
Second, we expanded the use of topology optimization to design problems where the structure is highly deformable and has a programmed deformation over time to move through intricate contact with the environment.
Our contributions include new modeling techniques, continuous relaxation of multi-indexed density using the softmax function, and the use of MPM and DP.
The proposed 4D topology optimization is conceptually relevant to the previous studies with space--time topology optimization~\cite{jensen2009space, wang2020space}, which dealt with spatial and temporal design domains to optimize material properties varying with time.
However, our method has the novelty of introducing design variables (actuation pulse density) into the time design domain as well, dealing with the design variables in the spatial and temporal design domains interlinked by the governing equations of time evolution.
Third, we applied Adam~\cite{kingma2014adam}, a state-of-the-art gradient-based optimization method, with the augmented Lagrangian method to solve constrained optimization problems.
Despite Adam being typically limited to unconstrained problems, we have shown that when used with the augmented Lagrangian method, it can handle constraints effectively.
This approach is beneficial in not only topology optimization, where there are often many constraints, but also in fields such as machine learning, where constraints are a challenge in optimization.

The rest of the paper is structured as follows:
Section~\ref{sec:simulation} describes the governing equations and the MPM used to solve them. 
Section~\ref{sec:4dtopopt} explains the concept of 4D topology optimization.
We formulate the optimization problem of the structure and self-actuation of soft bodies, represented by the multi-indexed and hierarchized densities distributed over the spatiotemporal domain.
We then describe the method used to optimize them under some constraints.
Section~\ref{sec:numer} presents five numerical examples where self-actuating soft bodies are designed for specific dynamic tasks using the proposed method.
The dynamic tasks considered include horizontal and vertical locomotion, posture control, and rotation, covering both 2D and 3D (in the spatial dimension) problems.
Section~\ref{sec:conclusion} summarizes the main achievements, findings, and conclusions.
%
\section{Dynamics simulation}\label{sec:simulation}
%
\subsection{Governing equation}\label{subsec:gov_eq}
Applying the conservation laws of mass and momentum to a continuum leads
\begin{equation}\label{eq:mass}
    \mdif{\rho} + \rho \nabla \cdot \vect{v} = 0,
\end{equation}
\begin{equation}\label{eq:momentum}
    \rho \mdif{\vect{v}} = \nabla \cdot \bs{\sigma} + \vect{f},
\end{equation}
where $t$ is the time, $\rho$ the density, $\vect{v}$ the velocity, $\bs{\sigma}$ the Cauchy stress tensor describing the internal force, $\vect{f}$ the external force, and the operator $D(\cdot)/Dt = \partial(\cdot)/\partial t + \vect{v} \cdot \nabla(\cdot)$ denotes the material derivative.
We assume that the Cauchy stress tensor in Eq.~\eqref{eq:momentum} is classified into two parts:
\begin{equation}\label{eq:cauchy}
    \bs{\sigma} = \bs{\sigma}_{\mathrm{mat}} + \bs{\sigma}_{\mathrm{act}},
\end{equation}
where $\bs{\sigma}_{\mathrm{mat}}$ is the stress related to strain through the constitutive equation for a material, and $\bs{\sigma}_{\mathrm{act}}$ is the offset stress to induce the deformation caused by actuation force.

Given the deformation gradient $\mat{F}$ and $J \defeq \det{\mat{F}}$, $\bs{\sigma}_{\mathrm{mat}}$ can be written as
\begin{equation}\label{eq:cauchy_mat}
    \bs{\sigma}_{\mathrm{mat}} = \dfrac{1}{J} \pdif{\Psi}{\mat{F}} \mat{F}^\top,
\end{equation}
where $\Psi$ denotes the potential energy.
A solid (soft body) is expected to be an isotropic and hyperelastic material, similar to rubber, and the neo-Hookean solid model is adopted to describe material properties.
For a neo-Hookean solid, the potential energy is given as follows:
\begin{equation}\label{eq:neohookean}
    \Psi(\mat{F}) = \dfrac{\mu}{2} \left( \mathrm{tr} (\mat{F}^\top \mat{F}) - d \right)
    - \mu \log{J} + \dfrac{\lambda}{2} \log^2{J},
\end{equation}
where $d$ ($= 2$ or $3$) is the spatial dimension, $\lambda$ and $\mu$ are the first and second Lam\'{e} constants, respectively.
Lam\'{e} constants are related to the Young's modulus $E$ and Poisson's ratio $\nu$ as follows:
\begin{subequations}\label{eq:lame}
\begin{alignat}{1}
    \lambda & = \dfrac{E \nu}{(1 + \nu)(1 - 2\nu)}, \\
    \mu & = \dfrac{E}{2(1 + \nu)}.
\end{alignat}
\end{subequations}
The stress $\bs{\sigma}_{\mathrm{mat}}$, obtained by substituting Eq.~\eqref{eq:neohookean} into Eq.~\eqref{eq:cauchy_mat}, exhibits a nonlinear relationship with strain at large deformation.

In addition, we assume $\bs{\sigma}_{\mathrm{act}}$ varies with time in the following form~\cite{hu2019chainqueen}:
\begin{equation}\label{eq:cauchy_act}
    \bs{\sigma}_{\mathrm{act}}(t) = - u(t) \, \mat{F} \mat{S} \mat{F}^\top,
\end{equation}
where $u(t)$ is the time-varying actuation signal, and $\mat{S}$ is the second Piola--Kirchhoff stress, which we assume to be the identity matrix of order $d$.
The actuation signal, $u(t)$, determines the sign and strength of actuation over time and is subject to design under parameterization as described in Sections~\ref{subsubsec:actuator} and \ref{subsubsec:actuation}.
Eq.~\eqref{eq:cauchy_act} applies an offset to the stress $\bs{\sigma}$ without changing the volume (i.e., strain), thereby generating an elastic restoring force.
For example, when $u(t) > 0$, $\bs{\sigma}_{\mathrm{act}}(t)$ adds an isotropic contraction stress offset to $\bs{\sigma}$, representing the current shape as being under compression.
Accordingly, this induces a future isotropic expansion of the solid due to its elastic restoring force, which can be regarded as an expansion by actuation.

For an external force, we only consider the gravity force $\vect{f} = \rho \vect{g}$ with the gravity acceleration $\vect{g} = [0, -9.8, 0]^{\top}$~\si{m.s^{-2}}.
Furthermore, when $d=3$, we assume the $y$-axis to be the axis parallel to gravity and the $xz$-plane to be perpendicular to gravity.
%
\subsection{Material point method}\label{subsec:mpm}
%
\subsubsection{Overview of computational procedures}\label{subsubsec:procedure}
\begin{figure}[tp]
    \centering
    \includegraphics[width=1\textwidth]{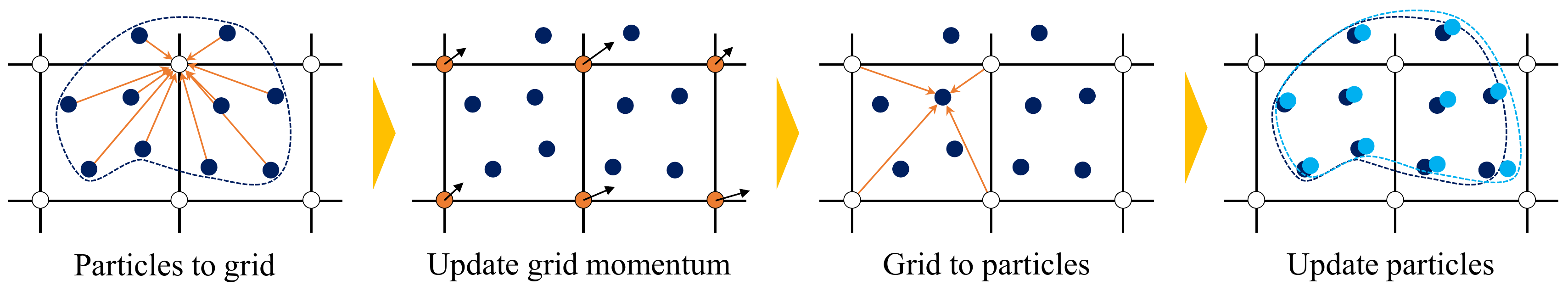}
    \caption{Schematic representation of the computational procedure of the material point method.}
    \label{fig:mpm}
\end{figure}

The governing equations of Eqs.~\eqref{eq:mass} and \eqref{eq:momentum} are solved using the MPM, which is a Lagrangian--Eulerian hybrid numerical scheme.
In the MPM, a solid is represented by a set of particles, and the governing equations are solved using the state variables at the background grid nodes.
The computational procedure for the time integration, using the symplectic Euler method, is outlined as follows (Fig.~\ref{fig:mpm}).
\begin{enumerate}[label=(\roman*), align=left, leftmargin=*, widest=iii, itemsep=0mm]
    \item \textbf{Particles to grid.}
    Transfer the mass and momentum of the particles representing the solid to the background grid nodes using a prescribed shape function.
    \item \textbf{Update grid momentum.}
    Solve the governing equations based on the mass and momentum at the grid nodes and obtain the momentum at the next time step for each of the grid nodes.
    The Dirichlet boundary condition can be enforced here by modifying the momentum at the grid nodes (Section~\ref{subsubsec:boundary}).
    \item \textbf{Grid to particles.}
    Transfer the updated momentum at the grid nodes back to the particles.
    \item \textbf{Update particles.}
    Update the position and velocity of each particle for the next time step according to the updated momentum.
\end{enumerate}

The MPM combines the merits of the Lagrangian approach, which can easily express the large deformation of soft bodies with particles, and the Eulerian approach, allowing for accurate and efficient computation of the time integration of the governing equations.
This hybrid nature is particularly advantageous in topology optimization because it contributes to stability in handling large deformations of void domains.
Even if the void particles are distorted or arranged in disorder, the governing equations can be solved stably using the non-moving grid nodes that contain the physical quantities transferred from the particles.
Furthermore, the MPM has advantageous features for soft body simulation in that it can automatically handle collisions between a solid and rigid boundary or between solids and extreme topological changes involving the split or merge of solids within the above computational procedures~\cite{jiang2016material, vaucorbeil2020chapter}.

In this study, we used the moving least square material point method (MLS-MPM)~\cite{hu2018moving}, which uses moving least squares for the local interpolation of the discretized state variables.
The simulation using the MLS-MPM was implemented in Taichi~\cite{hu2019taichi, hu2020difftaichi}.
Taichi is an open-source programming language for high-performance computing and supports parallel computing on multiple CPU cores or on GPU.
It also provides a noteworthy feature that supports AD.
Taichi can be embedded in Python, despite few unique implementation rules.
%
\subsubsection{Boundary condition}\label{subsubsec:boundary}
In the MPM, solid collisions with rigid boundaries can be accounted for by applying boundary conditions through a background Eulerian grid~\cite{jiang2016material, hu2018moving, vaucorbeil2020chapter}.
The boundary condition applies only to grid nodes inside the rigid boundary (e.g., floor and wall).
Let $\vect{v}_\mathrm{g}$ denote the grid node velocity, $\vect{v}_\mathrm{r}$ the rigid boundary velocity, and $\vect{n}_\mathrm{r}$ the outward-pointing normal vector of the rigid boundary.
The relative velocity of solids to the rigid boundary is given as $\vect{v}_\mathrm{rel} = \vect{v}_\mathrm{g} - \vect{v}_\mathrm{r}$.
If solids are approaching the rigid boundary ($\vect{v}_\mathrm{rel} \cdot \vect{n}_\mathrm{r} < 0$), the normal component of $\vect{v}_\mathrm{rel}$ is corrected to $0$ while the tangential component of $\vect{v}_\mathrm{rel}$ is decreased subject to the Coulomb friction:
\begin{equation}\label{eq:boundary}
    \tilde{\vect{v}}_\mathrm{g} =
    \begin{cases}
        \vect{v}_\mathrm{r} +
        \max(0, \lpnorm{\vect{v}_\mathrm{rel, t}}{2} - \mu_\mathrm{f} \lpnorm{\vect{v}_\mathrm{rel, n}}{2})
        \dfrac{ \vect{v}_\mathrm{rel, t} }{ \lpnorm{\vect{v}_\mathrm{rel, t}}{2} }
        & \text{if } \vect{v}_\mathrm{rel} \cdot \vect{n}_\mathrm{r} < 0, \\
        \vect{v}_\mathrm{g} & \text{otherwise},
    \end{cases}
\end{equation}
where $\tilde{\vect{v}}_\mathrm{g}$ is the velocity at the grid node after applying the boundary condition, $\vect{v}_\mathrm{rel, n} = (\vect{v}_\mathrm{rel} \cdot \vect{n}_\mathrm{r}) \, \vect{n}_\mathrm{r}$ and $\vect{v}_\mathrm{rel, t} = \vect{v}_\mathrm{rel} - \vect{v}_\mathrm{rel, n}$ are the normal and tangential component of $\vect{v}_\mathrm{rel}$, respectively, $\lpnorm{\cdot}{2}$ denotes the $l_2$-norm of a vector, and $\mu_\mathrm{f} \geq 0$ is the dynamic friction coefficient.
Eq.~\eqref{eq:boundary} is applied in the procedure ``Update grid momentum'' (Fig.~\ref{fig:mpm}) by correcting the momentum at the relevant grid nodes.

Assuming a very large friction coefficient ($\mu_{\mathrm{f}} \to \infty$), Eq.~\eqref{eq:boundary} can be rewritten into a no-slip condition:
\begin{equation}\label{eq:boundary_high_fric}
    \tilde{\vect{v}}_\mathrm{g} =
    \begin{cases}
        \vect{v}_\mathrm{r} & \text{if } \vect{v}_\mathrm{rel} \cdot \vect{n}_\mathrm{r} < 0, \\
        \vect{v}_\mathrm{g} & \text{otherwise}.
    \end{cases}
\end{equation}
In our numerical examples (Section~\ref{sec:numer}), we align the rigid floors and walls with the grid and use Eq.~\eqref{eq:boundary_high_fric} for simplicity.
The floors and walls are assumed to be stationary ($\vect{v}_\mathrm{r} = \vect{0}$), except for the \textit{Balancer} example (Section~\ref{subsec:2d_balancer}).
In the \textit{Balancer} example, $\tilde{\vect{v}}_\mathrm{g} = \vect{v}_\mathrm{r}$ is applied without conditional branching because the soft body is supposed to be embedded in the floor.

Collision treatment through the background grid favors topology optimization with design-dependent boundary conditions.
Applying boundary conditions to the grid allows all particles in the design domain to be subject to collision depending on the designed shape, regardless of whether they were positioned on the surface in the initial shape.
Void particles, introduced in density-based topology optimization, also collide with rigid boundaries but hardly interfere with solid particle collisions, as they are largely compressible.
In addition, void particle collisions have a negligible impact on the motion of the solid object due to their softness and small mass.

Eqs.~\eqref{eq:boundary} and \eqref{eq:boundary_high_fric} can be differentiated using AD, although the resulting derivative may exhibit discontinuities due to conditional branching.
Notably, when $\vect{v}_\mathrm{rel} \cdot \vect{n}_\mathrm{r} < 0$, Eq.~\eqref{eq:boundary_high_fric} replaces the grid velocity with $\tilde{\vect{v}}_\mathrm{g} = \vect{v}_\mathrm{r}$.
In the computational procedure for AD, this causes the new $\tilde{\vect{v}}_\mathrm{g}$ to become independent of the original $\vect{v}_\mathrm{g}$, thereby breaking the chain rule.
However, owing to the characteristic of the MPM, where the state variables of a single particle are distributed over multiple grid nodes and then back to the particle, most dependencies between variables persist across time steps, regardless of velocity overwriting at certain grid nodes.
%
\subsubsection{Memory efficient method for automatic differentiation}\label{subsubsec:checkpointing}
Applying AD to a time-evolving problem requires storing the computation history of all variables for the entire time step, which quickly overwhelms the memory resources.
Checkpointing is a technique used to save memory by re-computing the required variables during AD instead of storing the entire history.
The fundamental idea of checkpointing is to divide the whole time interval into multiple subintervals and store state variables only at the checkpoints of each subinterval.
Dependencies of state variables between adjacent checkpoints are restored by rerunning the forward simulation to acquire all state variables across one subinterval.
Various checkpointing algorithms have been proposed to reduce memory usage while suppressing the increase in cost for re-computing~\cite{griewank2000algorithm, yamaleev2010local, hascoet2013tapenade}.

Without checkpointing, the MPM simulation with AD requires a much memory to maintain the state variables on particles and grid nodes for all time steps.
Hu et~al.~\cite{hu2020difftaichi} presented two checkpointing schemes suited for the MPM.
One is to store only the state variables of the particles in memory and re-compute for the grid nodes (redo ``Particles to grid'' and ``Update grid momentum'' in Fig.~\ref{fig:mpm}), which saves memory proportional to the number of the grid nodes.
The other scheme is to divide the total $N_t$ time steps into segments of $S_t$ steps and re-compute for the segments (redo the entire procedure in Fig.~\ref{fig:mpm}).
In this case, the amount of required memory is only $\mathcal{O}(S_t + N_t / S_t)$.
Furthermore, the original required amount of memory $\mathcal{O}(N_t)$ can be reduced to $\mathcal{O}(\sqrt{N_t})$, by setting $S_t = \mathcal{O}(\sqrt{N_t})$.

To significantly reduce memory usage, we combine the two methods, i.e., storing only the state variables at the particles for each of the first time steps of the divided segments.
Although this requires re-computing the entire MPM procedures with the duplication of ``Particles to grid'' and ``Update grid momentum,'' for all segments, the increased computational cost remains $\mathcal{O}(N_t)$.
Such characteristics make the method favorable for simulations involving a large computational domain or many time steps but a moderate number of particles, e.g., when the soft body travels a long distance relative to its size or when the time step size needs to be small because of the large Young's modulus of the material.

A possible alternative is the local-in-time adjoint-based methods~\cite{yamaleev2010local, yaji2018large}, which compute the gradients in each of the divided subintervals and approximate the gradients over the entire time interval as a combination (e.g., average) of those.
The local-in-time methods can save memory usage while keeping the computational cost low.
However, they cannot capture the contributions of variables across subintervals.
In contrast, our checkpointing method does not cause a break in the chain rule because of memory savings, providing the same gradient as retaining all state quantities over the entire time interval on the memory.
%
\section{4D topology optimization}\label{sec:4dtopopt}
Let us consider the design of a self-actuating soft body that can execute a certain dynamic task such as locomotion.
The necessary steps to realize such a design using a gradient-based optimization method are as follows:
\begin{itemize}[itemsep=0mm]
    \item \textbf{Parameterization.} Represent both the structure and self-actuation of the soft body with design variables.
    \item \textbf{Defining objective function.} Define an objective function that reflects the performance for the dynamic task considered.
    \item \textbf{Gradient computation (or sensitivity analysis).} Compute the gradients (sensitivities) of the objective function to the design variables by dynamics simulation based on the governing equations.
    \item \textbf{Optimization.} Optimize the design variables under constraints to minimize or maximize the objective function based on the gradients.
\end{itemize}
In particular, parameterizing the structure and self-actuation on the same framework is crucial in achieving simultaneous optimization.
However, it is a challenging issue because of the enormous degrees of design freedom.

We propose to optimize both the structure and self-actuation of the soft body in the context of density-based topology optimization.
We regard the design problem as the optimization of the structures in the spatial and temporal design domain, i.e., the optimization of the 4D structure (Fig.~\ref{fig:schematic}).
In the spatial domain, we optimize the structure of a soft body and the layout of multiple actuators, which are represented as the multi-indexed density distributed over the design domain.
Meanwhile, in the temporal domain, we optimize the time-varying actuation of each actuator, which is represented as the density of sequential pulses.
The superposition of the material layout, actuator layout, and time-varying actuation represents the soft body with an optimized structure and movement schedule for a desired dynamic task.
The following subsections describe the details of the proposed framework, ``4D topology optimization,'' including the methods involved in the aforementioned steps.
%
\subsection{Structure in spatial domain}\label{subsec:spatial}
%
\subsubsection{Material layout}\label{subsubsec:material}
Let us assume a fixed spatial design domain $\mathrm{\Omega}_x$ in which solid materials can exist.
The design domain $\mathrm{\Omega}_x$ consists of a solid domain $\mathrm{\Omega}_{\mathrm{s}}$ and the complementary domain $\mathrm{\Omega}_x \setminus \mathrm{\Omega}_{\mathrm{s}}$ representing a void.
Following the concept of topology optimization~\cite{bendsoe1988generating}, we introduce a characteristic function $\chi_\mathrm{mat}(\vect{x})$, defined as
\begin{equation}\label{eq:chi_mat}
    \chi_\mathrm{mat}(\vect{x}) =
    \begin{cases}
        1 & \text{if } \vect{x} \in \mathrm{\Omega}_{\mathrm{s}}, \\
        0 & \text{if } \vect{x} \in \mathrm{\Omega}_x \setminus \mathrm{\Omega}_{\mathrm{s}},
    \end{cases}
\end{equation}
where $\vect{x}$ denotes a position in $\mathrm{\Omega}_x$.
Following this, the optimization problem of the solid structure can be replaced by the problem of optimizing the material distribution represented by $\chi_\mathrm{mat}(\vect{x})$.
Because $\chi_\mathrm{mat}(\vect{x})$ is a discrete function that takes $1$ (solid) or $0$ (void), it needs to be relaxed to a continuous function with interpolated material properties to apply gradient-based optimization methods.
We adopt a density-based approach~\cite{bendsoe2003topology}, where $\chi_\mathrm{mat}(\vect{x})$ is relaxed to a continuous scalar function, $\gamma(\vect{x}) \in [0, 1]$.
We specifically discretize the design domain $\mathrm{\Omega}_x$ with a finite number of particles and introduce a fictitious material density $\gamma^{(i)} \defeq \gamma(\vect{x}^{(i)})$ for particles.
Here, $\vect{x}^{(i)}$ denotes the position vector of the $i$-th particle for $i = 1, 2, \ldots, N_{\mathrm{par}}$, where $N_{\mathrm{par}}$ is the number of particles in $\mathrm{\Omega}_x$.

Material properties at each particle in $\mathrm{\Omega}_x$ are interpolated as a function of $\gamma^{(i)}$.
The mass density at initial volume without expansion or compression is interpolated as
\begin{equation}\label{eq:rho_particle}
    \rho^{(i)} = \{ (1 - \varepsilon) \gamma^{(i)} + \varepsilon \} \, \rho_0,
\end{equation}
where $\rho_0$ is the density of the solid material, and $\varepsilon$ is a small constant ($10^{-5}$ in this study) introduced to avoid the numerical instability of zero-density particles flying off with a very large velocity.
Additionally, the Young's modulus $E$ at each particle is interpolated as
\begin{equation}\label{eq:E_particle}
    E^{(i)} = \{ (1 - \varepsilon) \gamma^{(i)} + \varepsilon \} \, E_0,
\end{equation}
where $E_0$ is the Young's modulus of the material.
Eq.~\eqref{eq:E_particle} leads the interpolation of the Lam\'{e} constants in Eq.~\eqref{eq:neohookean} as
\begin{subequations}\label{eq:lame_particle}
\begin{alignat}{1}
    \lambda^{(i)} & = \{ (1 - \varepsilon) \gamma^{(i)} + \varepsilon \} \, \lambda_0, \\
    \mu^{(i)} & = \{ (1 - \varepsilon) \gamma^{(i)} + \varepsilon \} \, \mu_0,
\end{alignat}
\end{subequations}
where $\lambda_0$ and $\mu_0$ are the first and second Lam\'{e} constants of the material, respectively.
The interpolated material properties are used to solve the governing equations.

\begin{figure}[tp]
    \centering
    \includegraphics[width=0.45\textwidth]{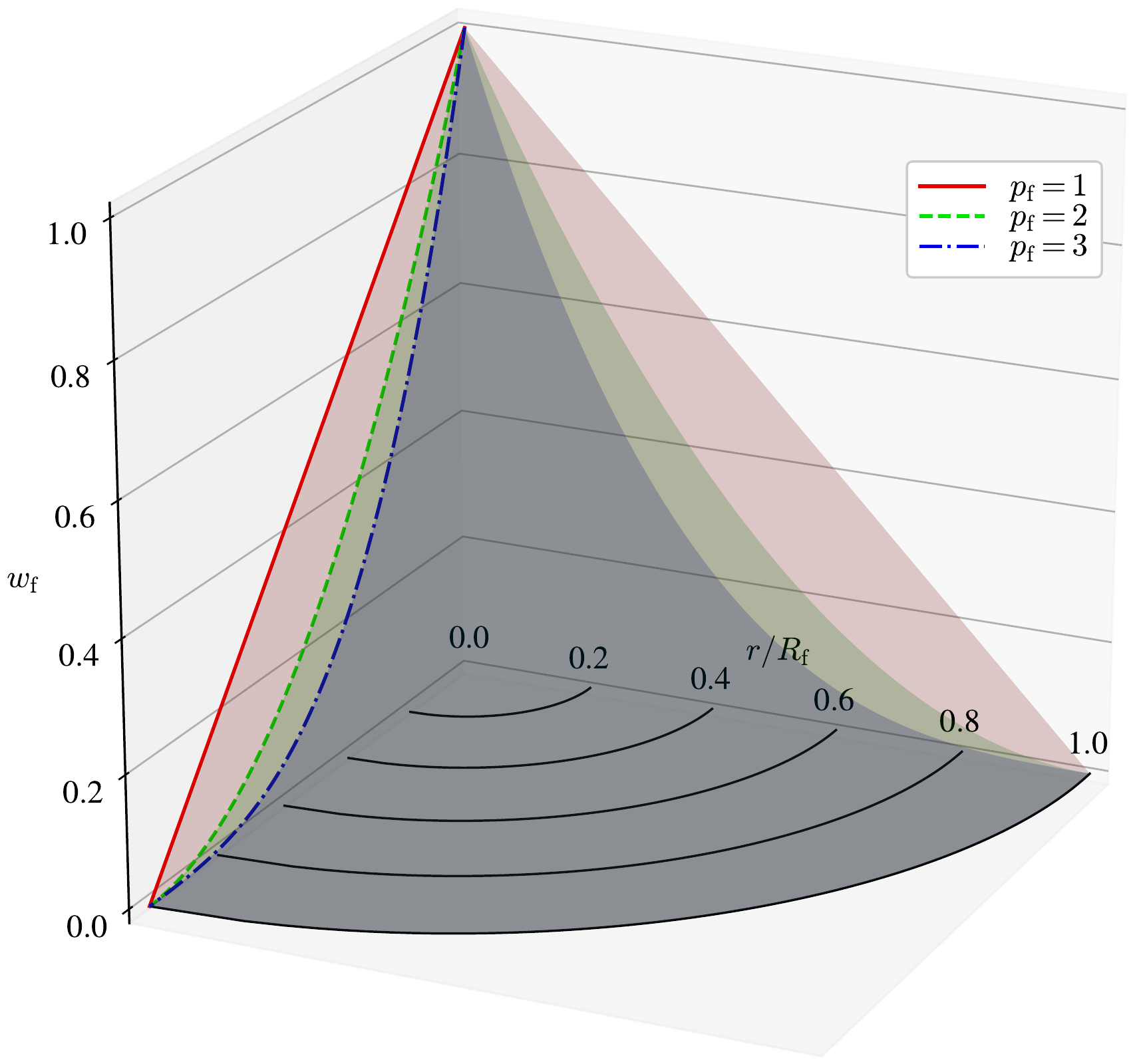}
    \caption{Visualization of the weighting function ($w_\mathrm{f}$) of the particle-wise filter with different values of $p_\mathrm{f}$. Here, $r$ is the distance between two particles, and $R_\mathrm{f}$ is a filter radius.}
    \label{fig:filter}
\end{figure}

Optimization of $\vect{\gamma} \defeq [\gamma^{(1)}, \gamma^{(2)}, \ldots, \gamma^{(N_\mathrm{par})}]$ in a discrete fashion leads to spatial discontinuities typified by checkerboard and mille-feuille patterns.
These patterns result from creating microstructures to control local stiffness and anisotropy, which is undesirable for spatial resolution dependency and manufacturing difficulty.
In topology optimization, a filtering scheme is commonly used to ensure spatial smoothness of the optimal structure~\cite{bourdin2001filters}.
There are several types of widely used filters, including convolution filters~\cite{bendsoe2003topology} and Helmholtz-type filters~\cite{kawamoto2011heaviside, lazarov2011filters}, which share the underlying concept of spatially smoothing the discretized density.
In this study, we applied a particle-wise convolution filtering scheme, where the fictitious material density is averaged between the neighborhood particles.
Let $\vect{\phi} \defeq [\phi^{(1)}, \phi^{(2)}, \ldots, \phi^{(N_\mathrm{par})}] \in [-1, 1]^{N_\mathrm{par}}$ denote a vector of design variables assigned to particles to determine $\vect{\gamma}$.
Then, the design variables are locally smoothed between particles whose distance is smaller than the filter radius, as follows:
\begin{equation}\label{eq:mat_filter}
    \tilde{\phi}^{(i)} = \dfrac{
        \sum_{i^\prime=1}^{N_\mathrm{par}}{w_\mathrm{f}( \lpnorm{\vect{x}^{(i^\prime)} - \vect{x}^{(i)}}{2} ) \, \phi^{(i^\prime)}}
    }{
        \sum_{i^\prime=1}^{N_\mathrm{par}}{w_\mathrm{f}( \lpnorm{\vect{x}^{(i^\prime)} - \vect{x}^{(i)}}{2} )}
    },
\end{equation}
where $\tilde{\phi}^{(i)} \in [-1, 1]$ is the filtered design variable of the $i$-th particle.
The weighting function $w_\mathrm{f}(\cdot)$ is defined as
\begin{equation}\label{eq:filter_weight}
    w_\mathrm{f}(r) = \rbrac{ 1 - \dfrac{\min(r, R_\mathrm{f})}{R_\mathrm{f}} }^{p_\mathrm{f}},
\end{equation}
where $R_\mathrm{f}$ is a filter radius, and $p_\mathrm{f}$ is a parameter determining the decay of the weight with respect to $r$ (Fig.~\ref{fig:filter}).
With $p_\mathrm{f} = 1$, the weight decreases linearly exhibiting a cone shape; the larger the value of $p_\mathrm{f}$ than $1$, the more rapidly the weight decays with an increase in $r$.
The weight is zero for particles outside the filter radius ($r > R_\mathrm{f}$).

\begin{figure}[tp]
    \centering
    \includegraphics[width=0.5\textwidth]{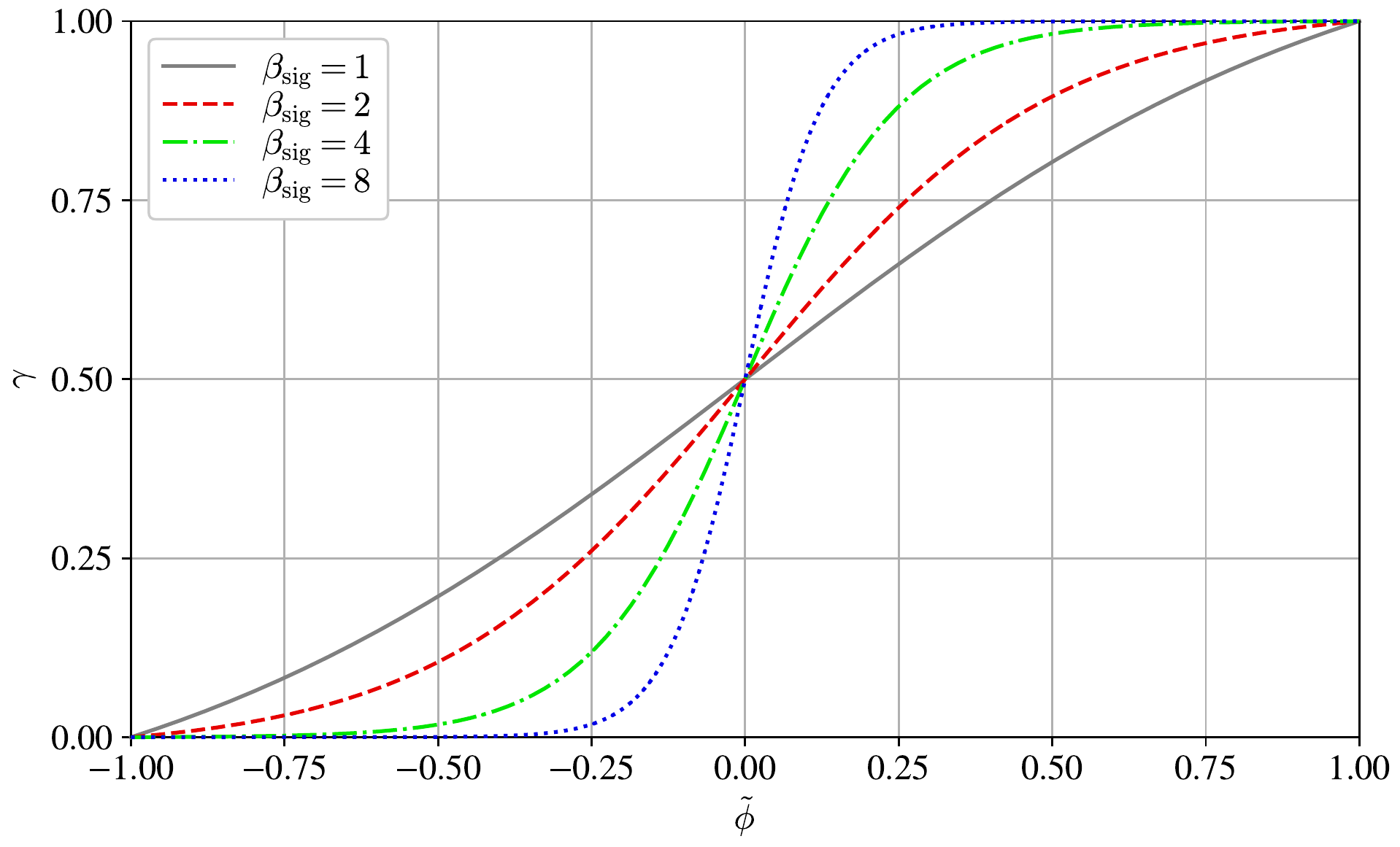}
    \caption{Visualization of the normalized sigmoid function with different values of shape parameter ($\beta_\mathrm{sig}$).}
    \label{fig:sigmoid}
\end{figure}

The optimized material layout may have regions of intermediate densities because of the continuous relaxation.
Moreover, applying the filtering scheme inevitably increases the intermediate-valued regions because it blurs the interface between solid and void.
Therefore, an additional projection scheme is applied to $\tilde{\phi}^{(i)}$ to allow a sharp transition between upper and lower limits.
We use the normalized sigmoid function for projection, where $\tilde{\phi}^{(i)} \in [-1, 1]$ is projected to $\gamma^{(i)} \in [0, 1]$ as
\begin{equation}\label{eq:sigmoid}
    \gamma^{(i)} = \dfrac{1}{2} \rbrac{ \dfrac{\tanh{( \beta_\mathrm{sig} \tilde{\phi}^{(i)} )}}{\tanh{( \beta_\mathrm{sig}} )} + 1 }.
\end{equation}
Here, $\beta_\mathrm{sig}$ is a shape parameter that adjusts the curvature of the sigmoid curve (Fig.~\ref{fig:sigmoid}).
%
\subsubsection{Actuator layout}\label{subsubsec:actuator}
Suppose that there are $N_\mathrm{act}$ actuators that are controlled by different actuation signals $\hat{u}_j(t)$ for $j = 1, 2, \ldots, N_\mathrm{act}$.
We now consider designing the layout of these actuators in $\mathrm{\Omega}_x$, in addition to the material layout.
The characteristic function representing the actuator layout can be defined as follows:
\begin{equation}\label{eq:chi_act}
    \chi_\mathrm{act}(\vect{x}) =
    \begin{cases}
        j & \text{if } \vect{x} \in \mathrm{\Omega}_{\mathrm{act}, j}, \\
        N_\mathrm{act} + 1 & \text{if } \vect{x} \in \mathrm{\Omega}_{\mathrm{noact}},
    \end{cases}
\end{equation}
where $\mathrm{\Omega}_{\mathrm{act}, j}$ is the domain of the $j$-th actuator, and $\mathrm{\Omega}_{\mathrm{noact}} \defeq \mathrm{\Omega}_x \setminus \bigcup_{j=1}^{N_\mathrm{act}}{\mathrm{\Omega}_{\mathrm{act}, j}}$ is the complementary domain that represents the domain without any actuators.
Optimizing $\chi_\mathrm{act}(\vect{x})$ in $\mathrm{\Omega}_x$, discretized by $N_\mathrm{par}$ particles, poses an integer programming problem with $(N_\mathrm{act}+1)^{N_\mathrm{par}}$ combinations and is intractable.
To efficiently handle the optimization problem using a gradient-based method, we first express the integer output of $\chi_\mathrm{act}(\vect{x})$ as a one-hot vector of dimension $N_\mathrm{act}+1$.
A one-hot vector, commonly used to represent categorical variables, is a binary vector with only one component having the value $1$ and the others having the value $0$.
For example, $[0, 1, 0, \ldots, 0]^\top$ expresses the integer ``$2$'' that represents the placement of the second actuator.
We then relax the design domain of actuators to the continuous space spanned by the one-hot vectors.
In the relaxed design domain, the actuation signal is given as a weighted linear combination of all candidate signals:
\begin{equation}\label{eq:u_relaxed}
    \bar{u}^{(i)}(t) = (\vect{\xi}^{(i)})^\top \hat{\vect{u}}(t) = \sum_{j=1}^{N_\mathrm{act}+1}{\xi^{(i)}_j \hat{u}_j(t)},
\end{equation}
where $\bar{u}^{(i)}$ is the relaxed actuation signal at the $i$-th particle; $\vect{\xi}^{(i)} \in [0, 1]^{N_\mathrm{act}+1}$ is a weight vector, which can be viewed as a multi-indexed density of multiple actuators; and $\hat{\vect{u}}$ is the vector of candidate actuation signals.
By definition of $\chi_\mathrm{act}(\vect{x})$, $\hat{u}_{N_\mathrm{act}+1}(t) = 0$ (no actuator).
Finally, we substitute $\bar{u}^{(i)}$ into the actuation signal $u$ in Eq.~\eqref{eq:cauchy_act} after interpolating it as a function of $\gamma^{(i)}$:
\begin{equation}\label{eq:u_particle}
    u^{(i)}(t) = \{ (1 - \varepsilon) (\gamma^{(i)})^3 + \varepsilon \} \, \bar{u}^{(i)}(t).
\end{equation}

Eq.~\eqref{eq:u_particle} relates the actuator layout to the material layout, attenuating the actuation on void particles.
Furthermore, the actuation strength relative to the interpolated density (i.e., mass) or Young's modulus is weakened at $0 < \gamma^{(i)} < 1$ (compare the order of $\gamma^{(i)}$ in Eqs.~\eqref{eq:u_particle} and \eqref{eq:rho_particle}).
Similar to the solid isotropic material with penalization (SIMP) method~\cite{bendsoe1999material} in the minimum compliance problem, penalizing the actuation strength reduces the contribution of particles with intermediate $\gamma^{(i)}$ to the mobility of the soft body.
This avoids numerical instability and also steers the optimized solution to $\gamma^{(i)} = 1$ (solid) or $\gamma^{(i)} = 0$ (void), although the latter effect may depend on the expected dynamic task.

Optimization of the actuator densities, $\mat{\Xi} \defeq [\vect{\xi}^{(1)}, \vect{\xi}^{(2)}, \ldots, \vect{\xi}^{(N_\mathrm{par})}] \in [0, 1]^{(N_\mathrm{act}+1) \times N_\mathrm{par}}$, allows particles to select a suitable actuator independently, which may result in a severe discontinuity of the optimized actuator layout.
This not only leads to numerical instability but also significantly decreases the manufacturing feasibility of the optimized solution.
To avoid such issues, we apply the filtering scheme to the actuator densities to loosely constrain the scale of the actuator domain from being too small.
Let $\mat{Z} \defeq [\vect{\zeta}^{(1)}, \vect{\zeta}^{(2)}, \ldots, \vect{\zeta}^{(N_\mathrm{par})}] \in \mathbb{R}^{(N_\mathrm{act}+1) \times N_\mathrm{par}}$ denote a matrix of design variables that determines $\mat{\Xi}$.
In the filtering scheme, $\mat{Z}$ is locally smoothed between particles as
\begin{equation}\label{eq:act_filter}
    \tilde{\vect{\zeta}}^{(i)} = \dfrac{
        \sum_{i^\prime=1}^{N_\mathrm{par}}{w_\mathrm{f}( \lpnorm{\vect{x}^{(i^\prime)} - \vect{x}^{(i)}}{2} ) \, \vect{\zeta}^{(i^\prime)}}
    }{
        \sum_{i^\prime=1}^{N_\mathrm{par}}{w_\mathrm{f}( \lpnorm{\vect{x}^{(i^\prime)} - \vect{x}^{(i)}}{2} )}
    },
\end{equation}
where $\tilde{\vect{\zeta}}^{(i)}$ is the filtered design variables of the $i$-th particle, and $w_\mathrm{f}(\cdot)$ is the weighting function (Eq.~\eqref{eq:filter_weight}) whose weight decays with the distance between the $i$-th and $i^\prime$-th particles.

\begin{figure}[tp]
    \centering
    \includegraphics[width=1\textwidth]{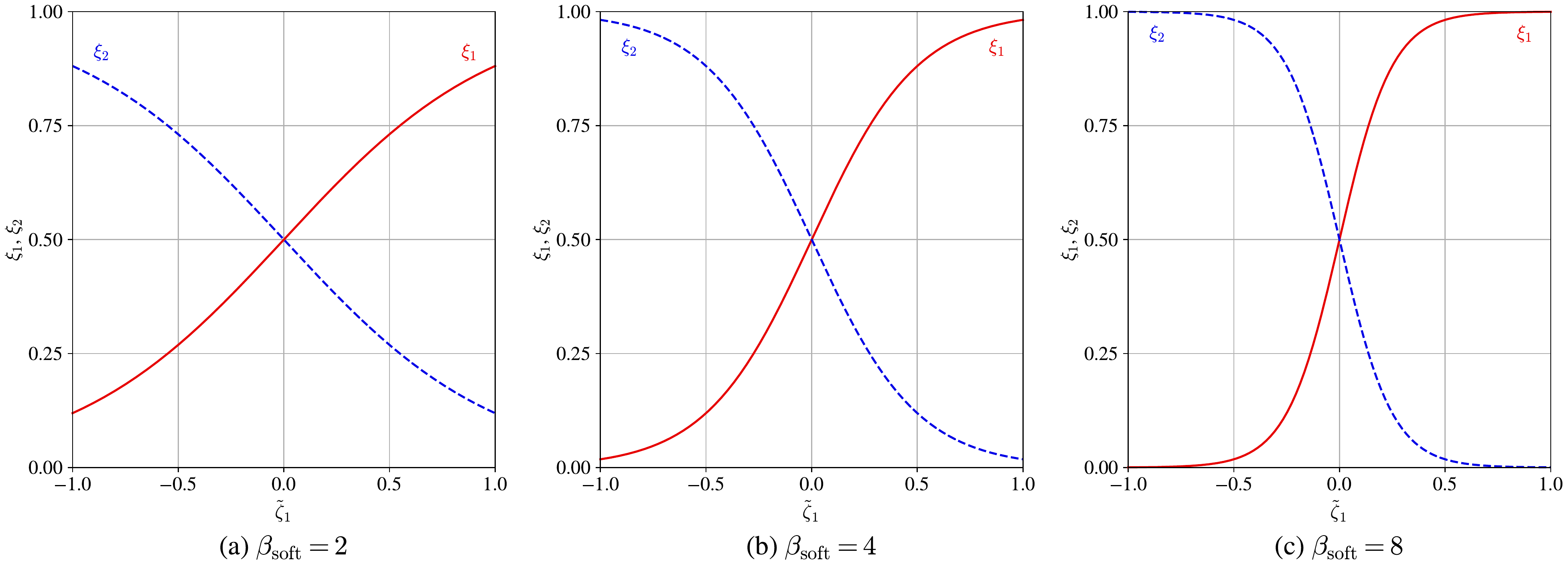}
    \caption{Visualization of the softmax function that projects the input vector $[\tilde{\zeta}_1, \tilde{\zeta}_2]^\top$ to $[\xi_1, \xi_2]^\top$ with different values of shape parameters ($\beta_\mathrm{soft}$). Here, $\tilde{\zeta}_2$ is fixed to $0$.}
    \label{fig:softmax}
\end{figure}

Furthermore, we apply a projection scheme to $\tilde{\mat{Z}} \defeq [\tilde{\vect{\zeta}}^{(1)}, \tilde{\vect{\zeta}}^{(2)}, \ldots, \tilde{\vect{\zeta}}^{(N_\mathrm{par})}]$ to facilitate sharp transitions between the domains of the actuators.
Using the softmax function as a projection function, we map $\tilde{\vect{\zeta}}^{(i)} \in \mathbb{R}^{N_\mathrm{act}+1}$ to $\vect{\xi}^{(i)} \in [0, 1]^{N_\mathrm{act}+1}$ in each particle as
\begin{equation}\label{eq:softmax}
    \xi^{(i)}_j = \dfrac{
        \exp(\beta_\mathrm{soft} \tilde{\zeta}^{(i)}_j)
    }{
        \sum_{j^\prime=1}^{N_\mathrm{act}+1}{ \exp(\beta_\mathrm{soft} \tilde{\zeta}^{(i)}_{j^\prime}) }
    },
\end{equation}
where $\beta_\mathrm{soft}$ is a shape parameter affecting the curvature of the transition curve (Fig.~\ref{fig:softmax}).
The softmax function is a generalization of the sigmoid function to multiple dimensions.
It serves as a smooth approximation of the $\mathop{\arg \max}$ operator (softmax converges to $\mathop{\arg \max}$ as $\beta_\mathrm{soft} \to \infty$), allowing for differentiation.
Applying the softmax projection provides a convenient property that the $l_1$-norm of the output vector equals $1$, which imposes a physically natural constraint $\lpnorm{\vect{\xi}^{(i)}}{1} = 1$ on the multi-indexed density.
%
\subsubsection{Constraints on binarization}\label{subsubsec:spat_bin}
Although applying the projection scheme of Eqs.~\eqref{eq:sigmoid} and \eqref{eq:softmax} reduces the transition bandwidth of the material and actuator densities when $\beta_\mathrm{sig}$ and $\beta_\mathrm{soft}$ are sufficiently large, it does not necessarily suppress the appearance of intermediate-valued regions in the optimized configuration.
Therefore, we impose an explicit constraint on the degree of binarization of the material and actuator densities.
A constraint function for the fictitious material density, $\vect{\gamma}$, is defined as
\begin{equation}\label{eq:mat_const}
    \mathcal{C}_\mathrm{mat}(\vect{\gamma}) = \dfrac{1}{N_\mathrm{par}} \sum_{i=1}^{N_\mathrm{par}} {\gamma^{(i)} (1 - \gamma^{(i)} )},
\end{equation}
which has the minimum value, $0$, if $\gamma^{(i)} \in \{0, 1\}$ for $\forall i = 1, 2, \ldots, N_\mathrm{par}$ and has the maximum value, $0.25$, if $\gamma^{(i)} = 0.5$ for $\forall i$.
We request that the optimized design variables satisfy the inequality constraint $\mathcal{C}_\mathrm{mat}(\vect{\gamma}) \leq \mathcal{C}_{\mathrm{mat}}^{*}$, imposing a quadratic penalty for $0 < \gamma^{(i)} < 1$.
Here, $\mathcal{C}_{\mathrm{mat}}^{*}$ denotes a tolerance of intermediate values.

Similarly, we define a constraint function for the actuator densities, $\mat{\Xi}$, as
\begin{equation}\label{eq:act_const}
    \mathcal{C}_\mathrm{act}(\mat{\Xi}) = \dfrac{1}{N_\mathrm{par}} \sum_{i=1}^{N_\mathrm{par}} {
        \rbrac{ 1 - \lpnorm{\vect{\xi}^{(i)}}{2}^{2} }
    },
\end{equation}
which is bounded within $[0, N_\mathrm{act} / (N_\mathrm{act} + 1)]$, given that $\lpnorm{\vect{\xi}^{(i)}}{1} = 1$ owing to the softmax projection (Eq.~\eqref{eq:softmax}).
The function attains its minimum if $\lpnorm{\vect{\xi}^{(i)}}{2} = 1$ for $\forall i = 1, 2, \ldots, N_\mathrm{par}$ and its maximum if all components of $\mat{\Xi}$ equal $1/(N_\mathrm{act}+1)$.
We impose an inequality constraint $\mathcal{C}_\mathrm{act}(\mat{\Xi}) \leq \mathcal{C}_{\mathrm{act}}^{*}$, where $\mathcal{C}_{\mathrm{act}}^{*}$ denotes a small tolerance.
As the constraint enforces $\lpnorm{\vect{\xi}^{(i)}}{2} \to 1$ and $\lpnorm{\vect{\xi}^{(i)}}{1} = 1$ is always satisfied, the optimization brings $\vect{\xi}^{(i)}$ to a one-hot vector.

In this study, $\mathcal{C}_{\mathrm{mat}}^{*}$ and $\mathcal{C}_{\mathrm{act}}^{*}$ were set to 5\% of the maximum of $\mathcal{C}_\mathrm{mat}(\vect{\gamma})$ and $\mathcal{C}_\mathrm{act}(\mat{\Xi})$, respectively.
Setting an extremely small value for the tolerance would result in an unintended perimeter constraint because intermediate densities inevitably exist at the interface between domains.
We experimentally observed that setting a tolerance of 5\% would yield well-binarized results while still allowing for complex material and actuator layouts.
%
\subsection{Structure in temporal domain}\label{subsec:temporal}
%
\subsubsection{Time-varying actuation}\label{subsubsec:actuation}
Next, we consider designing the time-varying actuation signals $\hat{u}_j(t)$ in Eq.~\eqref{eq:u_relaxed} for respective actuators $j = 1, 2, \ldots, N_\mathrm{act}$.
In the first step, we introduce a time-varying actuation model, which is inspired by the underlying mechanism of muscle contraction in creatures.
Notably, the basal form of muscle contraction is a pulse-like twitch that occurs with a single stimulus.
The pulses of individual twitches are additive because muscles are composed of multiple muscle cells.
Hence, if a subsequent stimulus is applied before the contracted muscle has fully relaxed, the contraction becomes more intense; a maximum contractile force is attained if the stimuli are applied in series for a certain duration.
As such, the temporal changes in the intensity of muscle contraction are controlled by the temporal sequence of the stimuli.
By emulating this elaborate mechanism of nature, we define the actuation over time as a convolution of Gaussian pulses:
\begin{equation} \label{eq:pulse}
    \hat{u}(t) = A_\mathrm{act} \tanh{ \sbrac{ \int_0^T{
        \chi_\mathrm{pul}(t) \, A_\mathrm{pul} \exp{ \rbrac{ -\frac{({t^\prime}-t)^2}{2\sigma_\mathrm{pul}^2} } } \, d{t^\prime}
    } } },
\end{equation}
where $[0, T]$ is the time duration subject to design, $A_\mathrm{pul}$ and $\sigma_\mathrm{pul}$ are the peak amplitude and standard deviation of a Gaussian pulse, and $A_\mathrm{act}$ is the strength of actuation.
In addition, $\chi_\mathrm{pul}(t)$ is the characteristic function to be optimized, defined as
\begin{equation}\label{eq:chi_pul}
    \chi_\mathrm{pul}(t) =
    \begin{cases}
         1 & \text{if } t \in \mathrm{\Omega}_{\mathrm{p}^+}, \\
        -1 & \text{if } t \in \mathrm{\Omega}_{\mathrm{p}^-}, \\
         0 & \text{if } t \in \mathrm{\Omega}_t \setminus (\mathrm{\Omega}_{\mathrm{p}^+} \cup \mathrm{\Omega}_{\mathrm{p}^-}),
    \end{cases}
\end{equation}
where $\mathrm{\Omega}_t$ denotes a temporal design domain of actuation, which composes a positive pulse domain $\mathrm{\Omega}_{\mathrm{p}^+}$, negative pulse domain $\mathrm{\Omega}_{\mathrm{p}^-}$, and complementary domain $\mathrm{\Omega}_t \setminus (\mathrm{\Omega}_{\mathrm{p}^+} \cup \mathrm{\Omega}_{\mathrm{p}^-})$ representing no pulse.
Additionally, we assume that the pulse can be both positive (expansion) and negative (contraction), which is different from muscles that can only contract.
Optimizing $\chi_\mathrm{pul}(t)$ involves the design of the timing and sign of Gaussian pulses in continuous time $t \in [0, T]$.
The resulting actuation signal $\hat{u}(t)$ can have complex shapes with a high degree of expression freedom.

\begin{figure}[tp]
    \centering
    \includegraphics[width=0.45\textwidth]{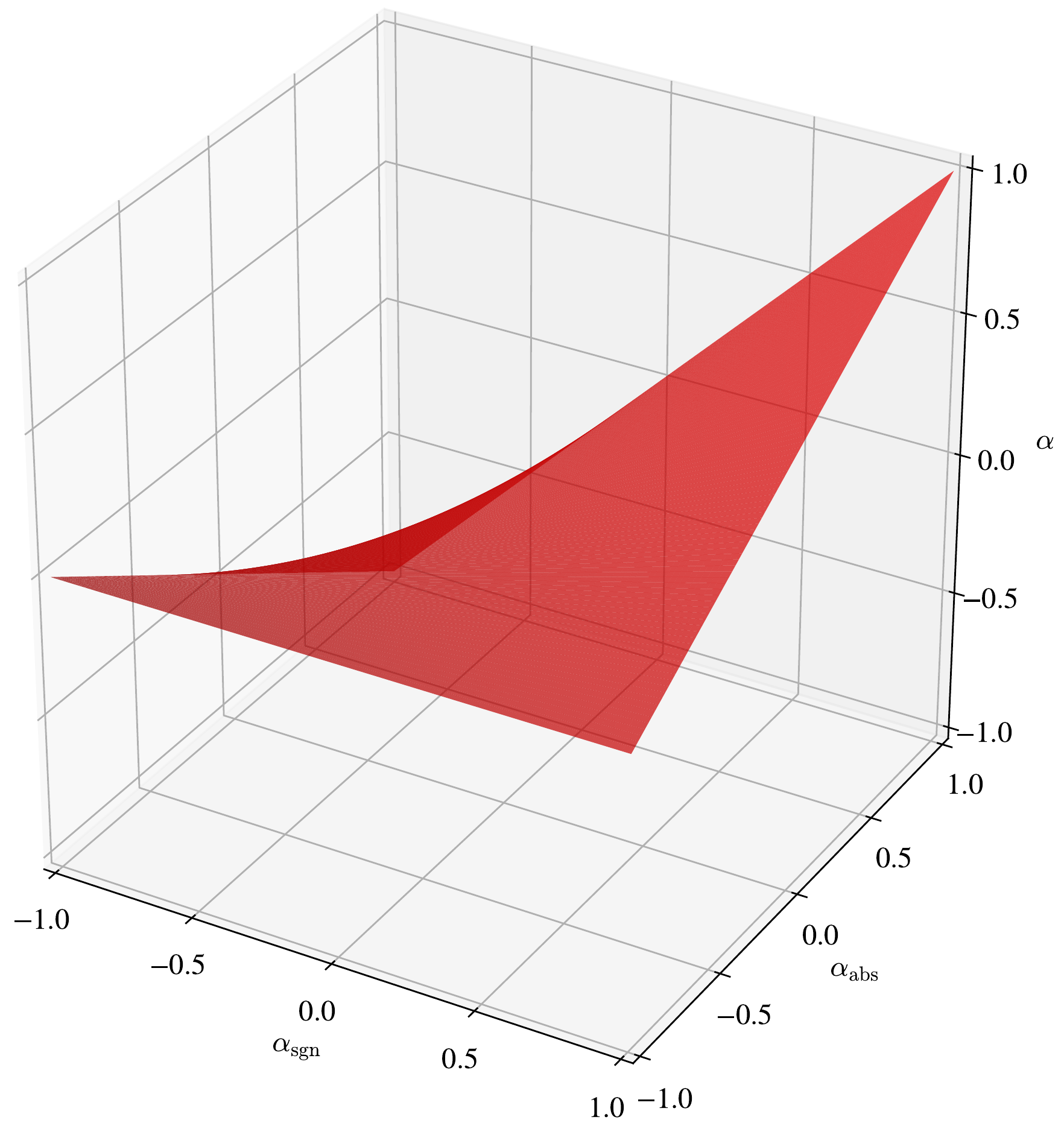}
    \caption{Visualization of $\alpha$ in Eq.~\eqref{eq:alpha} with respect to $\alpha_\mathrm{sgn}$ and $\alpha_\mathrm{abs}$.}
    \label{fig:alpha}
\end{figure}

To optimize the discrete function $\chi_\mathrm{pul}(t)$ with the gradient-based method, we relax it to a continuous function $\alpha(t) \in [-1, 1]$, as
\begin{equation}\label{eq:alpha}
    \alpha(t) = \alpha_\mathrm{sgn}(t) \rbrac{\dfrac{\alpha_\mathrm{abs}(t) + 1}{2}},
\end{equation}
where $\alpha_\mathrm{sgn}(t)$ and $\alpha_\mathrm{abs}(t)$ are the scalar functions both bounded within $[-1, 1]$.
As depicted in Fig.~\ref{fig:alpha}, the sign of $\alpha(t)$ is defined by $\alpha_\mathrm{sgn}(t)$, while the absolute value of $\alpha(t)$ is determined by $\alpha_\mathrm{abs}(t)$ and $\alpha_\mathrm{sgn}(t)$.
Using the two functions $\alpha_\mathrm{sgn}(t)$ and $\alpha_\mathrm{abs}(t)$ to solve the integer programming problem of $\alpha(t)$ to $-1$, $0$, and $1$ is more advantageous in gradient-based optimization than using only one function ($\alpha(t)$ itself).
If $\alpha(t)$ is directly used for optimization, it must pass through $0$ to go from $-1$ to $1$ (or from $1$ to $-1$), which causes the solution more likely to be trapped at $0$.
In contrast, optimizing the two functions avoids the above issue by placing the three local minima on a plane (Fig.~\ref{fig:alpha}) instead of a single line.
We further discretize the design domain $\mathrm{\Omega}_t$ to $N_\mathrm{pul}$ time steps with interval $\mathrm{\Delta} t_\mathrm{pul}$.
The optimization problem of $\chi_\mathrm{pul}(t)$ is then replaced by the optimization of pulse densities $\vect{\alpha} \defeq [\alpha^{(1)}, \alpha^{(2)}, \ldots, \alpha^{(N_\mathrm{pul})}]^\top \in [-1, 1]^{N_\mathrm{pul}}$, which are subject to the design variables $\vect{\alpha}_\mathrm{sgn} \defeq [\alpha_{\mathrm{sgn}}^{(1)}, \alpha_{\mathrm{sgn}}^{(2)}, \ldots, \alpha_{\mathrm{sgn}}^{(N_\mathrm{pul})}]^\top \in [-1, 1]^{N_\mathrm{pul}}$ and $\vect{\alpha}_\mathrm{abs} \defeq [\alpha_{\mathrm{abs}}^{(1)}, \alpha_{\mathrm{abs}}^{(2)}, \ldots, \alpha_{\mathrm{abs}}^{(N_\mathrm{pul})}]^\top \in [-1, 1]^{N_\mathrm{pul}}$.
We assume that each of the actuators can have unique pulse densities $\vect{\alpha}_j$ and design variables $\vect{\alpha}_{\mathrm{sgn}, j}$ and $\vect{\alpha}_{\mathrm{abs}, j}$ for $j = 1, 2, \ldots, N_\mathrm{act}$.
In the following, the pulse densities for all actuators are collectively denoted by $\mat{A} \defeq [\vect{\alpha}_1, \vect{\alpha}_2, \ldots, \vect{\alpha}_{N_\mathrm{act}}]$, and the corresponding design variables are denoted by $\mat{A}_\mathrm{sgn} \defeq [\vect{\alpha}_{\mathrm{sgn}, 1}, \vect{\alpha}_{\mathrm{sgn}, 2}, \ldots, \vect{\alpha}_{\mathrm{sgn}, N_\mathrm{act}}]$ and $\mat{A}_\mathrm{abs} \defeq [\vect{\alpha}_{\mathrm{abs}, 1}, \vect{\alpha}_{\mathrm{abs}, 2}, \ldots, \vect{\alpha}_{\mathrm{abs}}, N_\mathrm{act}]$.

\begin{figure}[tp]
    \centering
    \includegraphics[width=0.6\textwidth]{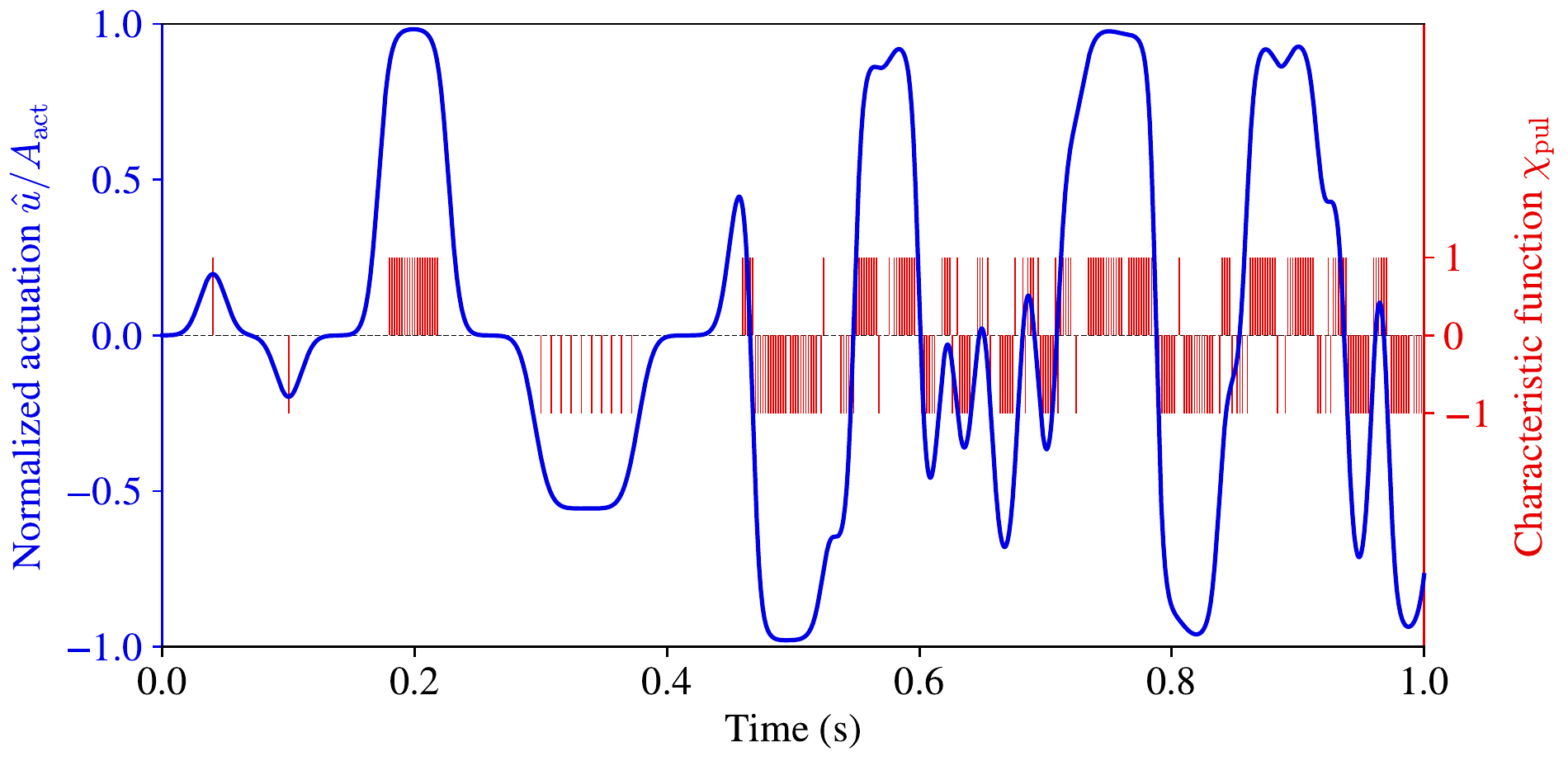}
    \caption{Example of the actuation signal given as the convolution of characteristic function ($\chi_\mathrm{pul}$) and Gaussian pulses as in Eq.~\eqref{eq:pulse}. The red bars represent the values of $\chi_\mathrm{pul}$ at discrete times, which can be $1$ (positive pulse), $-1$ (negative pulse), or $0$ (no pulse). The blue line represents the resulting actuation, which is continuous and can exhibit complex shapes. Here, $A_\mathrm{pul} = 0.2$, $\sigma_\mathrm{pul} = 0.01$~s, and the design domain is discretized with a time step size of $\mathrm{\Delta} t_\mathrm{pul} = 0.002$~s.}
    \label{fig:actuation}
\end{figure}

Notably, $\mathrm{\Delta} t_\mathrm{pul}$, the time interval between successive pulses, is independent of $\mathrm{\Delta} t$, which is the time step size of the simulation.
The value of $\mathrm{\Delta} t_\mathrm{pul}$, together with the value of $\sigma_\mathrm{pul}$, constrains the maximum frequency of $\hat{u}(t)$ in the frequency domain.
In this study, we set $A_\mathrm{pul} = 0.2$, $\sigma_\mathrm{pul} = 0.01$~s, and $\mathrm{\Delta} t_\mathrm{pul} = 0.002$~s for all actuators.
The bounds of $\hat{u}(t)$ is constrained within $(-A_\mathrm{act}, A_\mathrm{act})$ regardless of $A_\mathrm{pul}$, $\sigma_\mathrm{pul}$, and $\mathrm{\Delta} t_\mathrm{pul}$ because of the hyperbolic tangent in Eq.~\eqref{eq:pulse}.
An example of a pulse sequence and the resulting actuation is shown in Fig.~\ref{fig:actuation}.

Eqs.~\eqref{eq:u_relaxed}, \eqref{eq:u_particle}, \eqref{eq:pulse}, and \eqref{eq:alpha} complete the interpolation rule for the time-varying actuation signal $u(t)$, which is substituted into the governing equations through Eq.~\eqref{eq:cauchy_act}.
In the actuation representation, the densities for the material layout, actuator layout, and actuation pulse ($\vect{\gamma}$, $\mat{\Xi}$, and $\mat{A}$, respectively) have hierarchized relations.
The pulse density is multiplied by the actuator layout density to linearly combine all actuation signals (Eq.~\eqref{eq:u_relaxed}).
This combined actuation is further multiplied by the fictitious material density to assign the actuation to the particles (Eq.~\eqref{eq:u_particle}).
In this manner, we represent a self-actuating soft body as the material layout, actuator layout, and time-varying actuation superimposed by multiplication (Fig.~\ref{fig:schematic}).

While the optimization targets in both the spatial and temporal design domains are discrete characteristic functions, there is an essential distinction between the two.
In the spatial design domain, the characteristic functions ($\chi_\mathrm{mat}$ and $\chi_\mathrm{act}$) directly represent discrete physical quantities, specifically the structure and actuator layout.
Contrastingly, the characteristic function ($\chi_\mathrm{pul}$) in the temporal design domain does not directly represent a physical quantity.
Instead, it serves as the source that requires smoothing through convolution with a Gaussian pulse to obtain the continuous physical quantity, which is the actuation.
Consequently, the computation procedure for mapping the design variable to the physical quantity differs between the two domains.
In the spatial design domain, the procedure follows the order of design variable $\mapsto$ filtered design variable $\mapsto$ characteristic function (representing the discrete physical quantity).
On the other hand, the order in the temporal design domain is design variable $\mapsto$ characteristic function $\mapsto$ smoothed characteristic function (representing the continuous physical quantity).
Thus, a similar computational procedure can be used to design both discrete and continuous physical quantities, with the only difference being the order in which the smoothing process is applied.
%
\subsubsection{Constraints on binarization}\label{subsubsec:temp_bin}
The components of the pulse density matrix $\mat{A}$ must be optimized to $-1$, $0$, or $1$ despite continuous relaxation.
We enforce this in the optimization by imposing binarization constraints to the design variables $\mat{A}_\mathrm{sgn}$ and $\mat{A}_\mathrm{abs}$.
We define two constraint functions:
\begin{equation}\label{eq:pul_sgn_const}
    \mathcal{C}_\mathrm{pul, sgn}(\mat{A}_\mathrm{sgn}) =
        \dfrac{1}{N_\mathrm{pul}} \dfrac{1}{N_\mathrm{act}} \sum_{k=1}^{N_\mathrm{pul}} \sum_{j=1}^{N_\mathrm{act}} {(1 + \alpha^{(k)}_{\mathrm{sgn}, j}) (1 - \alpha^{(k)}_{\mathrm{sgn}, j})},
\end{equation}
and
\begin{equation}\label{eq:pul_abs_const}
    \mathcal{C}_\mathrm{pul, abs}(\mat{A}_\mathrm{abs}) =
        \dfrac{1}{N_\mathrm{pul}} \dfrac{1}{N_\mathrm{act}} \sum_{k=1}^{N_\mathrm{pul}} \sum_{j=1}^{N_\mathrm{act}} {(1 + \alpha^{(k)}_{\mathrm{abs}, j}) (1 - \alpha^{(k)}_{\mathrm{abs}, j})}
\end{equation}
which respectively evaluates the degree of binarization of $\mat{A}_\mathrm{sgn}$ and $\mat{A}_\mathrm{abs}$.
The ranges of $\mathcal{C}_\mathrm{pul, sgn}(\mat{A}_\mathrm{sgn})$ and $\mathcal{C}_\mathrm{pul, abs}(\mat{A}_\mathrm{abs})$ are both $[0, 1]$;
the minimum is attained when all components of the input matrix have upper or lower bound values.
To explicitly suppress the use of intermediate values, we impose quadratic inequality constraints $\mathcal{C}_\mathrm{pul, sgn}(\mat{A}_\mathrm{sgn}) \leq \mathcal{C}_\mathrm{pul, sgn}^{*}$ and $\mathcal{C}_\mathrm{pul, abs}(\mat{A}_\mathrm{abs}) \leq \mathcal{C}_\mathrm{pul, abs}^{*}$, where $\mathcal{C}_\mathrm{pul, sgn}^{*}$ and $\mathcal{C}_\mathrm{pul, abs}^{*}$ are the tolerances of intermediate values.
In contrast to $\vect{\gamma}$ and $\mat{\Xi}$, which are the densities for the spatial structures, filtering is not applied to obtain $\mat{A}$.
Therefore, setting small tolerances does not result in a perimeter constraint.
The tolerances $\mathcal{C}_\mathrm{pul, sgn}^{*}$ and $\mathcal{C}_\mathrm{pul, abs}^{*}$ are set stricter than $\mathcal{C}_{\mathrm{mat}}^{*}$ and $\mathcal{C}_{\mathrm{act}}^{*}$, to 1\% of their maximum values.
%
\subsection{Gradient-based optimization}\label{subsec:optimization}
%
\subsubsection{Optimization problem}\label{subsubsec:opt_problem}
In summary, we represent the structure and movement of a self-actuating soft body by the superposition of characteristic functions, namely $\chi_\mathrm{mat}(\vect{x})$, $\chi_\mathrm{act}(\vect{x})$, and $\chi_\mathrm{pul}(t)$, which respectively define the material layout, actuator layout, and time-varying actuation.
The characteristic functions are relaxed to the continuous densities in the discretized design domain: $\chi_\mathrm{mat}(\vect{x})$ to $\vect{\gamma}$, $\chi_\mathrm{act}(\vect{x})$ to $\mat{\Xi}$, and $\chi_\mathrm{pul}(t)$ to $\mat{A}$.
Furthermore, these densities are optimized with the design variables $\vect{\phi}$, $\mat{Z}$, $\mat{A}_\mathrm{sgn}$, and $\mat{A}_\mathrm{abs}$.
Recall that $\vect{\phi} \in [-1, 1]^{N_\mathrm{par}}$ represents the design variable for determining the fictitious material density $\vect{\gamma} \in [0, 1]^{N_\mathrm{par}}$ at $N_\mathrm{par}$ particles, while $\mat{Z} \in \mathbb{R}^{(N_\mathrm{act}+1) \times N_\mathrm{par}}$ denotes the design variable for determining the multi-indexed density $\mat{\Xi} \in [0, 1]^{(N_\mathrm{act}+1) \times N_\mathrm{par}}$, which defines the weights of $N_\mathrm{act}$ actuators at $N_\mathrm{par}$ particles.
The mapping of $\vect{\phi}$ and $\mat{Z}$ into $\vect{\gamma}$ and $\mat{\Xi}$, respectively, is achieved by applying the filtering and projection schemes.
Additionally, $\mat{A}_\mathrm{sgn} \in [-1, 1]^{N_\mathrm{pul} \times N_\mathrm{act}}$ and $\mat{A}_\mathrm{abs} \in [-1, 1]^{N_\mathrm{pul} \times N_\mathrm{act}}$ are the design variables that determine the sign and absolute value of $\mat{A} \in [-1, 1]^{N_\mathrm{pul} \times N_\mathrm{act}}$, representing the pulse density of $N_\mathrm{act}$ actuators at each of the $N_\mathrm{pul}$ discretized times.

Consequently, designing a soft body aimed at a certain dynamic task can be formulated as a continuous optimization problem as follows:
\begin{subequations} \label{eq:opt_problem}
\begin{alignat}{2}
    \underset{\mat{A}_\mathrm{sgn}, \mat{A}_\mathrm{abs}, \vect{\phi}, \mat{Z}}{\text{minimize}} & \quad && \mathcal{L}_{\mathrm{task}}, \label{eq:L_task} \displaybreak[1]\\
    \text{subject to} & \quad && \mat{A}_\mathrm{sgn} \in [-1, 1]^{N_\mathrm{pul} \times N_\mathrm{act}}, \label{eq:side_const_alpha_s} \displaybreak[1]\\
    & \quad && \mat{A}_\mathrm{abs} \in [-1, 1]^{N_\mathrm{pul} \times N_\mathrm{act}}, \label{eq:side_const_alpha_m} \displaybreak[1]\\
    & \quad && \vect{\phi} \in [-1, 1]^{N_\mathrm{par}}, \label{eq:side_const_phi} \displaybreak[1]\\
    & \quad && \mathcal{C}_\mathrm{pul, sgn}(\mat{A}_\mathrm{sgn}) \leq \mathcal{C}_\mathrm{pul, sgn}^{*}, \label{eq:bin_const_alpha_s} \displaybreak[1]\\
    & \quad && \mathcal{C}_\mathrm{pul, abs}(\mat{A}_\mathrm{abs}) \leq \mathcal{C}_\mathrm{pul, abs}^{*}, \label{eq:bin_const_alpha_m} \displaybreak[1]\\
    & \quad && \mathcal{C}_\mathrm{mat}(\vect{\gamma}) \leq \mathcal{C}_{\mathrm{mat}}^{*}, \label{eq:bin_const_gamma} \displaybreak[1]\\
    & \quad && \mathcal{C}_\mathrm{act}(\mat{\Xi}) \leq \mathcal{C}_{\mathrm{act}}^{*}, \label{eq:bin_const_xi} \displaybreak[1]\\
    & \quad && \text{Governing Eqs.~\eqref{eq:mass}--\eqref{eq:cauchy_act}}, \displaybreak[1]\\
    & \quad && \text{Interpolation Eqs.~\eqref{eq:rho_particle}, \eqref{eq:lame_particle}, \eqref{eq:u_relaxed}, \eqref{eq:u_particle}, \eqref{eq:pulse}, and \eqref{eq:alpha}} \displaybreak[1]\\
    & \quad && \text{Filtering and projection Eqs.~\eqref{eq:mat_filter}, \eqref{eq:sigmoid}, \eqref{eq:act_filter}, and \eqref{eq:softmax}}.
\end{alignat}
\end{subequations}
Here, $\mathcal{L}_{\mathrm{task}}$ is an objective function that reflects the performance for an expected dynamic task.
The objective function is a function of the state variables obtained from the governing equations.
It is defined in a task-specific manner in Section~\ref{sec:numer}.
Eqs.~\eqref{eq:side_const_alpha_s}--\eqref{eq:side_const_phi} are the side constraints that limit the intervals of design variables except for the unbounded $\mat{Z}$.
Moreover, Eqs.~\eqref{eq:bin_const_alpha_s}--\eqref{eq:bin_const_xi} are the constraints on binarization to suppress intermediate densities explicitly.
Notably, imposing a side constraint on $\mat{Z}$ is not straightforward because of the nature of the softmax function, whose domain is the set of all real numbers.
Even without side constraint, a feasible solution of $\mat{Z}$ can be obtained by optimization because of the small tolerance $\mathcal{C}_{\mathrm{act}}^{*}$ on the binarization constraint (Eq.~\eqref{eq:bin_const_xi}).
In the governing equations, $\rho(\vect{x})$, $\lambda(\vect{x})$, $\mu(\vect{x})$, and $u(t)$ are interpolated according to the design variables.
%
\subsubsection{Optimization method}\label{subsubsec:opt_method}
In this study, we used Adam~\cite{kingma2014adam} to solve the optimization problem of Eq.~\eqref{eq:opt_problem}.
Adam updates design variables according to the first-order derivatives while adapting the update amounts based on the momentums of the derivative and squared derivative.
It has demonstrated a robust performance in various optimization problems and is considered as the default choice, especially in deep learning.
However, Adam cannot be applied directly to Eq.~\eqref{eq:opt_problem} because it is a constrained optimization problem with four nonlinear inequality constraints of Eqs.~\eqref{eq:bin_const_alpha_s}--\eqref{eq:bin_const_xi}.
Therefore, we used the augmented Lagrangian method to replace Eq.~\eqref{eq:opt_problem} with an unconstrained optimization problem of minimizing the augmented Lagrangian function:
\begin{equation}\label{eq:aug_lag}
    \underset{\vect{p}}{\text{minimize}} \quad \mathcal{L}(\vect{p}, \vect{\kappa}, \vect{\tau}) \defeq \mathcal{L}_{\mathrm{task}} + \sum_{m=1}^{M}{\sbrac{
        -\kappa_m \mathcal{P}_m(\vect{p}) + \dfrac{1}{2} \tau_m \{\mathcal{P}_m(\vect{p})\}^2
    }},
\end{equation}
where the subscript $m$ stands for each of the $M$ ($= 4$) inequality constraints, $\vect{p}$ is the general notation for all design variables, $\vect{\kappa} \defeq [\kappa_1, \kappa_2, \ldots, \kappa_M]^\top$ is the Lagrangian multiplier, $\vect{\tau} \defeq [\tau_1, \tau_2, \ldots, \tau_M]^\top$ is the penalty coefficient, and $\mathcal{P}_m(\vect{p})$ is the penalty function defined for each of the constraints as
\begin{equation}\label{eq:pen}
    \mathcal{P}_m =
    \begin{cases}
        0 & \text{if } \mathcal{C}_m(\vect{p}) \leq \mathcal{C}_m^{*}, \\
        \mathcal{C}_m(\vect{p}) - \mathcal{C}_m^{*} & \text{otherwise}.
    \end{cases}
\end{equation}
Given a suitable $\vect{\kappa}$ that satisfies ${\partial \mathcal{L}}/{\partial \vect{\kappa}} = \vect{0}$ and any positive $\vect{\tau}$, the solution of Eq.~\eqref{eq:aug_lag} is also the solution of Eq.~\eqref{eq:opt_problem}.

The optimization problem of Eq.~\eqref{eq:aug_lag} is unconstrained (except for the side constraints) and can be solved with any optimization method.
However, it contains unknown Lagrange multipliers $\vect{\kappa}$.
In the augmented Lagrangian method, ${\vect{\kappa}}$ and ${\vect{\tau}}$ are updated sequentially starting from arbitrary initial values in the process of optimizing $\vect{p}$, thereby optimizing $\vect{p}$ and ${\vect{\kappa}}$ simultaneously.
The idea is to bring $\vect{p}$ close to a feasible solution by increasing $\vect{\tau}$ and then find $\vect{\kappa}$ satisfying ${\partial \mathcal{L}}/{\partial \vect{\kappa}} = \vect{0}$.
The specific algorithm used in this study is presented in Algorithm~\ref{alg:aug_lag}.
\begin{algorithm}[tp]
\caption{Augmented Lagrangian method} \label{alg:aug_lag}
\begin{algorithmic}[1]
    \Require
        \Statex $\vect{p}_0$: Initial values of the design variables
        \Statex $\vect{\tau}_0 \in (0, \infty)^M$: Initial values of the penalty coefficients
        \Statex $c \in (0, 1)$: Criteria for updating the Lagrange multipliers
        \Statex $a \in (1, \infty)$: Multiplier factor to increase the penalty coefficients
    \Ensure
        \Statex $\vect{p}$: Optimized design variables
        \Statex $\vect{\kappa}$: Lagrange multipliers at a stationary point

    \State $\vect{\kappa} \gets \vect{0}$, $\vect{\tau} \gets \vect{\tau}_0$, $\vect{V} \gets \vect{\mathcal{P}}(\vect{p}_0)$, $\vect{V}_{\mathrm{prev}} \gets \vect{V}$ \Comment{Initialize $\vect{\kappa}$, $\vect{\tau}$, and $\vect{p}$}
    \State $s \gets 0$

    \While {$\max_m{V_m} > 0$ \textbf{and} $s < s_\mathrm{max}$} \label{alg:line:outer_while}
        \While {$\vect{p}$ not converged \textbf{and} $s < s_\mathrm{max}$} \label{alg:line:inner_while}
            \State $s \gets s+1$ \Comment{Update iteration counter}
            \State $\vect{p} \gets \arg\min_{\vect{p}}{\mathcal{L}(\vect{p}, \vect{\kappa}, \vect{\tau})}$ \Comment{Update $\vect{p}$ by optimizer with fixed $\vect{\kappa}$ and $\vect{\tau}$} \label{alg:line:inner_update}
        \EndWhile
        \State $\vect{V} \gets \vect{\mathcal{P}}(\vect{p})$ \Comment{Evaluate the amounts of constraint violations}

        \For {$m = 1, 2, \ldots, M$}
            \If {$V_m > 0$} \Comment{If $\vect{p}$ violates the $m$-th constraint, then \ldots}
                \If {$V_m < c V_{\mathrm{prev}, m}$} \Comment{If the amount of violation is reduced to $c$ times, then \ldots}
                    \State $\kappa_m \gets \kappa_m - \tau_m V_m$ \Comment{Update $\vect{\kappa}$ toward a stationary point} \label{alg:line:outer_update}
                    \State $V_{\mathrm{prev}, m} \gets V_m$
                \Else
                    \State $\tau_m \gets a \tau_m$ \Comment{Increase $\vect{\tau}$ to bring $\vect{p}$ closer to a feasible solution}
                \EndIf
            \EndIf
        \EndFor
    \EndWhile
\end{algorithmic}
\end{algorithm}
The algorithm consists of the outer loop (line~\ref{alg:line:outer_while}) of updating ${\vect{\kappa}}$ and ${\vect{\tau}}$ and the inner loop (line~\ref{alg:line:inner_while}) of updating $\vect{p}$ with fixed ${\vect{\kappa}}$ and ${\vect{\tau}}$ using a standard optimizer.
Each time a converged solution of $\vect{p}$ is obtained by the inner loop, we decide whether to increase $\vect{\tau}$ or update $\vect{\kappa}$ according to the amount of violation of each constraint.
If the amount of violation is not reduced to $c$~times, we assume that the current $\vect{p}$ is too far from a feasible solution and thus increase $\vect{\tau}$ with a factor of $a$.
Otherwise, we update $\vect{\kappa}$ in the direction of the first-order derivative of $\mathcal{L}$ (line~\ref{alg:line:outer_update}), by which $\vect{\kappa}$ converges to a stationary point with a linear convergence rate~\cite{fletcher2000practical}.

We reset the accumulated momentums of Adam every time before entering the inner loop, making Adam forget the optimization history with previous $\vect{\kappa}$ and $\vect{\tau}$.
After each iteration of line~\ref{alg:line:inner_update}, we further update the design variables with the side constraints as $\vect{p} \gets \min(\vect{p}_\mathrm{u}, \max(\vect{p}_\mathrm{l}, \vect{p}))$, where the subscripts $\mathrm{u}$ and $\mathrm{l}$ denote the upper and lower bounds, respectively.
In addition, we compute two averages of $\mathcal{L}$ with different averaging intervals: $\bar{L}_\mathrm{cur}$, the average over the last $10$ iterations, and $\bar{L}_\mathrm{prev}$, the average from ($s-19$)-th to ($s-10$)-th iterations.
If $|\bar{L}_\mathrm{cur} - \bar{L}_\mathrm{prev}| / |\bar{L}_\mathrm{prev}| < 0.001$, we assume that $\vect{p}$ is converged (line~\ref{alg:line:inner_while}).
%
\subsubsection{Hyperparameters}\label{subsubsec:hyperparams}
The optimization problem that we solve includes some hyperparameters, which may largely affect the solution but cannot be determined through the optimization process.
Herein, we list the important hyperparameters, along with their effects on the solution and the values used in this study.
First, the filter parameters $R_\mathrm{f}$ and $p_\mathrm{f}$ in Eq.~\eqref{eq:filter_weight} determine the smoothness of the boundary of the material and actuator layouts.
The filter parameters affect the minimum size of each domain in the layout; the larger $R_\mathrm{f}$ and the smaller $p_\mathrm{f}$, the larger the minimum size.
Their optimal values may depend on the size of the design domain and the complexity of the optimal layout.
In this study, we searched for the best combination in each numerical example from $R_\mathrm{f} \in \{1.5\mathrm{\Delta} x, 2\mathrm{\Delta} x\}$ and $p_\mathrm{f} \in \{2, 3\}$, where $\mathrm{\Delta} x$ denotes a grid spacing.

Second, the shape parameters $\beta_\mathrm{sig}$ and $\beta_\mathrm{soft}$ in Eqs.~\eqref{eq:sigmoid} and \eqref{eq:softmax} govern the thickness (transition bandwidth) of the interface between domains; the larger value results in the thicker interface.
Under the binarization constraints, the shape parameters determine the amount of intermediate densities that increasing the interface will yield, i.e., the cost of creating the interface.
Therefore, assigning too small values to $\beta_\mathrm{sig}$ and $\beta_\mathrm{soft}$ may inhibit complex layouts with large interfaces or prevent the solution from converging to a feasible solution.
We set $\beta_\mathrm{sig} = 4$ and $\beta_\mathrm{soft} = 4$ in 2D examples, whereas we set $\beta_\mathrm{sig} = 8$ and $\beta_\mathrm{soft} = 8$ in 3D examples where the optimal layout tends to be more complex than that in 2D examples.

Third, the initial step size given to Adam governs the amount of update of design variables per iteration, which may alter the optimization path significantly.
Despite Adam adjusting the step size during the optimization, the maximum step size is approximately bounded by the initial step size~\cite{kingma2014adam}.
Therefore, it is reasonable to consider the intervals of the design variables in determining the initial step size.
In this study, we assigned $0.01$ to the initial step size, which corresponds to $0.5$\% of the interval $[-1, 1]$.
The other parameters for Adam were set to default values.

Finally, the parameters associated with the augmented Lagrangian method also affect the optimization path.
The initial penalty coefficients, $\vect{\tau}_0$, reflect how strongly the penalty is applied to the intermediate densities in the beginning of the optimization.
In this study, we started the optimization with a small value of $0.001$, which gives almost no penalty on any constraint.
The parameters $c$ and $a$ dictate the speed of updating $\vect{\kappa}$ and $\vect{\tau}$ in the optimization.
We set $c = 0.25$ and $a = 10$, i.e., $\vect{\kappa}$ is updated if the amount of violation decreased to 75\%, and $\vect{\tau}$ is increased $10$-fold otherwise.
%
\section{Numerical examples}\label{sec:numer}
\begin{figure}[tp]
    \centering
    \includegraphics[width=0.8\textwidth]{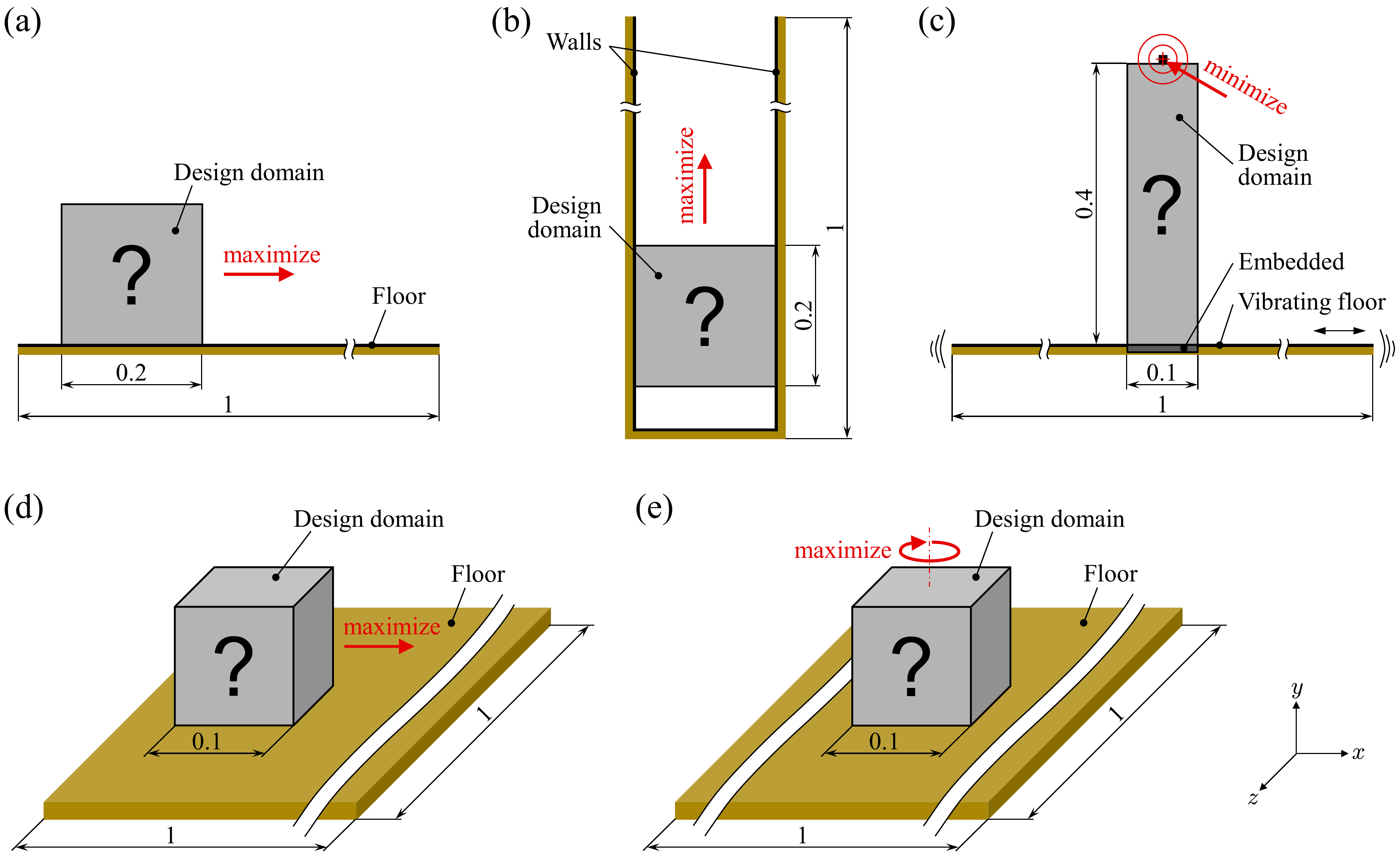}
    \caption{Design problems considered in this study. The units for the dimensions are meters. We refer to each problem as designing (a)~\textit{Walker}, (b)~\textit{Climber}, (c)~\textit{Balancer}, (d)~\textit{3D\,Walker}, and (e)~\textit{Rotator}. (a)--(c) are the design problems in 2D space and time, and (d) and (e) are the design problems in 3D space and time.}
    \label{fig:tasks}
\end{figure}

In this section, we present five numerical examples, wherein self-actuating soft bodies were designed for specific dynamic tasks by using the proposed method.
The dynamic tasks considered are horizontal locomotion on the floor, vertical locomotion up the walls against gravity, posture control to maintain the target position despite forced vibration, and rotation on the floor (Fig.~\ref{fig:tasks}).
The numerical examples cover both 2D and 3D (in spatial dimension) problems.

For all the examples, we assumed a square (in 2D) or cube (in 3D) computational domain with side length $L = 1$~\si{m}.
The computational domain was discretized into grids with an equal grid spacing $\mathrm{\Delta} x = 0.01$~\si{m}.
In the spatial design domain of the soft body, which was $0.2 \times 0.2$~\si{m^2} square in 2D and $0.1 \times 0.1 \times 0.1$~\si{m^3} cube in 3D, particles were placed with equal spacing at $0.5 \mathrm{\Delta} x$.
The material properties were set based on soft silicone rubber: density $\rho_0 = 1000$~\si{kg.m^{-3}}, Young's modulus $E_0 = 0.1$~\si{MPa}, and Poisson's ratio $\nu_0 = 0.4$.
The small constant $\varepsilon$ for the interpolation was set to $10^{-5}$.
The maximum value of the time step size $\mathrm{\Delta} t$ allowed for stability is determined by the Courant--Friedrichs--Lewy condition (CFL) condition.
Based on the velocity of the stress wave in an isotropic elastic body, $v_\mathrm{wave} = \sqrt{{(\lambda + 2 \mu)}/{\rho}}$, the CFL condition imposes a restriction: $\mathrm{\Delta} t \leq {\mathrm{\Delta} x}/{v_\mathrm{wave}} \sim 6.8 \times 10^{-4}$~\si{s}.
For this study, we used $\mathrm{\Delta} t = 1 \times 10^{-4}$~\si{s}, which is sufficiently large to reduce the computational costs while still fulfilling the CFL condition.

Furthermore, we set the initial values of the design variables as $\vect{\phi} = \vect{0}$ and $\mat{Z} = \mat{0}$, whereas we assumed the components of both $\mat{A}_\mathrm{sgn}$ and $\mat{A}_\mathrm{abs}$ initially to be independent Gaussian random variables whose mean and standard deviation are $0$ and $0.1$, respectively.
The number of actuators placed in the soft body, $N_\mathrm{act}$, was assumed to be $4$.

The simulations were performed on the CPU (Intel Core i9-11900K) using a single core for the 2D examples and on the GPU (NVIDIA GeForce RTX~3090) for the 3D examples.
The computation time required for one iteration of optimization, with a simulation time span of $1$~\si{s}, was approximately $22$~\si{s} for the 2D examples (using $1600$~particles and $10^4$~grid nodes) and $30$~\si{s} for the 3D examples (using $8000$~particles and $10^6$~grid nodes).

We provide a supplementary video (\ref{appendix:suppl}) to enhance the interpretation of the results presented in the following subsections.
The video includes an animated history of the optimization (changes in the material layout, actuator layout, and time-varying actuation with respect to the number of iterations) and a comparison of the movements between the initial, optimized, and post-processed (for binarization) soft bodies for each numerical example.
%
\subsection{Design in 2D space and time}\label{subsec:2d}
%
\subsubsection{Walker}\label{subsubsec:2d_walker}
The first example considers designing a soft body that moves in a horizontal direction on the floor, named \textit{Walker}.
We consider a spatial design domain of $0.2 \times 0.2$~\si{m^2} square placed on the floor, as shown in Fig.~\ref{fig:tasks}(a).
Expecting \textit{Walker} to travel as far as possible in the direction of the $x$-axis during the whole simulation time, we define the objective function (to be minimized) as follows:
\begin{equation}\label{eq:L_2d_walker}
    \mathcal{L}_\mathrm{task} = - \int_{0}^{T}{\vect{v}_\mathrm{g}(t) \cdot \vect{e}_x \, dt}.
\end{equation}
Here, $T$ is the total simulation time, $\vect{e}_x$ is the unit vector in the direction of the $x$-axis, and $\vect{v}_\mathrm{g}$ is the mass-weighted average velocity, defined as
\begin{equation}\label{eq:v_g}
    \vect{v}_\mathrm{g}(t) = \dfrac{
        \int_{\mathrm{\Omega}_x(t)}{\rho \vect{v} \, d\vect{x}}
    }{
        \int_{\mathrm{\Omega}_x(t)}{\rho \, d\vect{x}}
    },
\end{equation}
where $\mathrm{\Omega}_x(t)$ denotes the spatial design domain at time $t$, $\vect{v}$ is the velocity at each position in $\mathrm{\Omega}_x(t)$, and $\rho$ is the interpolated density.
In this example, we set the total simulation time $T = 0.5$~\si{s}, actuation strength as $A_\mathrm{act} = 1 \times 10^{4}$~\si{Pa}, filter parameters as $R_\mathrm{f} = 1.5\mathrm{\Delta} x$ and $p_\mathrm{f} = 2$, and shape parameters as $\beta_\mathrm{sig} = 4$ and $\beta_\mathrm{soft} = 4$.

\begin{figure}[tp]
    \centering
    \includegraphics[width=0.9\textwidth]{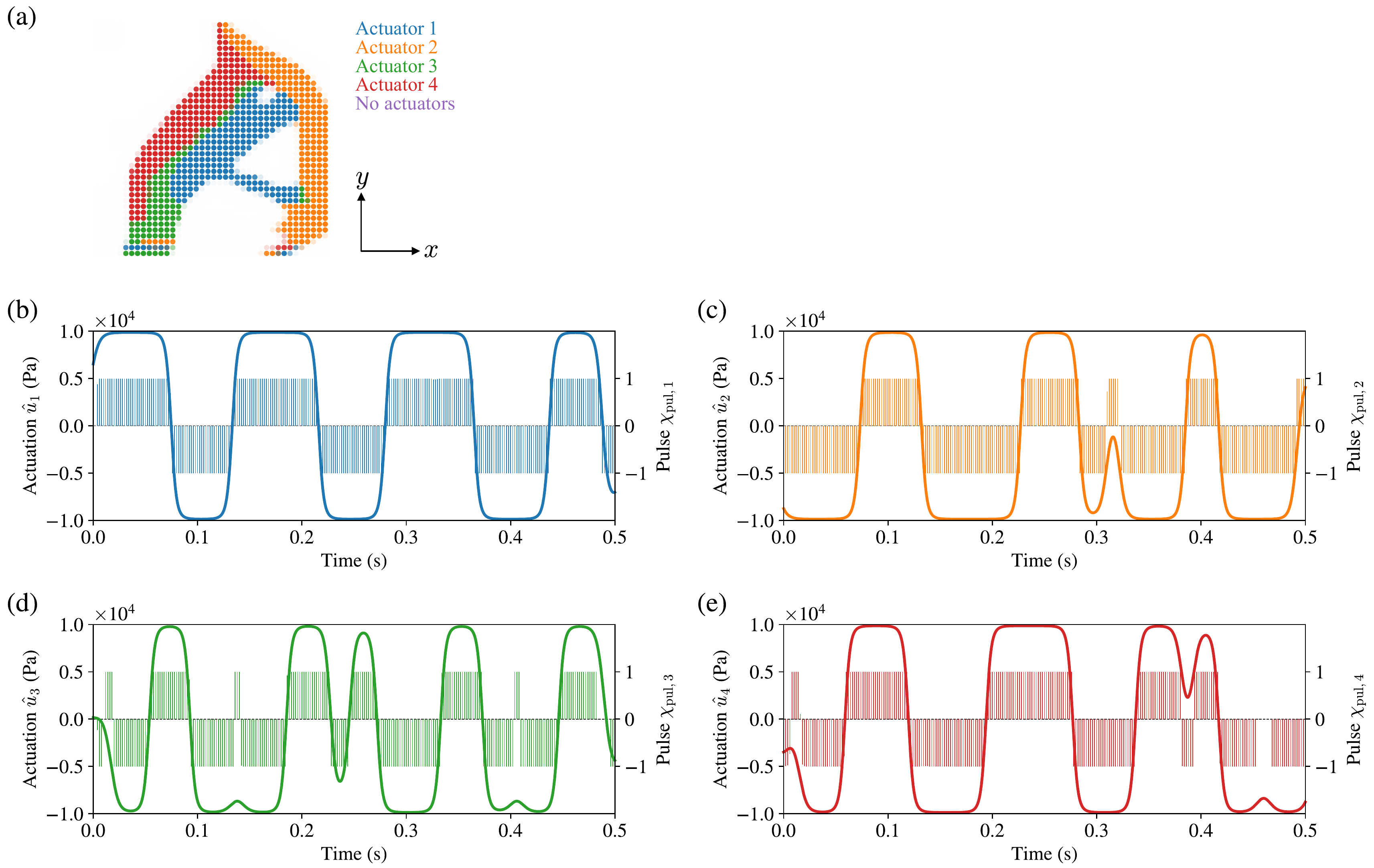}
    \caption{Optimized configuration of \textit{Walker}. (a)~Material and actuator layout. The color and transparency of the particles represent the placed actuator and fictitious material density, respectively. (b)--(e)~Changes in actuation of the respective actuators over the simulation duration. Positive and negative actuation values represent expansion and contraction, respectively.}
    \label{fig:2d_walker_opt}
\end{figure}

\begin{figure}[tp]
    \centering
    \includegraphics[width=1\textwidth]{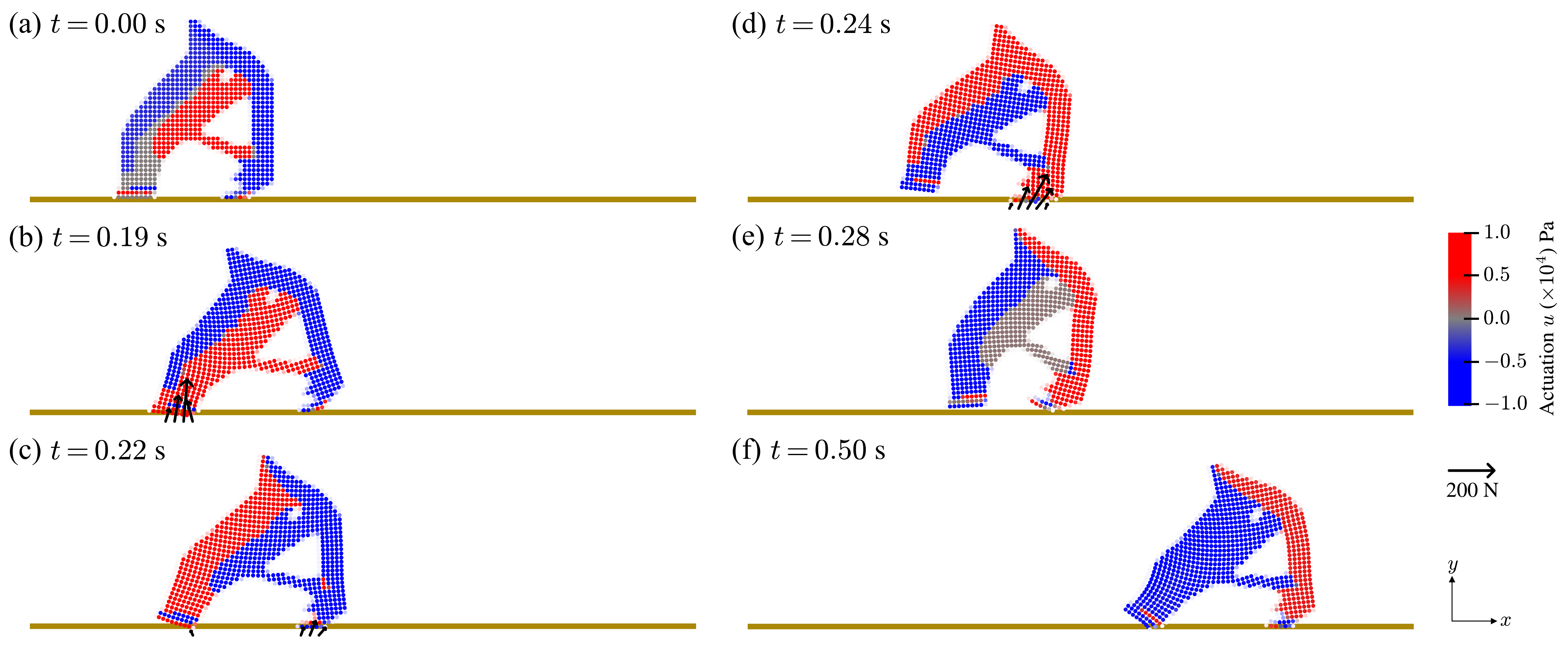}
    \caption{Movement of the optimized \textit{Walker} shown as time-series. Positive (red) and negative (blue) actuation values represent expansion and contraction, respectively. The black arrows on the floor indicate the magnitude and direction of the contact force computed at the background grid nodes. For visibility, particles with a very small fictitious material density ($\gamma < 0.01$) are not shown. See the supplementary video (\ref{appendix:suppl}) for \textit{Walker} in action.}
    \label{fig:2d_walker_per}
\end{figure}

The optimized configuration of \textit{Walker} is shown in Fig.~\ref{fig:2d_walker_opt}, and its movement is shown in Fig.~\ref{fig:2d_walker_per}.
The optimized \textit{Walker} exhibits a shape resembling the letter ``A'' with two leg-like structures extending perpendicular to the floor.
The front and rear legs are connected by the structure formed by the blue actuator extending in two mutually orthogonal directions.
When the blue actuator is expanded, the front leg is raised, being pushed forward and upward simultaneously (Fig.~\ref{fig:2d_walker_per}(b)).
Interestingly, \textit{Walker} utilizes an elongated shape to transform isotropic actuation forces into single-axis forces, resulting in a structure similar to muscle--tendon that is seen in animals.

The rear leg consists of multiple actuators; the actuator domain is divided into front and rear (blue and red) and further top and bottom (blue and green) domains.
The blue and red actuators arranged front and rear have opposite timing of expansion and contraction, showing a structure similar to the agonist and antagonist muscles of animals.
By bending the rear leg, \textit{Walker} kicks the floor with adequate changes in posture and contact angle with the floor.
Moreover, it makes a time gap between the motions of raising the front leg (Fig.~\ref{fig:2d_walker_per}(a)) and fully stretching the rear leg (Fig.~\ref{fig:2d_walker_per}(b)) by having the actuators further divided into upper (blue) and lower (green) domains, which ensures a smooth shift of weight from rear to front.

The noteworthy aspect of the result is that the optimization led to a periodic actuation signal without any assumptions on periodicity.
As shown in Fig.~\ref{fig:2d_walker_opt}(b)--(e), all actuators exhibit a roughly periodic actuation signal with a frequency of $6$~\si{Hz}, which is synchronized with the cyclic movement of the soft body (see \ref{appendix:periodicity} for more discussion on periodicity).
Consequently, the optimized \textit{Walker} demonstrates a skillful ``walk'' with the following sequence of motions as one cycle:
\begin{enumerate*}[label=(\roman*)]
    \item raise the front leg and propel it forward (Fig.~\ref{fig:2d_walker_per}(b)), 
    \item kick the floor by stretching the rear leg and shift the weight from the rear to the forward leg (Fig.~\ref{fig:2d_walker_per}(c)), 
    \item curl the body to land the rear leg forward and kick the floor by stretching the front leg (Fig.~\ref{fig:2d_walker_per}(d)), and 
    \item return to the initial posture by contracting the rear leg while maintaining the extension of the front leg (Fig.~\ref{fig:2d_walker_per}(e)).
\end{enumerate*}

\begin{figure}[tp]
    \centering
    \includegraphics[width=0.6\textwidth]{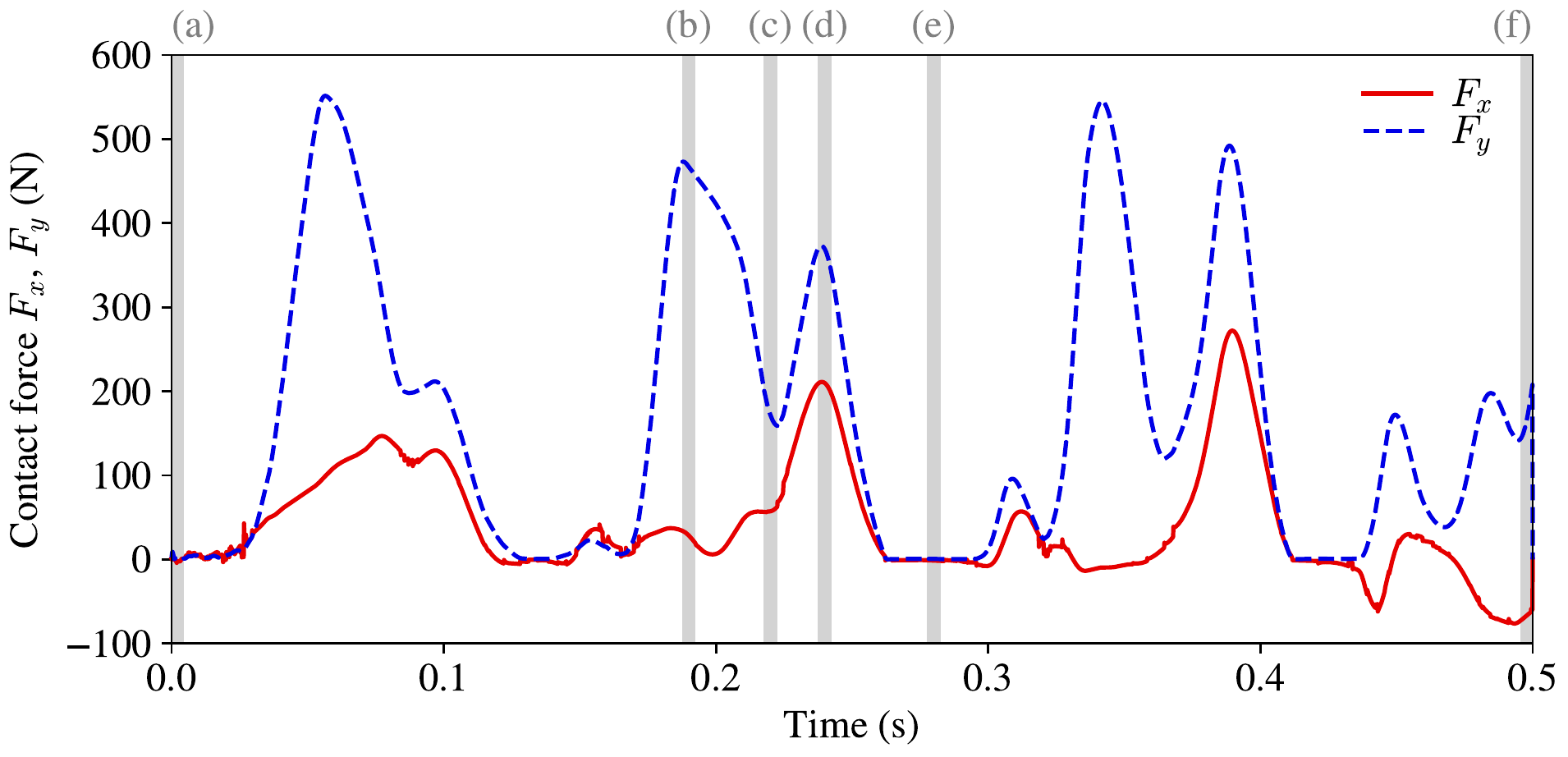}
    \caption{Change in the contact force that \textit{Walker} receives from the floor over time. $F_x$ and $F_y$ denote the $x$- and $y$-components of the contact force, respectively. The negative value indicates that the force direction is opposite to the axis. (a)--(f) represent the times shown in Fig.~\ref{fig:2d_walker_per}.}
    \label{fig:2d_walker_con}
\end{figure}

The change in contact force over time (Fig.~\ref{fig:2d_walker_con}) reveals more detail on how \textit{Walker} uses the two legs coordinately while distinguishing their roles.
The two peaks of the $y$-directional contact force during one motion cycle correspond to the kicking of the floor by the rear and front legs, respectively.
Comparing the magnitude of the $x$- and $y$-directional forces shows that the first kick by the rear leg is directed almost perpendicular to the floor.
The kick applies a clockwise torque, which rotates the body to fall forward.
With this postural change, the subsequent kick by the front leg can produce a large propulsive force, pulling the body forward.

Furthermore, the optimal solution successfully found a suitable frequency of actuation for walking.
Optimizing the actuation frequency along with the structure involves difficulty in that the optimal value varies widely depending on many design-dependent factors, such as the elastic compliance, size, and weight of the structure.
It may be for this difficulty that previous studies~\cite{cheney2013unshackling, vandiepen2022codesign, hu2020difftaichi, sato2022topology} using periodic functions in the actuation model fixed frequencies to the predetermined values rather than optimizing them.
Indeed, we experienced in our previous study~\cite{sato2022topology} that optimizing the frequency of periodic functions, introduced as a design variable, resulted in the frequency being trapped near the initial value.
One possible explanation for this is that frequency optimization is too under-parameterized a setup and thus has many local minima.
In contrast, the proposed method offers many degrees of design freedom in actuation by expressing it as pulse density over time dimension, the redundancy of which may be advantageous in the solution search with the gradient-based optimization method.

\begin{figure}[tp]
    \centering
    \includegraphics[width=0.9\textwidth]{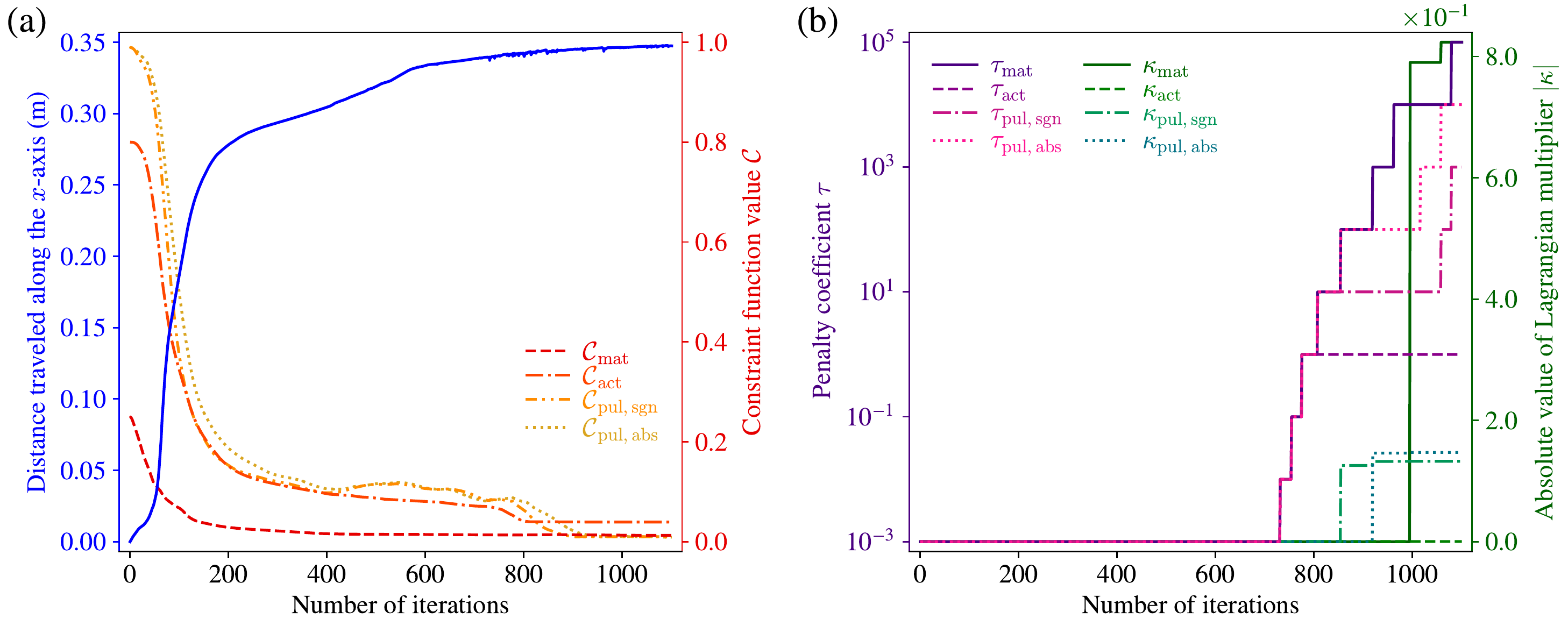}
    \caption{Optimization history of \textit{Walker}. (a)~Changes in the objective and constraint functions during iteration. (b)~Changes in the penalty coefficients and Lagrangian multipliers for each of the constraints during iterations.}
    \label{fig:2d_walker_cur}
\end{figure}

Fig.~\ref{fig:2d_walker_cur} shows the convergence curves of the objective function, constraint functions, penalty coefficients, and Lagrangian multipliers.
As described in Sections~\ref{subsubsec:spat_bin} and \ref{subsubsec:temp_bin}, we imposed four inequality constraints on the degree of binarization of the densities representing the material and actuator layouts in the spatial design domain and arrangement of pulses in the temporal design domain.
Optimization began with an initial condition that maximally violated all constraints but successfully led to a feasible solution while minimizing the objective function.
During the early iterations (0--200~iterations), the optimization of structure (material and actuator layouts) and movement (pulse density) proceeded at a similar rate, as indicated by the values of the four constraint functions decreasing at comparable slopes.
The subsequent several hundred iterations optimize the layout and signal of the actuators with the material layout almost fixed, and the last few hundred iterations are used to search for a feasible solution by updating the penalty coefficients $\vect{\tau}$ and Lagrangian multipliers $\vect{\kappa}$.

As visually presented in Fig.~\ref{fig:2d_walker_opt}, the optimization with constraints on binarization yielded well-binarized densities.
To confirm that the optimized \textit{Walker} hardly utilizes the intermediate densities or void regions with minimum stiffness, we further evaluated its performance when completely binarized by post-processing.
The forward simulation was performed with particles placed only in the solid region ($\gamma > 0.5$), similar to the body-fitted mesh in the FEM.
The optimized material density, actuator layout, and actuation pulses were binarized completely, with the threshold set to the mid-value of the defined range.
As presented in \ref{appendix:suppl}, the post-processed \textit{Walker} still exhibits performance comparable to the optimized \textit{Walker}.
The objective function value evaluated with the post-processed \textit{Walker} was $-0.3466$ (vs. $-0.3474$ with the optimized \textit{Walker}).

In summary, we confirmed that the proposed method successfully achieved simultaneous optimization of the structure, actuator layout, and time-varying actuation of a soft body.
Compared to the designs for horizontal locomotion in previous studies~\cite{cheney2013unshackling, cheney2014evolved, bhatia2021evolution, vandiepen2019spatial, vandiepen2022codesign, hu2020difftaichi, sato2022topology}, the optimized \textit{Walker} shows more complexity in its structure and elaborate movement, with the essence resembling an animal in nature.
The advances in co-design with the proposed method can be attributed to allowing many degrees of design freedom for both structure and self-actuation, which are efficiently optimized using the gradient-based method.
%
\subsubsection{Climber}\label{subsubsec:2d_climber}
Next, we consider designing \textit{Climber} that travels in a vertical direction up the walls against gravity.
As shown in Fig.~\ref{fig:tasks}(b), a spatial design domain of $0.2 \times 0.2$~\si{m^2} square, sandwiched between the left and right walls, was considered.
Similar to the previous example, we expect \textit{Climber} to travel as far as possible in the direction of the $y$-axis during the total simulation time and define the objective function as
\begin{equation}\label{eq:L_2d_climber}
    \mathcal{L}_\mathrm{task} = - \int_{0}^{T}{\vect{v}_\mathrm{g}(t) \cdot \vect{e}_y \, dt},
\end{equation}
where $\vect{e}_y$ is the unit vector in the direction of the $y$-axis.
We set $T = 1$~\si{s}, $A_\mathrm{act} = 1 \times 10^{4}$~\si{Pa}, $R_\mathrm{f} = 1.5\mathrm{\Delta} x$, $p_\mathrm{f} = 2$, $\beta_\mathrm{sig} = 4$, and $\beta_\mathrm{soft} = 4$.
\textit{Climber} with the initial design variables slides down because of gravity, and we experienced that the gradient-based optimization steers \textit{Climber} to hold the initial position (rather than climb) using actuation force.
To avoid the optimized solution being trapped in such local minima, we imposed a break-in period where the gravity acceleration was set to $0$~\si{m.s^{-2}} and increased by $0.98$~\si{m.s^{-2}} every $20$~iterations until it reached $9.8$~\si{m.s^{-2}}.

\begin{figure}[tp]
    \centering
    \includegraphics[width=0.9\textwidth]{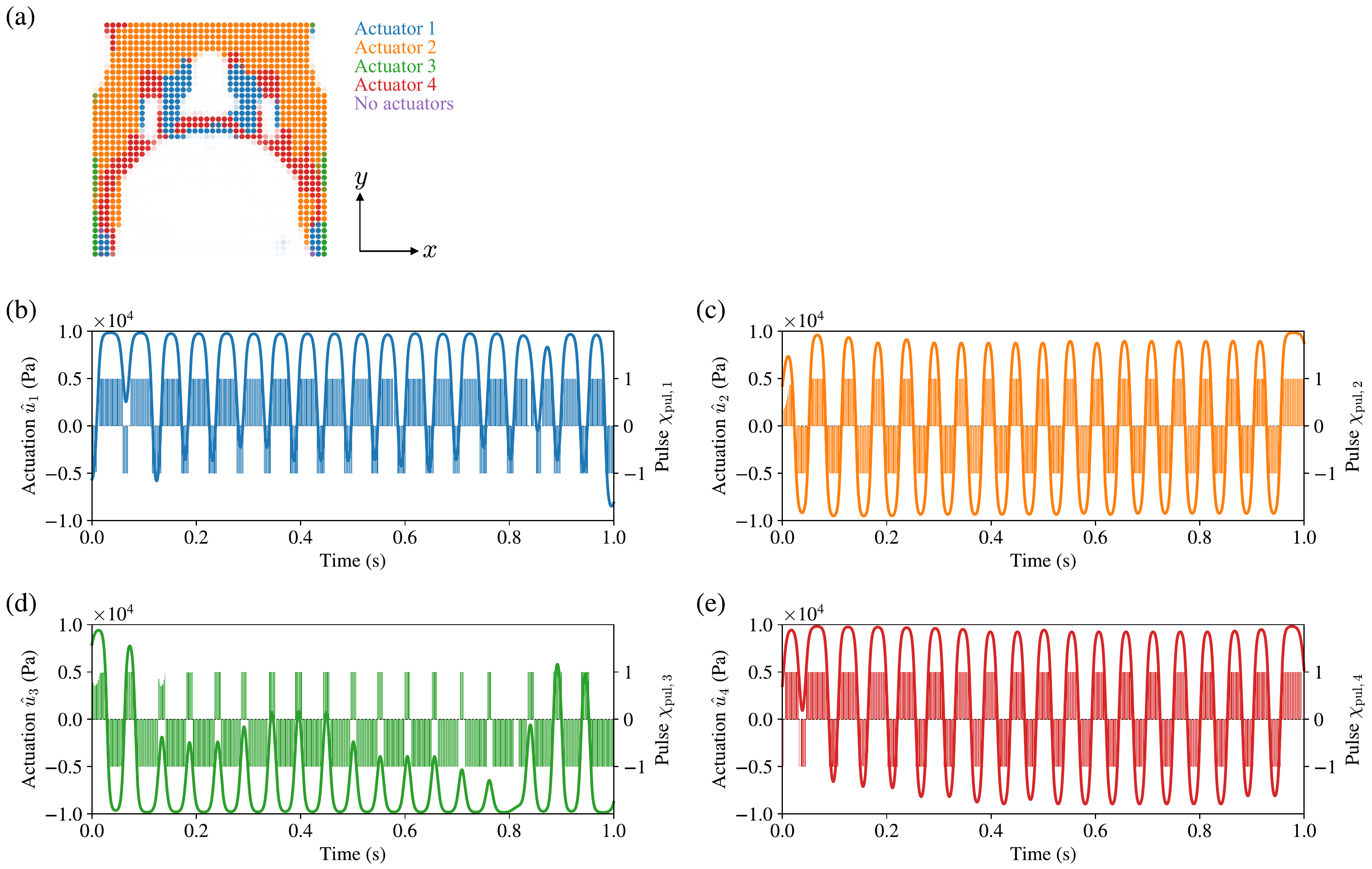}
    \caption{Optimized configuration of \textit{Climber}. (a)~Material and actuator layout. The color and transparency of the particles represent the placed actuator and fictitious material density, respectively. (b)--(e)~Changes in actuation of the respective actuators over the simulation duration. Positive and negative actuation values represent expansion and contraction, respectively.}
    \label{fig:2d_climber_opt}
\end{figure}

\begin{figure}[tp]
    \centering
    \includegraphics[width=1\textwidth]{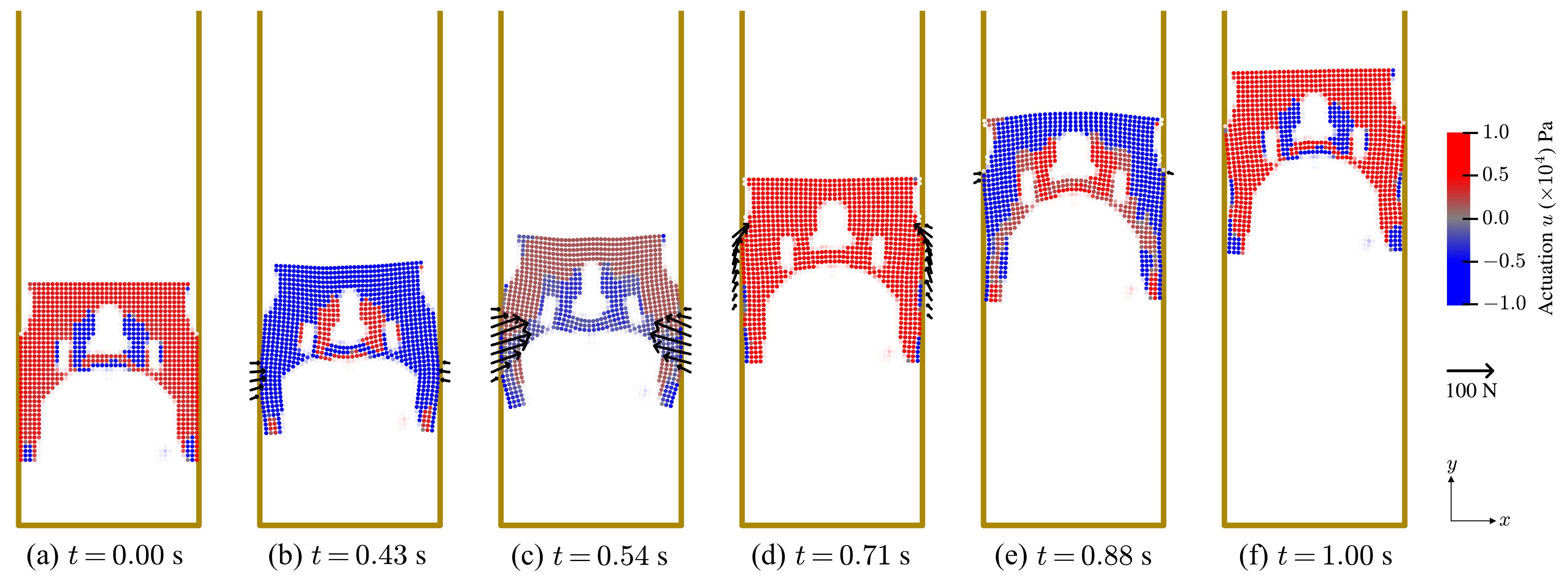}
    \caption{Movement of the optimized \textit{Climber} shown as time-series. Positive (red) and negative (blue) actuation values represent expansion and contraction, respectively. The black arrows on the walls indicate the magnitude and direction of the contact force computed at the background grid nodes. For visibility, particles with a very small fictitious material density ($\gamma < 0.01$) are not shown. See the supplementary video (\ref{appendix:suppl}) for \textit{Climber} in action.}
    \label{fig:2d_climber_per}
\end{figure}

\begin{figure}[tp]
    \centering
    \includegraphics[width=0.6\textwidth]{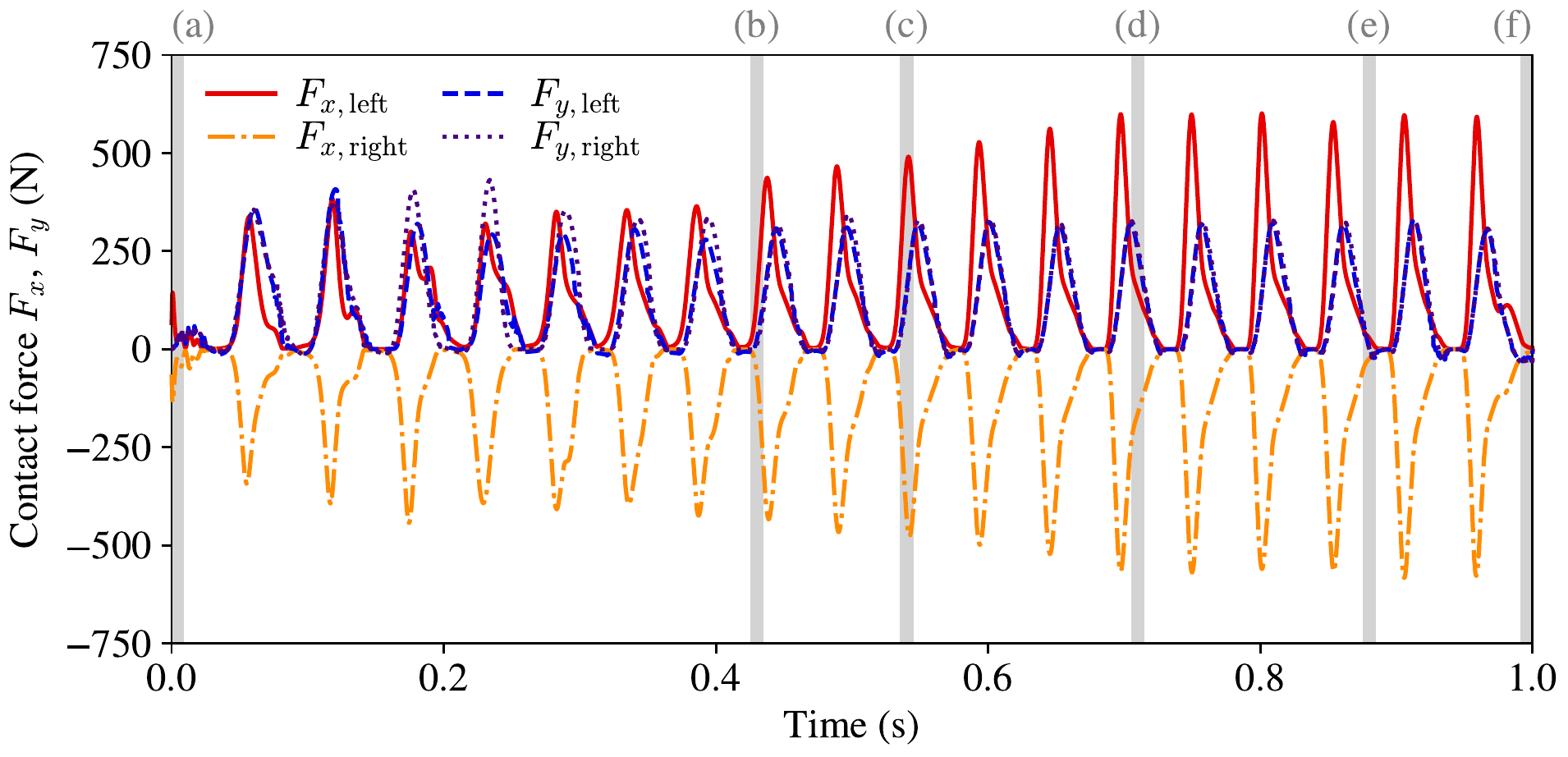}
    \caption{Change in the contact force that \textit{Climber} receives from the walls over time. $F_x$ and $F_y$ denote the $x$- and $y$-components of the contact force, respectively, and are shown separately for those received from the left and right walls. The negative value indicates that the force direction is opposite to the axis. (a)--(f) represent the times shown in Fig.~\ref{fig:2d_climber_per}.}
    \label{fig:2d_climber_con}
\end{figure}

Figs.~\ref{fig:2d_climber_opt} and \ref{fig:2d_climber_per} show the optimized configuration of \textit{Climber} and its movements during simulation time, respectively.
The shape of the optimized \textit{Climber} resembles the letter ``M'';
the orange and red actuators exhibit an inverted U-shape, which is connected by two vertical structures composed of the blue actuator.
\textit{Climber} has multiple structures whose expansion and contraction timings are reversed, allowing it to change shape flexibly.
The placement of three holes within the shape allows for the effective variation of the direction of the actuation forces and the local flexibility such as bending stiffness.
With such an elaborate structure, \textit{Climber} exhibits more complex movements than \textit{Walker}.
\textit{Climber} bends the two vertical structures on the left and right sides that contact the wall inward (Fig.~\ref{fig:2d_climber_per}(c)) and backward (Fig.~\ref{fig:2d_climber_per}(d)) depending on the timing of the actuation.
Accordingly, it travels upward by varying the contact area with the walls, while keeping the body contacted with the walls to prevent sliding down.
The change in contact force over time (Fig.~\ref{fig:2d_climber_con}) shows that the climber receives approximately the same force from both walls, except for the asymmetry observed in $0.1$--$0.4$~\si{s} to restore the disrupted balance.
The slight asymmetry observed in the optimized solution may be attributed to random effects caused by numerical errors, including rounding errors.

Similar to \textit{Walker}, the actuation signals of the optimized \textit{Climber} are almost periodic (Fig.~\ref{fig:2d_climber_opt}(b)--(e)).
The orange and red actuators expand and contract evenly within one cycle, whereas the blue and green actuators expand and contract most of the time, respectively.
These four actuators function in a coordinated manner to accomplish complex movements.
The frequency of actuation is approximately $19$~\si{Hz}, which is far higher than that of \textit{Walker}.
\textit{Climber} automatically found a favorable strategy for climbing up the walls without sliding down as a result of 4D topology optimization, which is to limit the travel distance per actuation cycle while increasing the number of cycles within the given time span.

\begin{figure}[tp]
    \centering
    \includegraphics[width=0.9\textwidth]{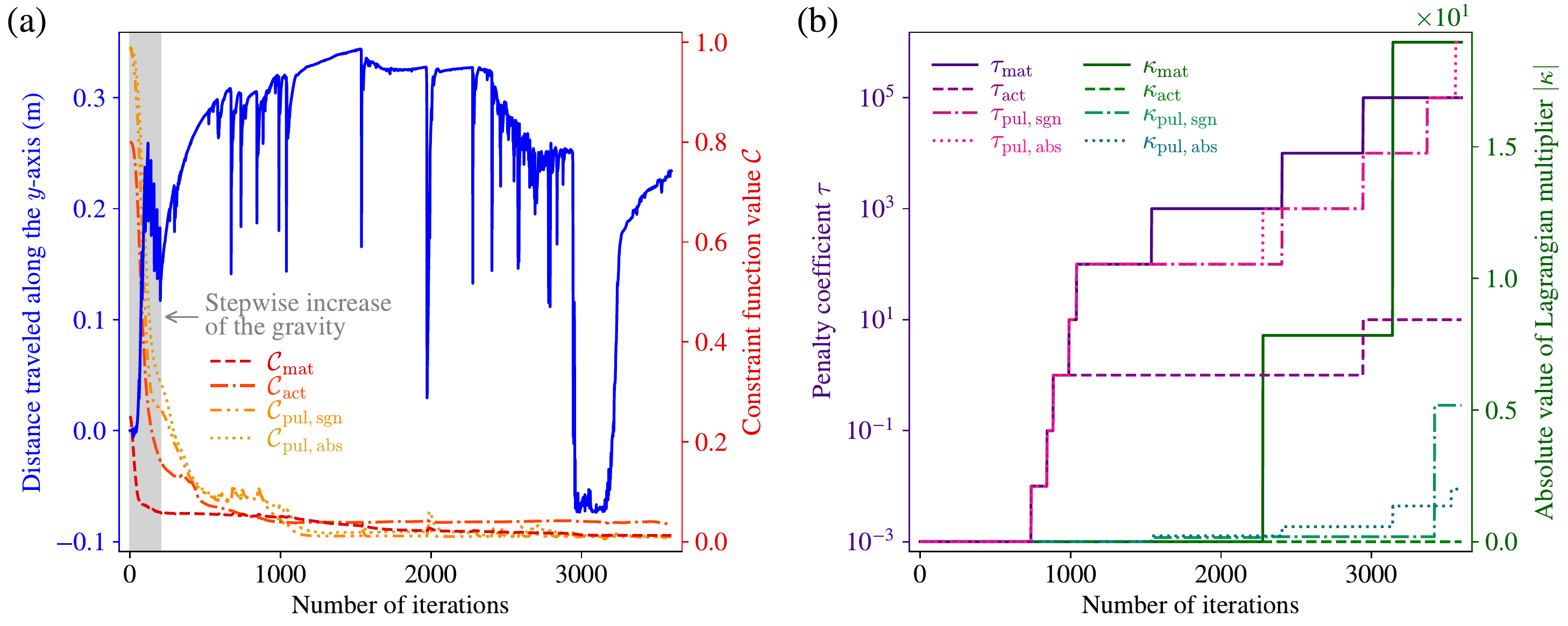}
    \caption{Optimization history of \textit{Climber}. (a)~Changes in the objective and constraint functions during iteration. The shaded area represents the break-in period where the gravity acceleration was increased stepwise. (b)~Changes in the penalty coefficients and Lagrangian multipliers for each of the constraints during iterations.}
    \label{fig:2d_climber_cur}
\end{figure}

The changes in the objective function, four constraint functions, penalty coefficients, and Lagrangian multipliers during the optimization are shown in Fig.~\ref{fig:2d_climber_cur}.
The optimization of \textit{Climber} involves a severe discontinuity in the solution space such that a small change in the design variables can result in a drastic increase in the objective function (sliding down to the floor).
For this reason, the convergence curve of the travel distance (blue line in Fig.~\ref{fig:2d_climber_cur}(a)) demonstrates abrupt drops mainly after updating $\vect{\tau}$ or $\vect{\kappa}$.
Nevertheless, Adam, each time managed to recover performance rapidly and eventually led to a feasible solution that can climb up the walls.
Interestingly, \textit{Climber} recovered the performance even after it slid down entirely to the floor (showing a negative distance) around $3000$~iterations.

The optimization resulted in a well-binarized solution, as visualized in Fig.~\ref{fig:2d_climber_opt}.
The additional evaluation of \textit{Climber}'s performance when completely binarized by post-processing (\ref{appendix:suppl}) confirms that the post-processed \textit{Climber} does not lose its ability to climb walls, although the objective function value increases to $-0.1512$ (vs. $-0.2340$ with the optimized solution).
%
\subsubsection{Balancer}\label{subsec:2d_balancer}
To demonstrate the effectiveness of our approach for a variety of dynamic tasks, not limited to locomotion, we consider designing \textit{Balancer} that controls its posture to maintain the position regardless of the forced vibration of the floor.
In this example, the spatial design domain is considered to be a vertical rectangle ($0.1 \times 0.4$~\si{m^2}) embedded in the floor, as shown in Fig.~\ref{fig:tasks}(c).
The floor is assumed to be forced to vibrate in the $x$-axis direction according to $0.03 \sin{(\omega t)}$~\si{m} with an angular frequency $\omega$ set to $40$~\si{rad.s^{-1}}.
We expect \textit{Balancer} to maintain the initial position of the tip, which leads to the objective function defined as the time-averaged deviation:
\begin{equation}\label{eq:L_2d_balancer}
    \mathcal{L}_\mathrm{task} = \dfrac{1}{T} \int_{0}^{T}{\lpnorm{\vect{x}_\mathrm{tip}(t) - \vect{x}_\mathrm{tip}(0)}{2} \, dt},
\end{equation}
where $\vect{x}_\mathrm{tip}(t)$ is the center of the particles representing the tip at time $t$.
We set $T = 1$~\si{s}, $A_\mathrm{act} = 2 \times 10^{4}$~\si{Pa}, $R_\mathrm{f} = 2\mathrm{\Delta} x$, $p_\mathrm{f} = 3$, $\beta_\mathrm{sig} = 4$, and $\beta_\mathrm{soft} = 4$.

\begin{figure}[tp]
    \centering
    \includegraphics[width=0.9\textwidth]{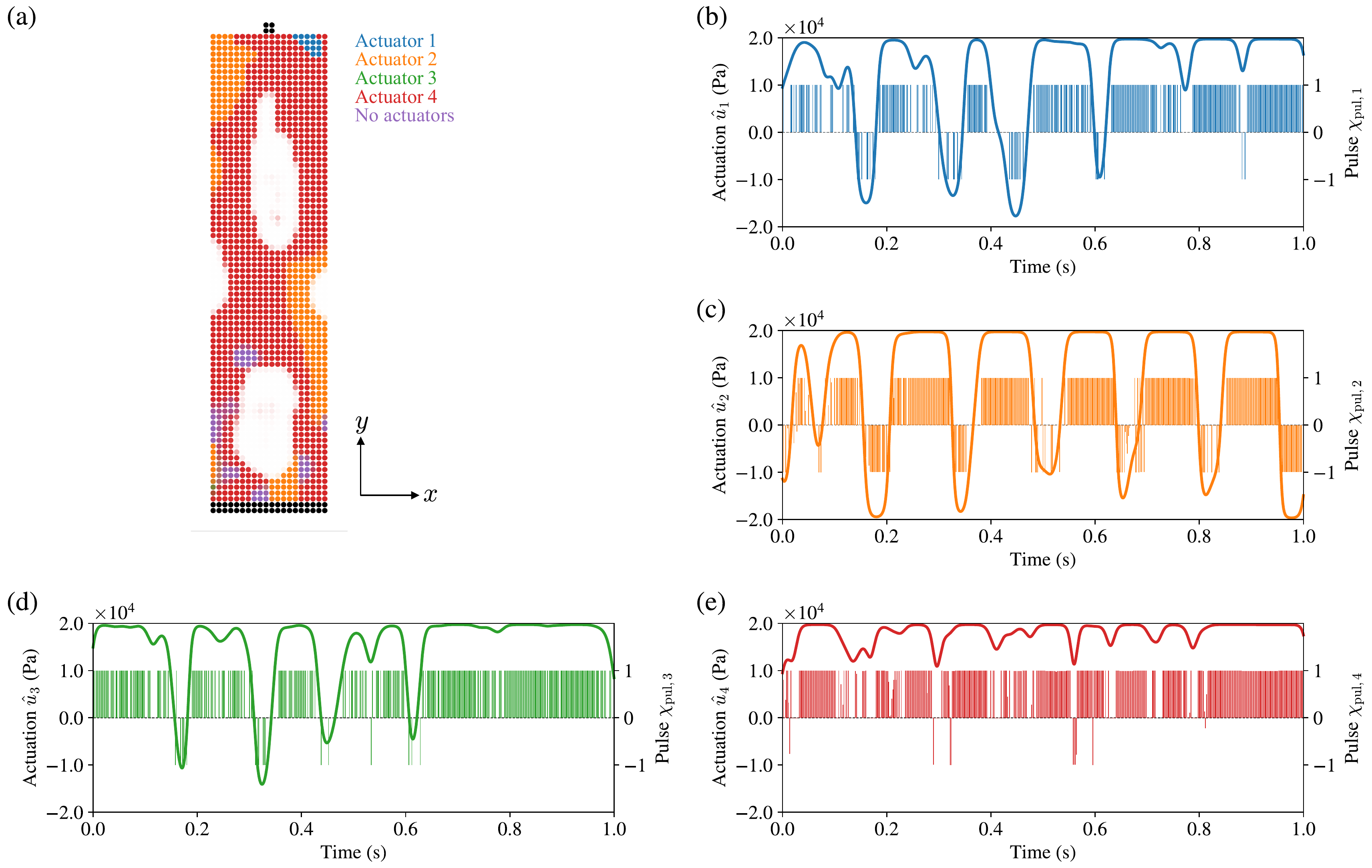}
    \caption{Optimized configuration of \textit{Balancer}. (a)~Material and actuator layout. The color and transparency of the particles represent the placed actuator and fictitious material density, respectively. (b)--(e)~Changes in actuation of the respective actuators over the simulation duration. Positive and negative actuation values represent expansion and contraction, respectively.}
    \label{fig:2d_balancer_opt}
\end{figure}
\begin{figure}[tp]
    \centering
    \includegraphics[width=1\textwidth]{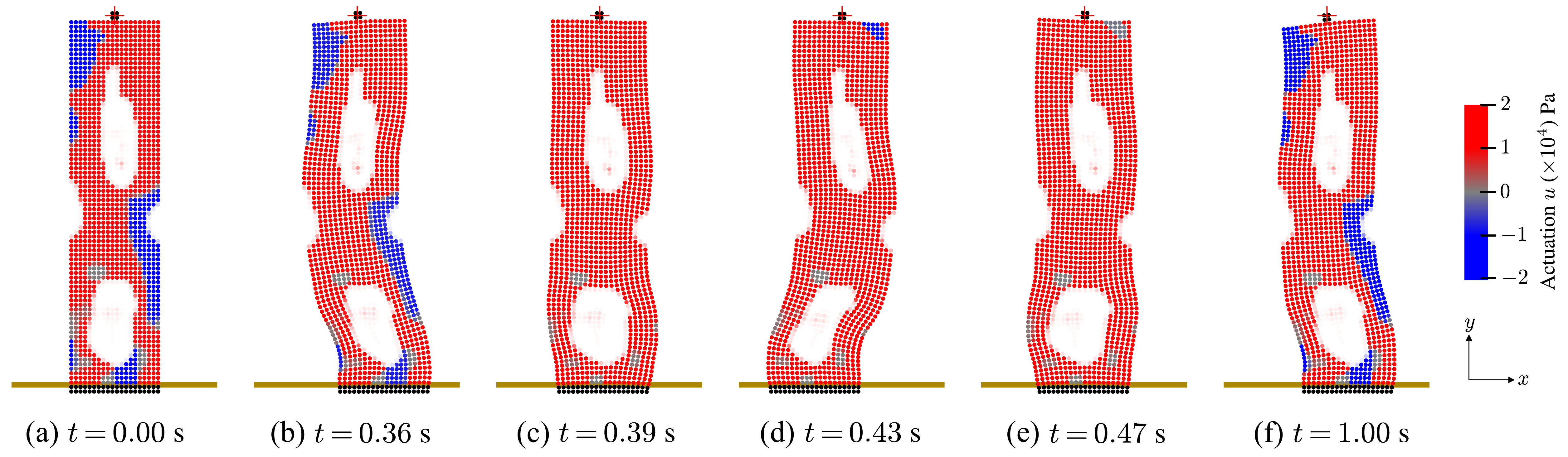}
    \caption{Movement of the optimized \textit{Balancer} shown as time-series. Positive (red) and negative (blue) actuation values represent expansion and contraction, respectively. For visibility, particles with a very small fictitious material density ($\gamma < 0.01$) are not shown. See the supplementary video (\ref{appendix:suppl}) for \textit{Balancer} in action.}
    \label{fig:2d_balancer_per}
\end{figure}
The optimized configuration of \textit{Balancer} and its movement are shown in Figs.~\ref{fig:2d_balancer_opt} and \ref{fig:2d_balancer_per}, respectively.
The optimized \textit{Balancer} has two large holes aligned vertically within the shape, resembling the digit ``8.''
Most of the structure is occupied by the red actuator, and the orange actuator is arranged symmetrically about the center of the shape.
The domain of the blue actuator is only limited to a small portion.
In addition, \textit{Balancer} does not use the green actuator, indicating that the number of actuators used is optimized favorably to the task.
In contrast to the two numerical examples above, there is a domain (purple) with no actuators placed, which moves passively in response to the deformation of the neighboring material.

The optimized \textit{Balancer} exhibits a periodic movement.
The sequence of motions comprising one cycle (Fig.~\ref{fig:2d_balancer_per}(b)--(e)) depicts that \textit{Balancer} holds its tip position by periodically bending its structure, primarily the lower half, in response to floor vibrations.
The red actuator, occupying most of the structure, constantly expands to counteract the longitudinal compression caused by self-weight.
The bending of the structure in the opposite direction to the floor movement is accomplished by the contraction of the orange actuator.
The blue actuator is responsible for the fine-grained control of the tip position.
In analogy to animals, the red and orange actuators are large muscles, such as the muscles exercising the thighs or trunk, while the blue actuator is a small muscle that regulates subtle forces, such as the muscles exercising the fingers.
The actuation signals have a frequency of approximately $6$~\si{Hz}, successfully reflecting the floor vibration frequency $40~\si{rad.s^{-1}} \sim 6.4~\si{Hz}$ as a result of the optimization of the pulse density in the temporal design domain.

\begin{figure}[tp]
    \centering
    \includegraphics[width=0.9\textwidth]{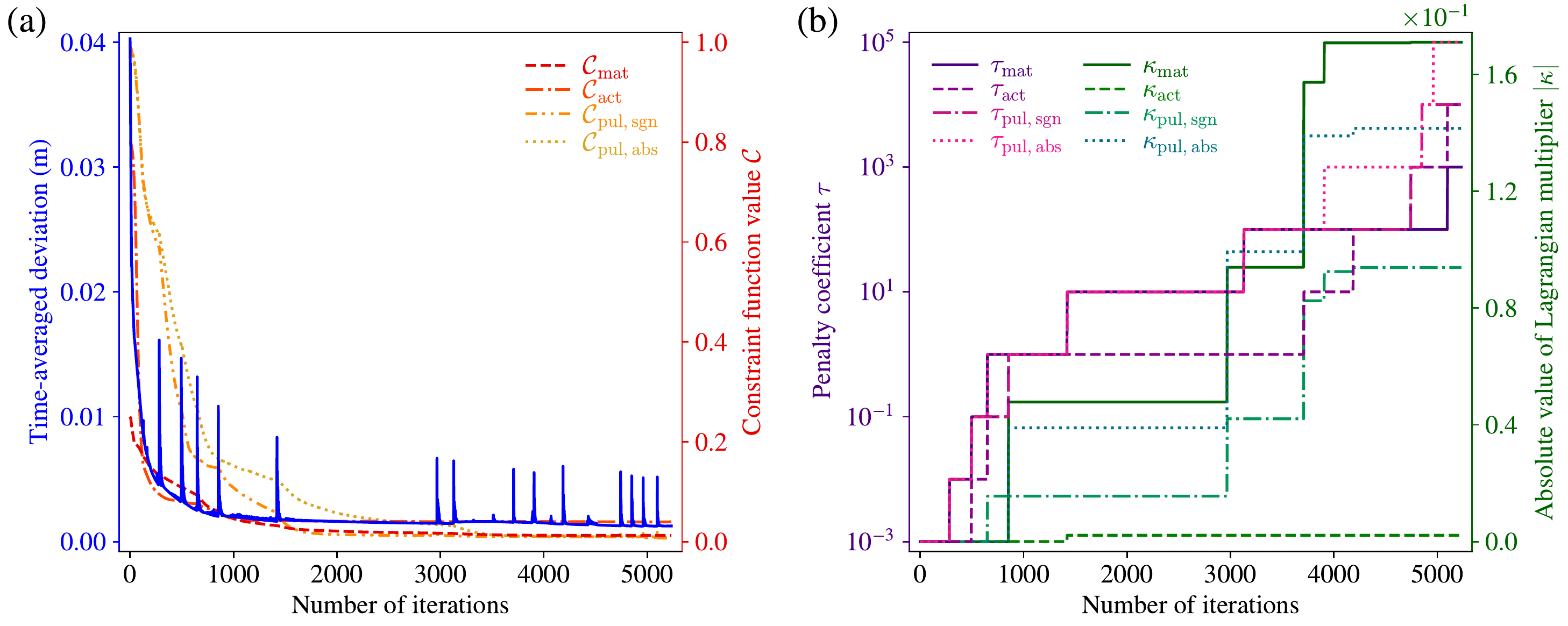}
    \caption{Optimization history of \textit{Balancer}. (a)~Changes in the objective and constraint functions during iteration. (b)~Changes in the penalty coefficients and Lagrangian multipliers for each of the constraints during iterations.}
    \label{fig:2d_balancer_cur}
\end{figure}

As shown in Fig.~\ref{fig:2d_balancer_cur}, the objective and constraint functions showed a stable convergence performance, despite some spike-like jumps in the objective function at the timing of the $\vect{\tau}$ and $\vect{\kappa}$ updates.
Compared to the material and actuator layouts ($\mathcal{C}_\mathrm{mat}$ and $\mathcal{C}_\mathrm{act}$, respectively), the pulse density ($\mathcal{C}_\mathrm{pul, sgn}$ and $\mathcal{C}_\mathrm{pul, abs}$) required a larger number of iterations to satisfy the constraint.
One explanation for this is the nonlinearity of the passive waving movement of the bar-shaped body caused by the floor vibration.
Enforcing the binarization constraint would require modification of each part of the actuation signal represented by the intermediate pulse density, which affects the posture of the body at a later time nonlinearly, thus requiring many iterations to modify the pulse density over the entire simulation time.

We provide in \ref{appendix:suppl} the result of an additional forward simulation conducted to evaluate the performance of \textit{Balancer} completely binarized by post-processing.
Despite some worsening of the objective function value to $0.02263$ (vs. $0.001257$ with the optimized \textit{Balancer}), \textit{Balancer} maintains the initial position of the tip for a long duration even after post-processing, confirming that the performance of the optimized \textit{Balancer} does not largely depend on intermediate densities.
%
\subsection{Design in 3D space and time}\label{subsec:3d}
%
\subsubsection{3D Walker}\label{subsubsec:3d_walker}
We further apply our approach to design problems with spatial dimensions of 3D.
The first example is the design of \textit{3D\,Walker}, a 3D version of \textit{Walker} (Section~\ref{subsubsec:2d_walker}).
Adding a $z$-axis in the depth direction, we consider a cubic spatial design domain of $0.1 \times 0.1 \times 0.1$~\si{m^3}, as shown in Fig.~\ref{fig:tasks}(d).
The objective function is set as in Eq.~\eqref{eq:L_2d_walker}, i.e., we design a soft body that travels as far as possible in the direction of the $x$-axis during the simulation time span.
We set $T = 0.5$~\si{s}, $A_\mathrm{act} = 2 \times 10^{4}$~\si{Pa}, $R_\mathrm{f} = 2\mathrm{\Delta} x$, $p_\mathrm{f} = 3$, $\beta_\mathrm{sig} = 8$, and $\beta_\mathrm{soft} = 8$.

\begin{figure}[tp]
    \centering
    \includegraphics[width=0.85\textwidth]{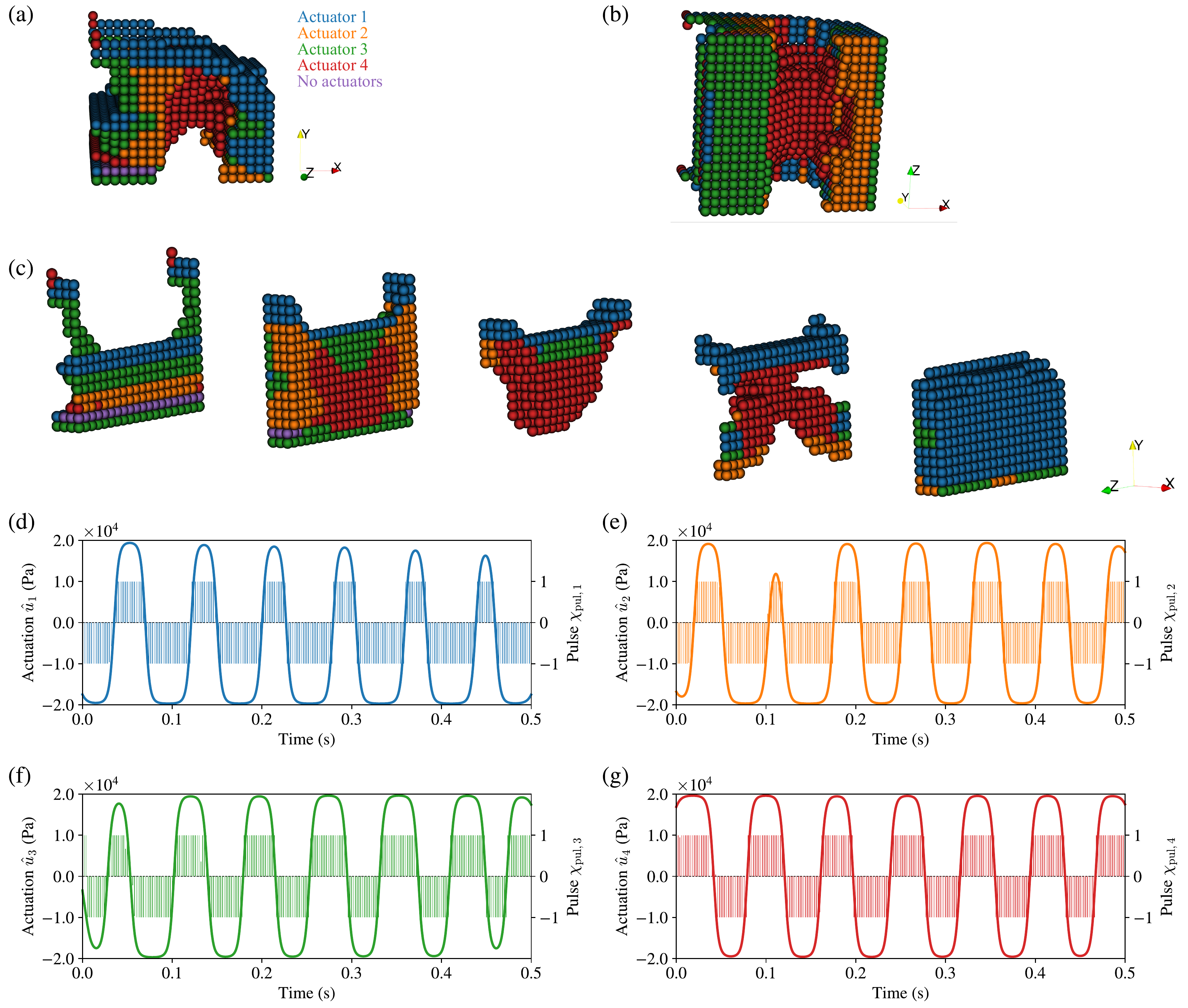}
    \caption{Optimized configuration of \textit{3D\,Walker}. Material and actuator layout: (a)~side view, (b)~bottom view, and (c)~cross-sections parallel to the $yz$-plane. The color of the particles represents the placed actuator. For visibility, only particles with fictitious material densities larger than the mid-value ($\gamma > 0.5$) are shown. (d)--(g)~Changes in actuation of the respective actuators over the simulation duration. Positive and negative actuation values represent expansion and contraction, respectively.}
    \label{fig:3d_walker_opt}
\end{figure}

\begin{figure}[tp]
    \centering
    \includegraphics[width=1\textwidth]{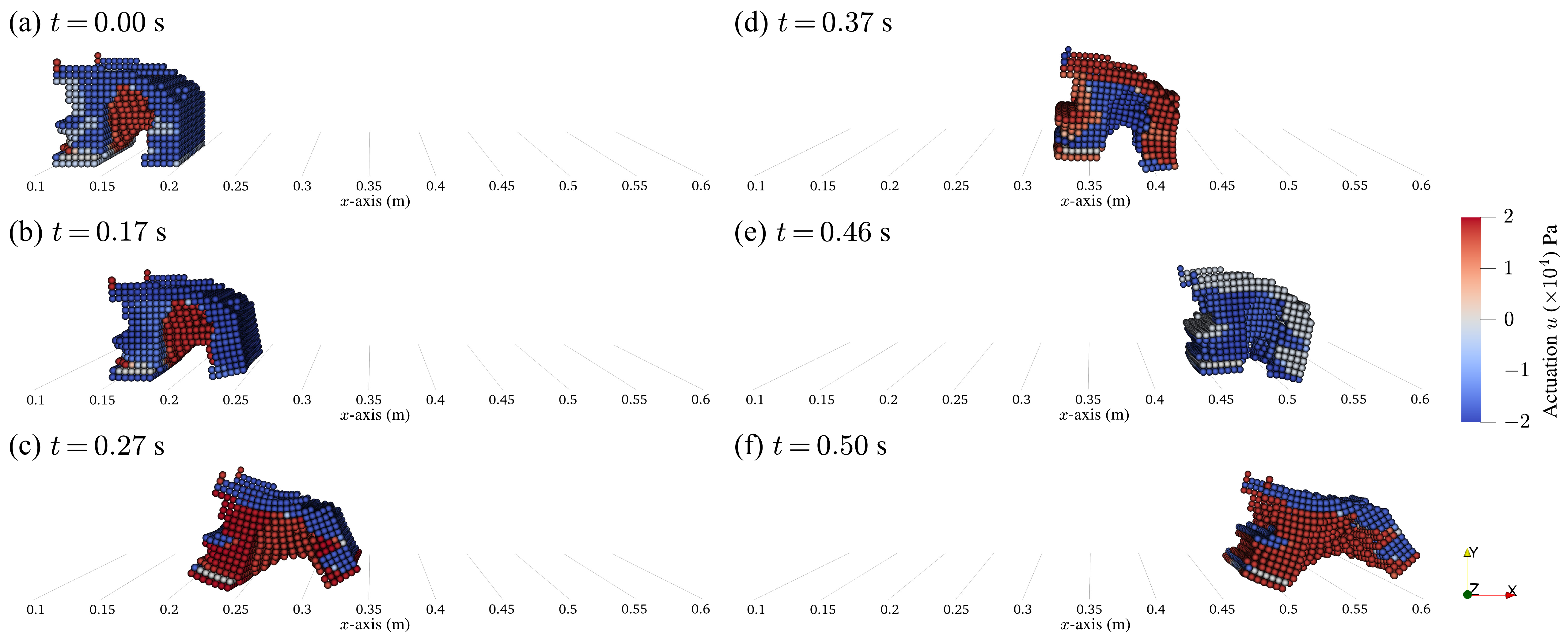}
    \caption{Movement of the optimized \textit{3D\,Walker} shown as time-series. Positive (red) and negative (blue) actuation values represent expansion and contraction, respectively. For visibility, only particles with fictitious material densities larger than the mid-value ($\gamma > 0.5$) are shown. See the supplementary video (\ref{appendix:suppl}) for \textit{3D\,Walker} in action.}
    \label{fig:3d_walker_per}
\end{figure}

Figs.~\ref{fig:3d_walker_opt} and \ref{fig:3d_walker_per} show the optimized configuration of \textit{3D\,Walker} and its movements within the simulation time span, respectively.
The optimized \textit{3D\,Walker} exhibits a car-like shape.
Its side view (Fig.~\ref{fig:3d_walker_opt}(a)) is similar to the shape of \textit{Walker}.
However, the structure is not uniform with respect to the $z$-axis direction.
The structure is bilaterally symmetric when viewed in the direction of travel.
\textit{3D\,Walker} has two leg-like structures with a large contact area with the floor.
The front and rear legs are joined by the red actuator that widens or shortens the distance between the legs, resembling an abdominal muscle of animals.
The red actuator is Y-shaped, extending from the rear leg and joining to the front leg after bifurcation.
Two wing-like structures on either side of the top appear to be counterweighted.

\textit{3D\,Walker} travels in the designed direction by stepping periodically with its two legs.
Its dexterous locomotion, executed according to multiple actuation signals synchronized in frequency but out of phase, resembles a gallop, which is the fastest animal gait.
The frequency of the optimized actuation signal is approximately $12$~\si{Hz}.

\begin{figure}[tp]
    \centering
    \includegraphics[width=0.9\textwidth]{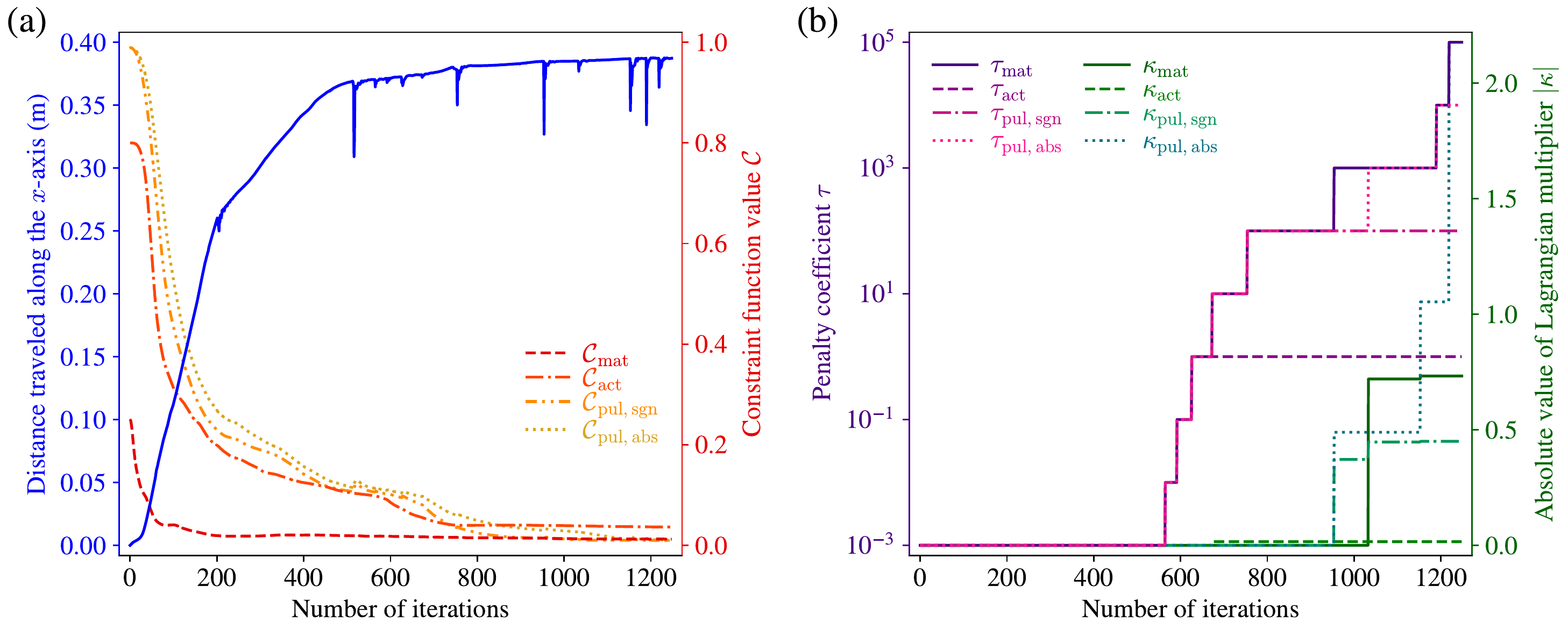}
    \caption{Optimization history of \textit{3D\,Walker}. (a)~Changes in the objective and constraint functions during iteration. (b)~Changes in the penalty coefficients and Lagrangian multipliers for each of the constraints during iterations.}
    \label{fig:3d_walker_cur}
\end{figure}

The convergence curves of \textit{3D\,Walker} (Fig.~\ref{fig:3d_walker_cur}) show two phases: a phase where the values of the objective and constraint functions simultaneously decrease considerably and the subsequent phase where the constraint function values continue to decrease without worsening the objective function value by updating $\vect{\tau}$ and $\vect{\kappa}$.
The feasible solution obtained by optimization is sufficiently binarized, as supported by the result that the travel distance of \textit{3D\,Walker} remains almost the same even when completely binarized by post-processing (\ref{appendix:suppl}).
The objective value evaluated with the post-processed \textit{3D\,Walker} was $-0.3893$, showing a slight improvement over that with the optimized \textit{3D\,Walker}, $-0.3873$.
%
\subsubsection{Rotator}\label{subsubsec:3d_rotator}
The last example is to design \textit{Rotator} whose desired task is rotation in 3D space.
As shown in Fig.~\ref{fig:tasks}(e), we consider a spatial design domain of $0.1 \times 0.1 \times 0.1$~\si{m^3} cube placed on the floor.
In this example, we assume the desired direction of rotation to be clockwise about the positive $y$-axis of the global coordinates and maximize the angle of rotation around the center of gravity over the entire simulation time.
The objective function is defined as
\begin{equation}\label{eq:L_3d_rotator}
    \mathcal{L}_\mathrm{task} = \int_{0}^{T}{ \dfrac{
        \int_{\mathrm{\Omega}_x(t)}{\rho (\vect{x} - \vect{x}_g) \times (\vect{v} - \vect{v}_g) \, d\vect{x}}
    }{
        \int_{\mathrm{\Omega}_x(t)}{\rho \lpnorm{\vect{x} - \vect{x}_g}{2}^2 \; d\vect{x}}
    } \cdot \vect{e}_y \, dt},
\end{equation}
where $\vect{x}_g$ is the center of gravity at time $t$, given as
\begin{equation}\label{eq:x_g}
    \vect{x}_\mathrm{g}(t) = \dfrac{
        \int_{\mathrm{\Omega}_x(t)}{\rho \vect{x} \, d\vect{x}}
    }{
        \int_{\mathrm{\Omega}_x(t)}{\rho \, d\vect{x}}
    },
\end{equation}
and $\vect{v}_g$ is the mass-weighted average velocity, defined as Eq.~\eqref{eq:v_g}.
The numerator and denominator of the Eq.~\eqref{eq:L_3d_rotator} are the angular momentum and moment of inertia about the center of gravity, respectively.
The objective function represents the angle of rotation about the positive $y$-axis (positive when counterclockwise); minimizing it maximizes the clockwise angle of rotation.
In this example, we set $T = 1$~\si{s}, $A_\mathrm{act} = 2 \times 10^{4}$~\si{Pa}, $R_\mathrm{f} = 1.5\mathrm{\Delta} x$, $p_\mathrm{f} = 2$, $\beta_\mathrm{sig} = 8$, and $\beta_\mathrm{soft} = 8$.

\begin{figure}[tp]
    \centering
    \includegraphics[width=0.85\textwidth]{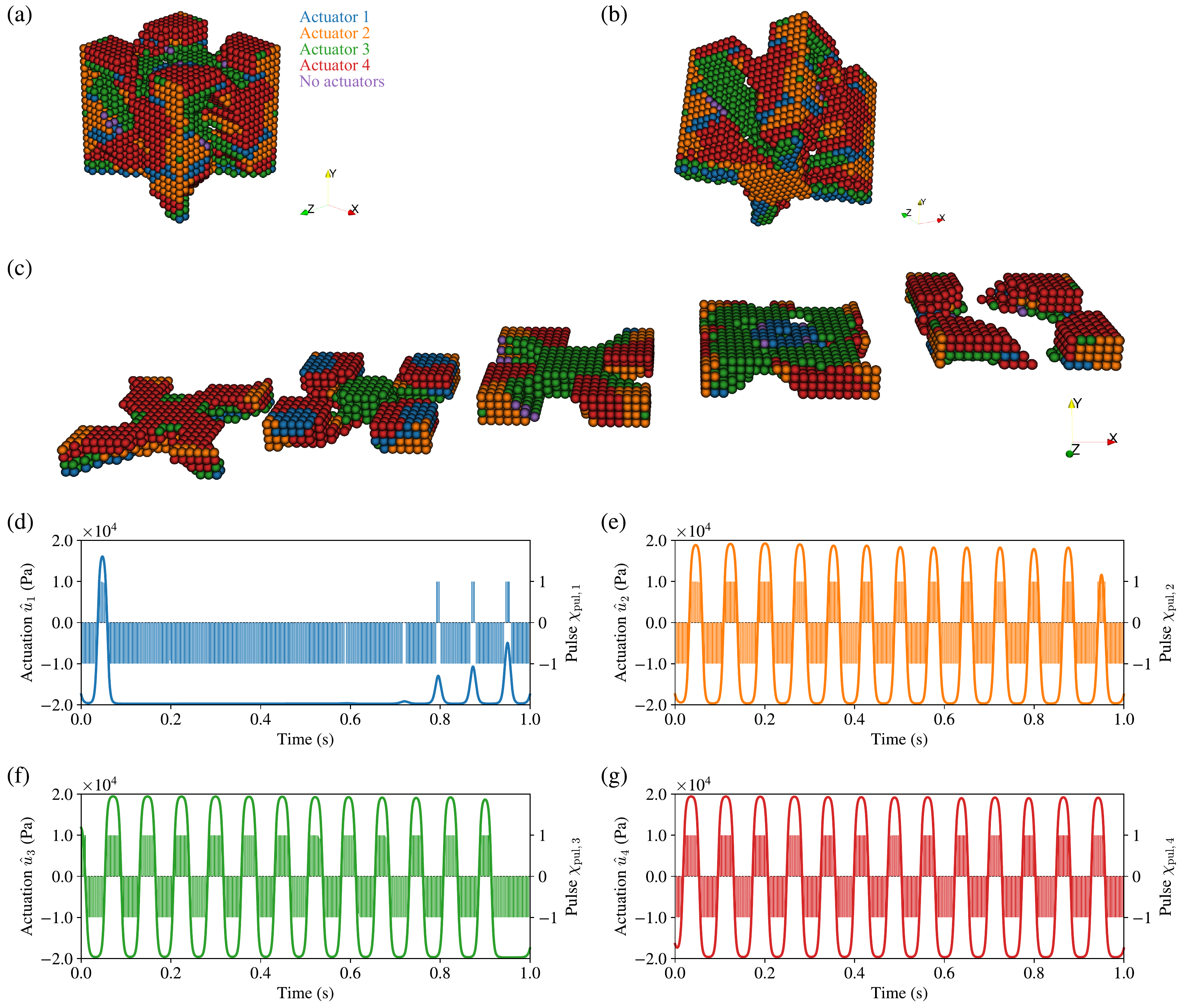}
    \caption{Optimized configuration of \textit{Rotator}. Material and actuator layout: (a)~top diagonal view, (b)~bottom diagonal view, and (c)~cross-sections parallel to the $xz$-plane. The color of the particles represents the placed actuator. For visibility, only particles with fictitious material densities larger than the mid-value ($\gamma > 0.5$) are shown. (d)--(g)~Changes in actuation of the respective actuators over the simulation duration. Positive and negative actuation values represent expansion and contraction, respectively.}
    \label{fig:3d_rotator_opt}
\end{figure}

\begin{figure}[tp]
    \centering
    \includegraphics[width=1\textwidth]{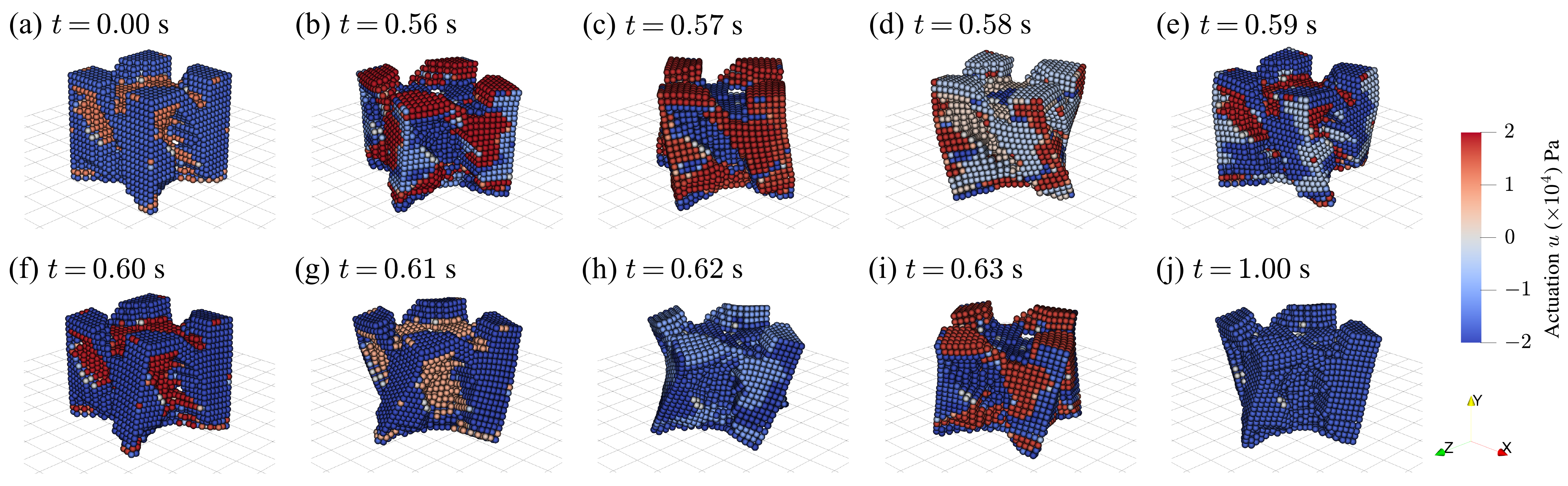}
    \caption{Movement of the optimized \textit{Rotator} shown as time-series. Positive (red) and negative (blue) actuation values represent expansion and contraction, respectively. For visibility, only particles with fictitious material densities larger than the mid-value ($\gamma > 0.5$) are shown. See the supplementary video (\ref{appendix:suppl}) for \textit{Rotator} in action.}
    \label{fig:3d_rotator_per}
\end{figure}

The optimized configuration of \textit{Rotator} and its movement in the simulated time span are depicted in Figs.~\ref{fig:3d_rotator_opt} and \ref{fig:3d_rotator_per}, respectively.
The optimized \textit{Rotator} exhibits lateral faces resembling the letter ``N,'' having four legs perpendicular to the floor.
Its cross-sections parallel to the $xz$-plane are shaped similarly to a fan blade.
\textit{Rotator} rotates periodically by jumping in the vertical direction with its four legs and twisting the body.
The ground-kicking action is alternated by the four legs and the cylindrical structure on the axis of rotation, respectively, during one cycle.
During the latter, the four legs are twisted back in the air.

As seen in Fig.~\ref{fig:3d_rotator_opt}(d)--(g), the actuation signals are periodic, except for that of the blue actuator, with a frequency of approximately $13$~\si{Hz}.
The blue actuator is placed in small amounts throughout the body, such as the feet bottoms in contact with the ground.
The actuation signal of the blue actuator is time-varying during the first jump for acceleration and the last few jumps for subtle balancing adjustments, and otherwise remains fully contracted throughout.

\begin{figure}[tp]
    \centering
    \includegraphics[width=0.9\textwidth]{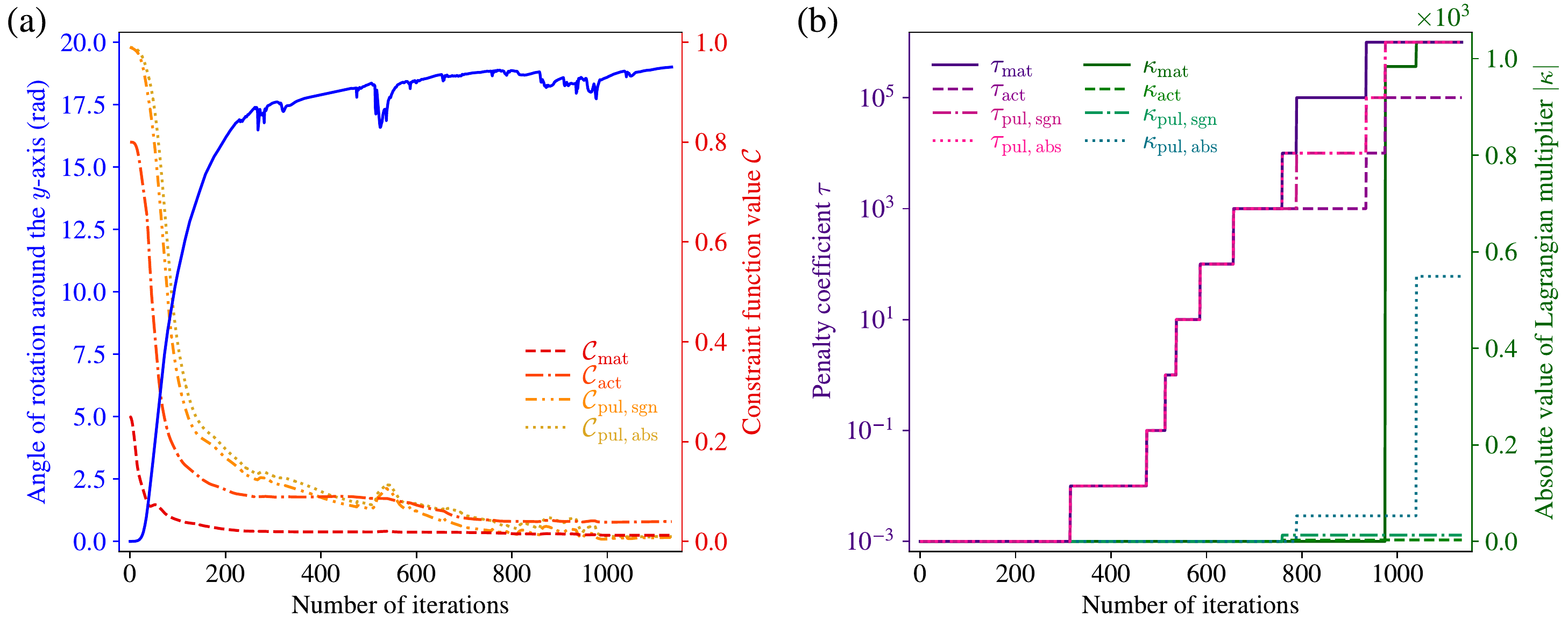}
    \caption{Optimization history of \textit{Rotator}. (a)~Changes in the objective and constraint functions during iteration. (b)~Changes in the penalty coefficients and Lagrangian multipliers for each of the constraints during iterations.}
    \label{fig:3d_rotator_cur}
\end{figure}

We observe in this example as well that the augmented Lagrangian method with Adam shows stable convergence performance by steadily decreasing the objective and constraint functions with adequate updates of $\vect{\tau}$ and $\vect{\kappa}$ (Fig.~\ref{fig:3d_rotator_cur}).
In addition, the optimized \textit{Rotator} demonstrates a good performance even when completely binarized by post-processing (\ref{appendix:suppl}), verifying that the imposed constraints lead to sufficiently binarized densities.
The respective objective function values evaluated with the optimized and post-processed \textit{Rotator} were $-19.01$ and $-17.25$.
%
\section{Conclusion}\label{sec:conclusion}
In this study, we devised 4D topology optimization, a method for simultaneously optimizing the mutually influencing structure and control of self-actuating soft bodies.
By representing the structure and self-actuation of soft bodies as multi-indexed and hierarchized densities distributed over the spatiotemporal domain, the co-design problem was efficiently solved using the gradient-based optimization algorithm without reducing the degrees of design freedom.
We provided five numerical examples of designing self-actuating soft bodies for various dynamic tasks, including locomotion, posture control, and rotation.
The optimized soft bodies exhibited structures similar to those of animals, such as legs, muscles, and tendons, and performed the desired task by dexterously expanding and contracting their flexible structure according to multiple actuation signals.
In particular, the soft bodies showed periodic movements, resembling animal locomotion, at a frequency suitable for maximizing performance on each task, despite no assumptions being made regarding periodicity.
In addition, we confirmed that the optimized solution is well-binarized despite continuous relaxation by imposing explicit constraints on the binarization and solving the constrained optimization problem with the augmented Lagrangian method.
These results demonstrate that our method successfully designs self-actuating soft bodies with complex shapes and biomimetic movements, utilizing its many degrees of design freedom.

One potential direction for future exploration is to extend the application of the proposed method to fabricate real-world solutions.
The numerical examples provided in this paper were primarily intended to demonstrate the effectiveness of 4D topology optimization in discovering the essence of optimal structure and movement within a given scenario.
In line with this goal, the modeling of the physical system was kept simple.
However, it is worth noting that the proposed optimization method can also accommodate designs with specific conditions by incorporating more intricate models tailored to practical applications.
For example, one could consider incorporating advanced boundary conditions for collision and contact or formulating actuation that takes into account specific types of actuators, such as pneumatic actuators.
Moreover, it is possible to address more complex dynamic tasks or introduce constraints to further refine the solution.
At present, there exists a limitation in directly fabricating the soft body designed in this study due to the absence of soft actuators capable of expanding and contracting according to predefined profiles over time.
Nevertheless, we are optimistic that future advancements in soft actuator technologies and the development of bio-robots utilizing living cells will help overcome this limitation.
%
\section*{Acknowledgments}
The authors would like to thank Dr.~Noboru Kikuchi for the fruitful discussion.
%
\appendix
\setcounter{figure}{0}
\renewcommand\thefigure{A.\arabic{figure}}
%
\section{Supplementary video}\label{appendix:suppl}
We provide the supplementary video on YouTube (\url{https://youtu.be/sPY2jcAsNYs}) to enhance the interpretation of the results presented in Section~\ref{sec:numer}.
The video includes an animated history of the optimization (changes in the material layout, actuator layout, and time-varying actuation with respect to the number of iterations) and a comparison of the movements between the initial, optimized, and post-processed (for binarization) soft bodies for each numerical example.
It also contains the results of additional investigations described in the following appendices.
%
\section{Periodicity of optimized actuation}\label{appendix:periodicity}
The optimized actuation signals are roughly periodic but not precisely, as can be seen from the \textit{Walker} example (Section~\ref{subsubsec:2d_walker}).
The walking motion of \textit{Walker} is synchronized with the blue actuator that exhibits a clear periodic signal (Fig.~\ref{fig:2d_walker_opt}(b)).
The signals of the other actuators (Fig.~\ref{fig:2d_walker_opt}(c)--(e)) are roughly synchronized with this signal but disturbed by spike-like pulses with the opposite sign to the periodic background signal.
To investigate how intermittently disturbing the actuation signal affects performance, we conducted a forward simulation with the signal manually compensated for a neater periodicity (Fig.~\ref{fig:periodicity_opt}).
We provide the result in the supplementary video (\ref{appendix:suppl}), where \textit{Walker} still exhibits good performance with compensated actuation, demonstrating that its walking motion is achieved by synchronizing the four periodic actuation signals.
The slightly worsened objective function value to $-0.3237$ (vs. $-0.3474$) suggests that disturbing the periodic background signal with instantaneous pulses is an attempt to fine-tune the posture (i.e., contact angle and timing) for better performance.
\begin{figure}[tp]
    \centering
    \includegraphics[width=0.9\textwidth]{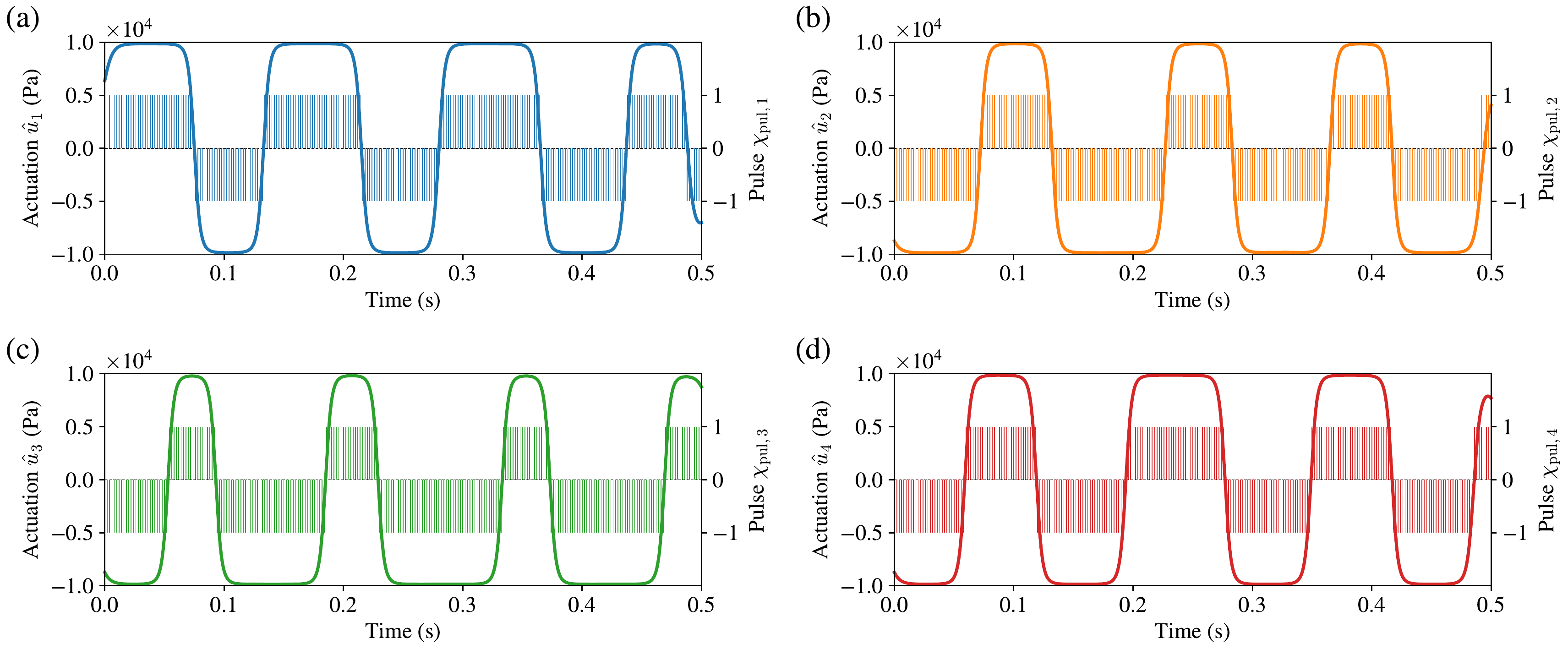}
    \caption{Actuation signals of \textit{Walker} manually compensated for neater periodicity.}
    \label{fig:periodicity_opt}
\end{figure}

In summary, the optimized actuation signals are generally synchronized with the periodic movement of the soft body.
Still, they are disturbed intentionally at some times to maintain cyclic motion throughout the transient period where the body accelerates from rest.
The design of such complicated actuation with flexible frequency and phase changes can be attributed to utilizing a high degree of design freedom expressed as pulse density over time dimension.
%
\section{Effect of total simulation time and number of actuators}\label{appendix:effect}
Exploring the impact of varying hyperparameters on optimization results is crucial for understanding and applying the method.
In addition to the hyperparameters discussed in Section~\ref{subsubsec:hyperparams}, we conducted further investigations to understand the influence of parameters included in the physical system.
Notably, we found that the total simulation time, $T$, considerably impacts the optimization outcomes.
Furthermore, we also explored the effects of varying the number of actuators, $N_\mathrm{act}$, which was fixed at $4$ in the main text.
To discuss the impact of these parameters, we revisit the \textit{Walker} example in Section~\ref{subsubsec:2d_walker}.

First, we optimized \textit{Walker} using shorter ($T = 0.25$~\si{s}) and longer ($T = 1$~\si{s}) durations, in contrast to the $0.5$~\si{s} duration mentioned in the main text.
All remaining parameters were kept the same, including $\mathrm{\Delta} t$ (the time step size) and $\mathrm{\Delta} t_\mathrm{pul}$ (the time interval between successive pulses).
The resulting soft bodies are depicted in Figs.~\ref{fig:eff_params}(a) and \ref{fig:eff_params}(b), and their movements are shown in the supplementary video (\ref{appendix:suppl}).
The objective function values evaluated for soft bodies optimized with durations of $0.25$~\si{s}, $0.5$~\si{s}, and $1$~\si{s} were $-0.1525$, $-0.3474$, and $-1.056$, respectively.
\begin{figure}[tp]
    \centering
    \includegraphics[width=0.65\textwidth]{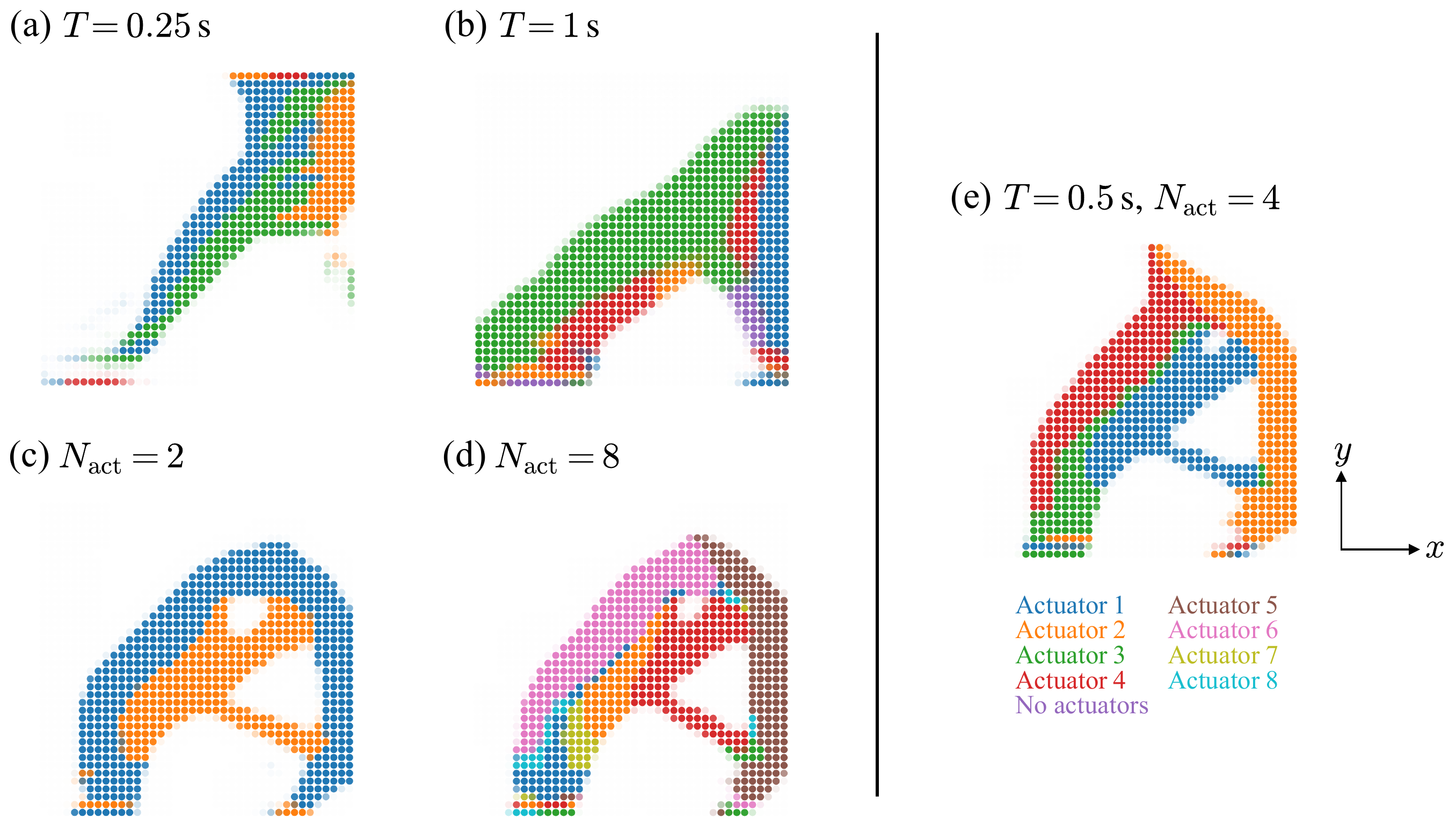}
    \caption{Structure and actuator layout of the optimized \textit{Walker} with different total simulation times $T$ (a, b) and different numbers of actuators $N_\mathrm{act}$ (c, d). The reference design (e) with $T = 0.5$~\si{s} and $N_\mathrm{act} = 4$ is shown for comparison. Refer to the supplementary video (\ref{appendix:suppl}) to see the soft bodies in action.}
    \label{fig:eff_params}
\end{figure}

Interestingly, the optimized soft body demonstrates structures and movements specialized in achieving the longest travel distance during each time interval.
With $T = 0.25$~\si{s}, the optimized soft body (Fig.~\ref{fig:eff_params}(a)) slides by kicking the ground backward with one leg, allowing it to rapidly convert gravitational potential energy into kinetic energy.
Consequently, it travels a longer distance over the assumed $0.25$~\si{s} than the others optimized with $T = 0.5$~\si{s} (Fig.~\ref{fig:eff_params}(e)) and $1$~\si{s} (Fig.~\ref{fig:eff_params}(b)).

The movement of the soft body optimized with $T = 1$~\si{s} (Fig.~\ref{fig:eff_params}(b)) resembles the running of a sea lion.
It achieves powerful forward propulsion by simultaneously kicking the ground with two legs while bending at the waist.
In comparison to the reference \textit{Walker} (Fig.~\ref{fig:eff_params}(e)), optimized with $T = 0.5$~\si{s}, this optimized soft body attains a higher maximum velocity through a longer duration of acceleration.
Accordingly, while the distance covered within the first $0.5$~\si{s} is shorter than the reference \textit{Walker}, the distance traveled over $1$~\si{s} exceeds the reference \textit{Walker}'s distance in $0.5$~\si{s} by over three times.

Next, we examined the impact of reducing ($N_\mathrm{act} = 2$) or increasing ($N_\mathrm{act} = 8$) the number of actuators from the default value of $4$.
Figs.~\ref{fig:eff_params}(c) and \ref{fig:eff_params}(d) provide a comparison of the optimized solutions.
The soft bodies optimized with different $N_\mathrm{act}$ display highly similar structures and movements.
Their performances are also nearly identical, although increasing $N_\mathrm{act}$ leads to slight improvements, as evidenced by objective function values of $-0.3282$, $-0.3474$, and $0.3502$ for $2$, $4$, and $8$ actuators, respectively.
This result yields two key findings:
\begin{enumerate*}[label=(\roman*)]
    \item the walking motion of \textit{Walker} can be achieved using only two actuators, and 
    \item incorporating multiple actuators allows for fine-tuning of the movement by subdividing the actuator layout, despite the potential introduction of redundancy.
\end{enumerate*}
\bibliography{refs_4dtopopt_R2}

\end{document}